\documentclass[prl,reprint,aps,twocolumn,nopacs,superscriptaddress,notitlepage]{revtex4-2}
\usepackage[utf8]{inputenc}
\usepackage[american,]{babel}
\usepackage[T1]{fontenc}
\usepackage[pdftex]{graphicx}  
\usepackage{graphicx, xcolor}
\usepackage{dcolumn}
\usepackage{bm}
\usepackage{amsmath,amsthm,amssymb}
\usepackage{hyperref}
\usepackage[T1,T2A]{fontenc}
\usepackage{xcolor}
\hypersetup{colorlinks,bookmarksopen,bookmarksnumbered,
    citecolor=blue,
    linkcolor=blue,
    pdfstartview=false,
    urlcolor=blue}
\usepackage{graphicx}
\usepackage{braket}
\usepackage{wrapfig}
\usepackage{soul}
\usepackage{mathtools}
\usepackage{float}
\usepackage{tabularx}

\renewcommand{\vec}{\mathbf}

\begin{document}
\title{Controlling entanglement at absorbing state phase transitions in random circuits
}

\author{Piotr Sierant}
\affiliation{ICFO-Institut de Ci\`encies Fot`oniques, The Barcelona Institute of Science and Technology, Av. Carl Friedrich Gauss 3, 08860 Castelldefels (Barcelona), Spain}
\author{Xhek Turkeshi}
\affiliation{JEIP, USR 3573 CNRS, Coll\`{e}ge de France, PSL Research University, 11 Place Marcelin Berthelot, 75321 Paris Cedex 05, France}
\date{\today}
\begin{abstract}
Many-body unitary dynamics interspersed with repeated measurements display a rich phenomenology hallmarked by measurement-induced phase transitions. Employing feedback-control operations that steer the dynamics toward an absorbing state, we study the entanglement entropy behavior at the absorbing state phase transition. For short-range control operations, we observe a transition between phases with distinct sub-extensive scalings of entanglement entropy. In contrast, the system undergoes a transition between volume-law and area-law phases for long-range feedback operations. The fluctuations of entanglement entropy and of the order parameter of the absorbing state transition are fully coupled for sufficiently strongly entangling feedback operations. In that case,  entanglement entropy inherits the universal dynamics of the absorbing state transition. This is, however, not the case for arbitrary control operations, and the two transitions are generally distinct. We quantitatively support our results by introducing a 
framework based on stabilizer circuits with classical flag labels. Our results shed new light on the problem of observability of measurement-induced phase transitions.
\end{abstract}

\maketitle

\paragraph{Introduction.} The technological surge in quantum simulators~\cite{Fraxanet22} and devices ~\cite{preskill2018quantumcomputingin,ferris2022quantumsimulationon} motivated the investigation of how monitoring affects the otherwise unitary quantum dynamics.
Repeated measurements enrich the evolution of many-body systems introducing non-unitary effects~\cite{ashida2020nonhermitianphysics,fisher2022randomquantumcircuits,potter2022entanglementdynamicsin,lunt2022quantumsimulationusing,rossini2021coherentanddissipative,chen2020emergentconformalsymmetry,chen2021nonunitaryfree}, including the celebrated measurement-induced phase transitions (\textbf{MIPTs})~\cite{cao2019entanglementina,skinner2019measurementinducedphase,li2018quantumzenoeffect,li2019measurementdrivenentanglement,chan2019unitaryprojective}.
These critical phenomena
leave fingerprints on non-linear functions of quantum trajectories but are unobservable on the average state level~\cite{vasseur2019entanglementtransitionsfrom,jian2020measurementinducedcriticality,nahum2921measurementandentanglement,bao2020theoryofthe,choi2020quantumerrorcorrection,gullans2020dynamicalpurificationphase,gullans2020scalableprobesof}. 
Extensive analytical and numerical investigations~\cite{
czischek2021simulating,han2022entanglementstructure,fidkowski2021howdynamicalquantum,altland2022dynamicsofmeasured,fuji2020measurementinducedquantum,biella2021manybodyquantumzeno,gopalakrishnan2021entanglementandpurification,jian2021yangleeedge,ippoliti2021entanglementphasetransitions,ippoliti2022fractallogarithmicand,lang2020entanglementtransitionin,li2021robustdecodingin,li2021statisticalmechanicsmodel,li2021entanglementdomainwalls,li2021statisticalmechanicsof,jian2021measurementinducedphase,lopezpiqueres2020meanfieldentanglement,jin2022KPZ,willsher2022measurementinducedphase,pizzi2022bridgingthegap,lyons2022auniversalcrossover,zhang2020nonuniversalentanglementlevel,zhang2021emergentreplica,zhang2022universalentanglementtransitions,zhou2021nonunitaryentanglementdynamics,bentsen2021measurementinducedpurification,yang2022entanglementphasetransitions,rossini2020measurementinduceddynamics,medina2021entanglementtransitionsfrom,lunt2020measurementinducedentanglement,nahum2020entanglementanddynamics,kelly2022coherencerequirementsfor,szyniszewski2019entanglementtransitionfrom,szyniszewski2020universalityofentanglement,parveen2020quantumzenoeffect,tang2020measurementinducedphase,tang2021quantumcriticalityin,shtanko2020classicalmodelsof,van2021entanglemententropyscaling}, in particular on random circuits~\cite{vijay2020measurementdrivenphase,zabalo2020criticalpropertiesof,fan2021selforganizederror,sang2021entanglementnegativityat,shi2020entanglementnegativityat,weinstein2022measurementinducedpower,li2021conformal,
liu2022measurementinducedentanglement,bao2021finitetimeteleportation,agrawal2022longrangebell,garratt2022measurementsconspirenonlocally,turkeshi2022measurementinducedcriticality,barratt2022transitions,dehgani2022neuralnetworkdecoders,iaconis2020measurementinducedphase,han2022measurementinducedcriticality,sierant2022universalbehaviorbeyond,zabalo2022operatorscalingdimensions,iaconis2021multifractalityinnonunitary,kalsi2022threefoldway,zabalo2022infiniterandomnesscriticality,weinstein2022scramblngtransitionin} and free fermion models~\cite{alberton2021entanglementtransitionin,buchhold2021effectivetheoryfor,turkeshi2021measurementinducedentanglement,turkeshi2022entanglementtransitionsfrom,turkeshi2022entanglementandcorrelation,gal2022volumetoarea,turkeshi2022enhancedentanglementnegativity,ladewig2022monitoredopenfermion,coppola2022growth,boorman2022diagonisticsofentanglement,botzung2021engineereddissipationinduced,
fleckenstein2022nonhermitiantopology,kells2021topologicaltransitionswith,szyniszewski2022disorderedmonitoredfree,piccitto2022entanglementtransitionsin,kawabata2022entanglementphasetransition,
minoguchi2022continuousgaussianmeasurements,muller2022measurementinduceddark}, outlined a phase diagram with a quantum error-correcting phase and a quantum Zeno phase.
To experimentally observe a MIPT, one needs to perform measurements on a fixed final state of the system. This requires a post-selection of experimental data for a registry of measurements performed throughout the system evolution. The registry is exponentially large in the spatio-temporal volume of the system, which leads to a problem of post-selection that hinders the possibilities of experimental observation of MIPTs. Apart from the brute-force experiment~\cite{koh2022experimentalrealizationof}, and fine-tuned setups~\cite{ippoliti2021postselectionfreeentanglement,lu2021spacetimeduality,noel2022measurementinducedquantum,claeys2022exactdynamicsin,li2022crossentropybenchmark,feng2022measurementinducedphase}, a systematic solution to this problem is yet missing. 

Feedback-control operations that condition the unitary operations on the measurement results have been recently proposed as a way of circumventing the post-selection problem.
Such operations induce non-trivial dynamics of the average state that could be linked with MIPT~\cite{iadecola2022dynamicalentanglementtransition, buchhold2022revealingmeasurementinduced}.
However, this is not generally the case, as either the control operations can suppress the MIPT~\cite{wang2022absenceofentanglement} or yield a distinct transition point~\cite{ravindranath2022entanglementsteeringin, odea2022entanglementandabsorbing}.
Thus, it is crucial to understand if and when the average dynamics can effectively encode the MIPT. Besides, feedback-controlled monitored systems are a largely unexplored many-body framework ~\cite{friedman2022measurementinducedphases} worthy of study on its own due to their potential applications.

\begin{figure}[t!]
    \centering
    \includegraphics[width=\columnwidth]{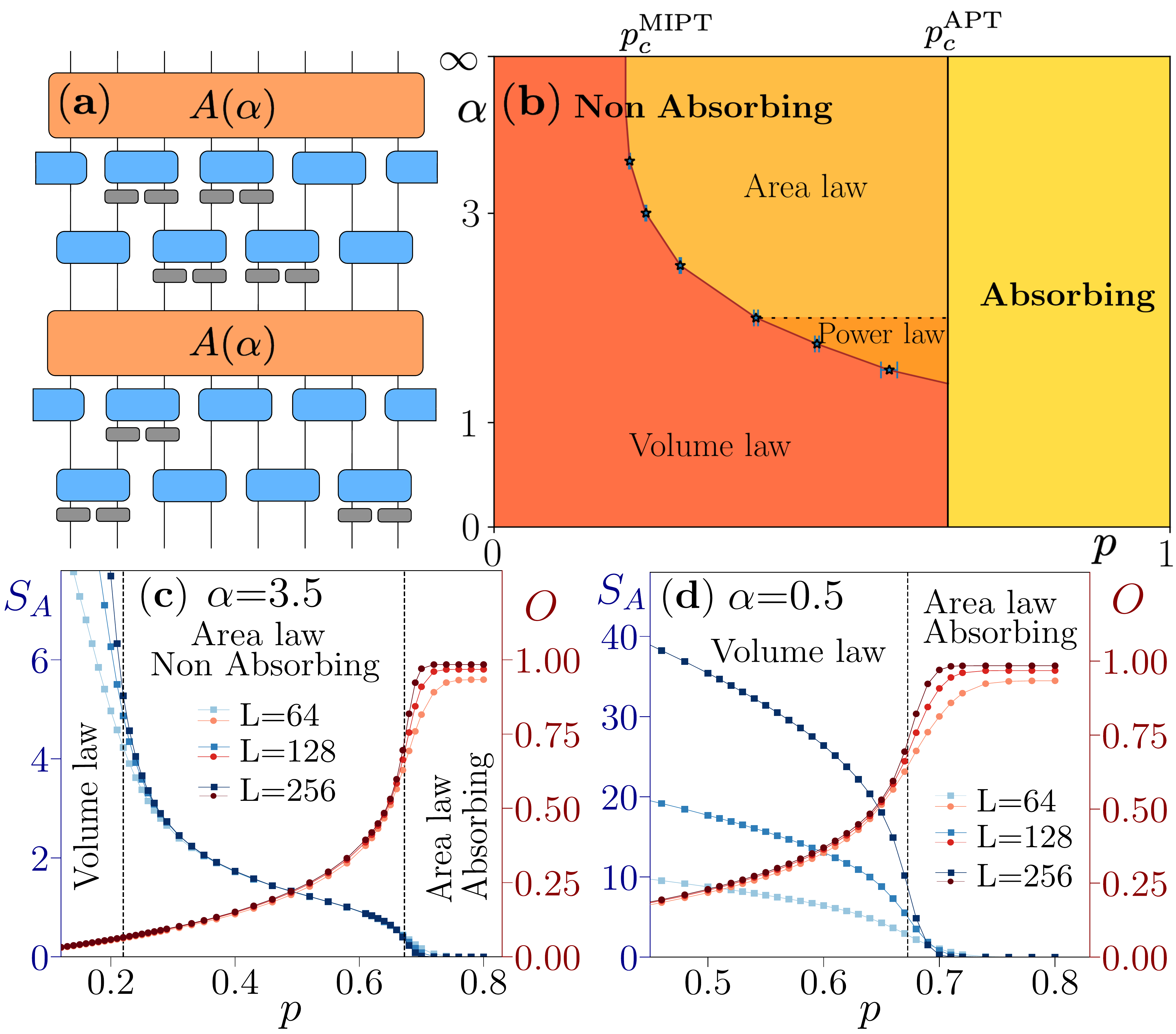}
    \caption{
    (\textbf{a}) Quantum circuit consisting of measurement operations (in grey), two-qubit unitary gates (in blue) preserving the $\ket{\uparrow \uparrow}$ state, and control operations $A(\alpha)$, built from two-body gates with range distributed according to a power-law with an exponent $\alpha$.
    (\textbf{b}) Dynamical phases of the circuit as a function of $\alpha$ and the measurement rate $p$. In blue are the critical points for the flagged Clifford circuits. 
    (\textbf{c}) The entanglement entropy $S_A$ and the order parameter $O=1-n_\mathrm{def}$ at a time $t=4L$ indicate, for $\alpha=3.5$, the presence of a volume-law, an area-law non-absorbing, and an area-law absorbing phases.
    (\textbf{d}) For long-range interactions, $\alpha=0.5$, we observe a direct transition between volume-law and area-law phases that coincides with the absorbing state phase transition.
    }
    \label{fig:cartoon_fig}
\end{figure}

This work demonstrates how feedback-control operations of variable range affect the dynamical phase diagram of a monitored quantum circuit with an absorbing stationary state. 
Using feedback-control, we show that the system generally exhibits two distinct, critical points: a MIPT and an absorbing state phase transition (\textbf{APT}). Signatures of the APT are visible in the behavior of entanglement entropy even when the control operations are short-range. Upon increase of the range, the MIPT approaches the APT. Eventually, for sufficiently long-range interactions, the two transitions coincide; see Fig.~\ref{fig:cartoon_fig}. 
The entanglement entropy inherits the universal dynamical behavior of the APT, provided the control operations do not present additional symmetries and are sufficiently entangling. In this case, the entanglement entropy reflects the average state dynamics, undergoing a transition between phases with distinct sub-extensive scalings or a volume-to-area-law transition,
depending on the range of control operations. 
Otherwise, the control operations may act as a relevant operator (in the renormalization group sense): then MIPT and APT
 belong to distinct universality classes.
 We support these arguments by introducing and numerically analyzing a class of stabilizer circuits with feedback-control.

\paragraph{Variable-range feedback-control of measurement-induced transitions.} 
We consider a quantum circuit on a 1D lattice, composed of interspersing layers of feedback-control unitary operations $\mathcal{A}(\alpha)$ and of projective measurements $\mathcal{M}$ of a complete set of commuting local hermitian operators $\{ M_i \}$. The parameter $\alpha$ encodes the range of $\mathcal{A}(\alpha)$, with ${\alpha=0}$ and ${\alpha=\infty}$ corresponding to all-to-all and short-range interactions, respectively.
The measurements act stochastically with a rate ${p\in[0,1]}$. 
Without feedback, the entangling unitary operations compete with disentangling local measurements. This yields an MIPT that is invisible on the level of the average state  $\rho_\mathrm{ave}$ that remains maximally mixed. 
The control operations condition the unitary gates on the results of the measurements in such a way that there exists an absorbing state $|\Psi_\mathrm{abs}\rangle$, left invariant by the circuit dynamics~\cite{buchhold2022revealingmeasurementinduced,odea2022entanglementandabsorbing,ravindranath2022entanglementsteeringin}. 
This introduces non-trivial dynamics of $\rho_\mathrm{ave}(t)$ as the system eventually relaxes to the state $\rho_\mathrm{ave}(\infty) = \ket{\Psi_\mathrm{abs}} \bra{\Psi_\mathrm{abs}} $.

The timescale at which $|\Psi_\mathrm{abs}\rangle$ is reached reveals a dynamic transition in the system. At small measurement rates, $p\simeq 0$, the unitary evolution scrambles the degrees of freedom, and the approach towards the absorbing state starts at timescales exponentially large 
in the system size $L$. In the opposite limit, $p\simeq 1$, the measurements steer the dynamics towards $\ket{\Psi_\mathrm{abs}}$, which is reached at times independent of $L$. The system undergoes the APT, which separates the two types of dynamical behavior at a measurement rate $p=p^{\mathrm{APT}}_c$.
The MIPT between a volume-law and an area-law (or sub-volume, power-law) phase occurs at $p=p^{\mathrm{MIPT}}_c$. The absorbing state is left invariant by the local measurements; hence it has, at most, an area-law entanglement. It follows that  $p_c^\mathrm{MIPT}\le p_c^\mathrm{APT}$.

In our setting, varying the range $\alpha$ of $\mathcal{A}(\alpha)$ changes the entanglement in the system. Still, it \textit{does not} affect the dynamics of $\rho_\mathrm{ave}(t)$. This allows us to control and investigate the entanglement at the APT.

\begin{figure}[t!]
    \centering
    \includegraphics[width=\columnwidth]{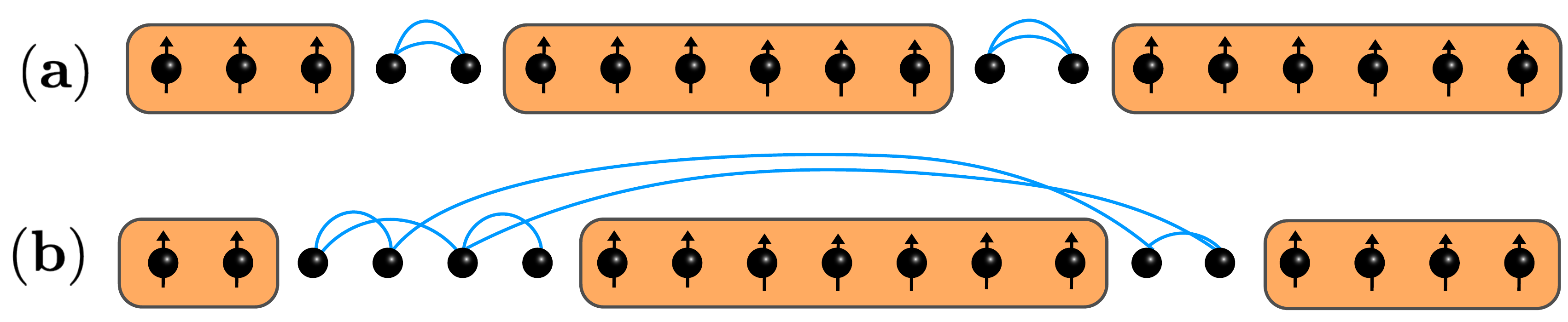}
    \caption{
        Short-range control operations generate entanglement locally between defects (\textbf{a}); long-range operations entangle clusters of distant defects (\textbf{b}).
    }
    \label{fig:domains}
\end{figure}

Above the APT, at $p> p_c^\mathrm{APT}$, the state of the system $\ket{\Psi_t}$ approaches $\ket{\Psi_\mathrm{abs}}$, which is a common eigenstate of $M_i$ with eigenvalues $m_i$. Below and at the APT, $p \leq p_c^\mathrm{APT}$, the state $\ket{\Psi_t}$ hosts a pattern of ordered domains at which the local measurements yield results the same as in $\ket{\Psi_\mathrm{abs}}$: $M_i\ket{\Psi_t} = m_i \ket{\Psi_t}$, interlayered with defect domains in which $\bra{\Psi_t}M_i\ket{\Psi_t} \neq m_i$, see Fig.~\ref{fig:domains}. 
The average density of defects $n_{\mathrm{def}}$, determined solely by $\rho_\mathrm{ave}(t)$, allows tracking down the APT.

In our construction, the feedback-control unitary operations $\mathcal{A}(\alpha)$ act non-trivially only on the clusters of defects and their immediate surrounding. 
Local unitary operations (${\alpha=\infty}$) can entangle only neighboring defects.
Indeed, the probability of collisions of defect domains is exponentially small in their separation. 
Thus, $\ket{\Psi_t}$ is a product state of  entangled defect clusters and ordered domains. 
Such a state follows an area-law since the entanglement is generated among the defects of a cluster of typical size $\xi_\mathrm{typ}$. 
In this case, therefore, ${p_c^\mathrm{MIPT}< p_c^\mathrm{APT}}$.
Conversely, long-range feedback-control unitary operations (${\alpha\to 0}$) may correlate arbitrarily distant defects, generating extensive entanglement. In particular, a global unitary operation acting on the defects will generate a volume-law entanglement entropy proportional to $n_\mathrm{def}$~\cite{supmat}. 
In that case, the APT separates a state with volume-law entanglement from the area-law entangled $\ket{\psi_\mathrm{abs}}$, and thus ${p_c^\mathrm{MIPT}= p_c^\mathrm{APT}}$. Hence, the entanglement entropy dynamics reflects $\rho_\mathrm{ave}(t)$ for any $\alpha$.
Nevertheless, MIPT and APT may reveal equivalent or different universal content depending on $\mathcal{A}(\alpha)$. 
Below, we provide examples where $\mathcal{A}(\alpha)$ plays the role of an irrelevant field leading to coinciding critical exponents. We also find setups for which $\mathcal{A}(\alpha)$ is a relevant field.

\begin{figure*}
    \centering
    \includegraphics[width=1\linewidth]{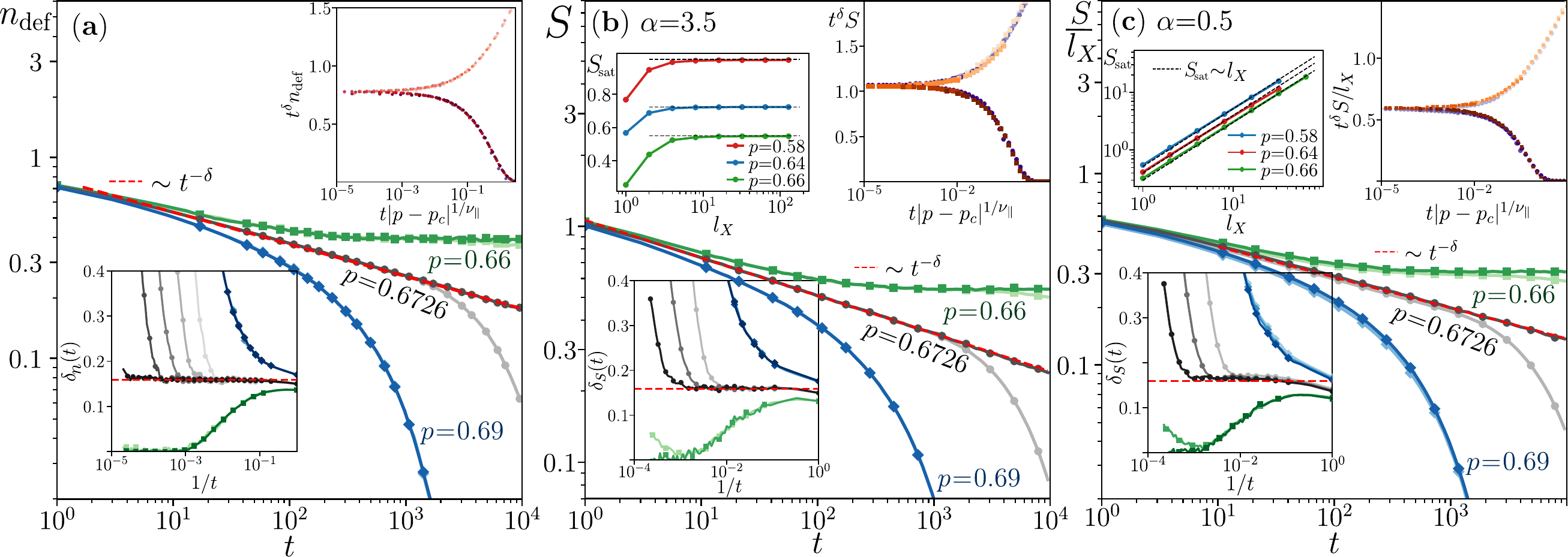}
    \caption{Average density of defects, $n_\mathrm{def}$,  (\textbf{a}), and average entanglement entropy $S$ in the short-range (\textbf{b}) and long-range regime (\textbf{c}) close to the APT. Darker colors reflect larger system sizes in the range $L=128\div 1024$, while the red dashed lines indicate the DP universality. In (\textbf{a}), the bottom inset illustrates the development of the DP critical exponent $\delta_{DP}\simeq 0.1595(1)$, while the top-right inset shows a data collapse with $\delta_n=0.159(1)$, $p_c=p_c^\mathrm{APT}=0.6726(2)$ and $\nu_\parallel=1.73(1)$. 
    Similarly, (\textbf{b}) and (\textbf{c}) present the behavior of $\delta_S(t)$ for an area-to-area and a volume-to-area transitions for subsystem size $l_X=8$, respectively. In (\textbf{b}), the collapse is for $\delta_S=0.16(1)$, $p_c=0.6722(5)$ and $\nu_\parallel=1.73(8)$. In (\textbf{c}), the critical values are $\delta_S=0.16(1)$,  $p_c=0.672(1)$ and $\nu_\parallel=1.70(5)$.
    The top-left insets in (\textbf{b}) and (\textbf{c}) show the saturation value $S_{\mathrm{sat}}$ of $S(t)$ for
    $p<p_c^\mathrm{APT}$ with $l_X$. For $\alpha=3.5$, $S_{\mathrm{sat}}$ becomes independent of $l_X$ (an area-law), whereas for $\alpha=0.5$, $S_{\mathrm{sat}} \sim l_X$ (a volume-law). 
    }
    \label{fig2}
\end{figure*}

\paragraph{Implementations on stabilizer circuits.} 
To illustrate the above ideas, we consider 1D stabilizer circuits, built of Clifford unitary gates and projective measurements, efficiently simulable using the Gottesmann-Knill theorem~\cite{aaronson2004improvedsimulationof,gidney2021stimfaststabilizer,nielsen00, Gross2021}. 
We consider projective measurements of the on-site magnetization, $M_i\equiv Z_i$ \footnote{We denote the Pauli matrices by $X_i$, $Y_i$,  $Z_i$; $\ket{\uparrow_i}$ and $\ket{\downarrow_i}$ are the +1 and -1 eigenvectors of $Z_i$. } and $\mathcal{A}(\alpha)$ built of two-body unitary gates. For convenience, we unpack the unitary and measurement layers as $\mathcal{A}(\alpha)\mathcal{M}= A(\alpha)U_\mathrm{e} M_\mathrm{e} U_{o} M_{o}$, with $A(\alpha)=\prod_{(i,j)\in I_\alpha} U_{i,j} $ and $U_{\mathrm{e/o}}=\prod_{i \in \mathrm{e/o}} U_{i,i+1}$, and $M_{\mathrm{e/o}}=\prod_{i \in \mathrm{e/o}} (P^z_i P^z_{i+1})^{r_{i}}$, see Fig.~\ref{fig:cartoon_fig}. 
Here $U_{i,j} $ are two body Clifford gates, $I_\alpha$ is a set of indices, e,o denote the sets of even and odd lattice sites, $P^z_i$ is a projector on eigenstates of $Z_i$, and $r_{i}=1$ with probability $p$ (otherwise $r_i=0$). Such a circuit structure allows for a neat separation of MIPT and APT within the short-range regime.

We fix $|\Psi_\mathrm{abs}\rangle=\bigotimes_{i} |\uparrow_i\rangle$ as the absorbing state. To that end, we require the unitary gates to act trivially on ${|\uparrow\uparrow\rangle}$ state. Two body Clifford gates that preserve this state do not generate entanglement (i.e., do not create Bell pairs) from eigenstates of $Z_i$. To overcome this limitation, we introduce a classical label, a \textit{flag}, $f_i=0,1$ for each site. If $f_i f_j = 0$, the $U_{i,j}$ is randomly chosen from the full Clifford group, and after its action, both $f_i$ and $f_j$ are set to $0$. Otherwise, i.e. when $f_i=f_j=1$, the $U_{i,j}$ is an identity operation and $f_{i/j}$ remains untouched. Such a feedback mechanism ensures that our stabilizer circuit can generate extensive entanglement in the presence of $P^z_i$, while preserving the $|\Psi_\mathrm{abs}\rangle$ state. Initially, the state of the system is $\bigotimes_i |\downarrow_i\rangle $ and all $f_i=0$. 
After each measurement, we set $f_i=1$ if the result is $+1$, in line with $|\Psi_\mathrm{abs}\rangle$, and put $f_i=0$ otherwise. 
The unitary gates are organized into layers $U_\mathrm{e/o}$ of local gates and into a control operation $A(\alpha)$. The set $I_\alpha$ consists of $N_{\mathrm{flag}}$ pairs $(i,j)$ drawn with probability $P(r) \sim 1/r^\alpha $ among the sites at which $f_i=0$, where $r=\mathrm{min}\{|i-j|, L-|i-j\}$ (periodic boundary conditions are assumed) with $N_{\mathrm{flag}}$ being the number of sites at which $f_i=0$.

In~\cite{supmat}, we prove that the dynamics of the average state  under our stabilizer circuit is independent of $\alpha$ and fully captured by a probabilistic cellular automaton (similarly to \cite{odea2022entanglementandabsorbing}), enabling an efficient simulation of the density of defects $n_{\mathrm{def}} = \sum_{i=1}^L \langle \Psi_t| (1-Z_i)|\Psi_t\rangle/(2L)$. 
Additionally, we study the entanglement entropy \cite{hamma2004,Hamma_2005} defined as $S(t) = -\mathrm{tr}(\rho_X\log_2\rho_X)$ where $\rho_X = \mathrm{tr}_Y |\Psi_t\rangle\langle\Psi_t|$ and $X\cup Y$ is a bipartition of the lattice.
\paragraph{Numerical results.}
Without feedback, MIPTs in the presence of long-range interactions were studied in~\cite{minato2022fateofmeasurementinduced,block2022measurementinducedtransition,sharma2022measurementinducedcriticality,hashizum2022measurementinducedphase}. Here, we analyze the phase diagram of the stabilizer circuit with feedback-control, varying $\alpha$ and $p$. 
In Fig.~\ref{fig2}(a), we observe the APT reflected in dynamics of $n_{\mathrm{def}}$. For $p=0.66$, $n_\mathrm{def}$ attains a non-zero stationary value, whereas, for $p=0.69$, $n_\mathrm{def}$ decays exponentially in time. At $p=0.6726$ we observe a power-law behavior $n_\mathrm{def}(t) \sim t^{-\delta_{DP}}$ with exponent $\delta_{DP}$ characteristic for the direct percolation (\textbf{DP}) universality class \cite{Wang13}. We consider $\delta_n(t)=\log_b({n_\mathrm{def}(bt)}/{n_\mathrm{def}(t)})$ \cite{Mendon11} with $b=4$, to quantify $n_\mathrm{def}(t)$ at the APT. We find a plateau in $\delta_n(t)$ at $\delta_{DP}$ that extends to longer times with increasing $L$; see the lower inset in Fig.~\ref{fig2}(a). To confirm the DP universality class of the dynamics, we show, in the upper inset of  Fig.~\ref{fig2}(a), a collapse of $n_{\mathrm{def}} t^{\delta_{DP}}$ as a function of $|p-p^{\mathrm{APT}}_c|t^{\nu_{\parallel}}$, where $p^{\mathrm{APT}}_c=0.6726(2)$ and $\nu_{\parallel}=1.73(1)$.

We find that the entanglement entropy $S(t)$ behaves analogously for short-range interactions, $\alpha=3.5$, see Fig.~\ref{fig2}(b): $S(t)$ saturates at $p< p^{\mathrm{APT}}_c$, decays according to a power-law at $p\approx p^{\mathrm{APT}}_c$ and decays exponentially in time for $p> p^{\mathrm{APT}}_c$.
Plotting $\delta_S(t)=\log_b({S(bt)}/{S(t)})$, we observe a development of the characteristic plateau at 
$\delta_{DP}$ for $ p = p^{\mathrm{APT}}_c$ which indicates that $S(t)$ inherits the universal behavior of DP class. This is further confirmed by the collapse of $t^{\delta_{DP}} S(t)$ versus $|p-p^{\mathrm{APT}}_c|t^{\nu_{\parallel}}$. Importantly, similar behavior is also observed for long-range control operation, $\alpha=0.5$, as shown in Fig.~\ref{fig2}(c). For $\alpha=3.5$, $S(t)$ follows an area-law, saturating to constant independent of $l_X$ (the size of the subsystem $X$), whereas for $\alpha=0.5$, we observe a volume-law, $S\propto l_X$.

\begin{figure}
    \centering
    \includegraphics[width=1\columnwidth]{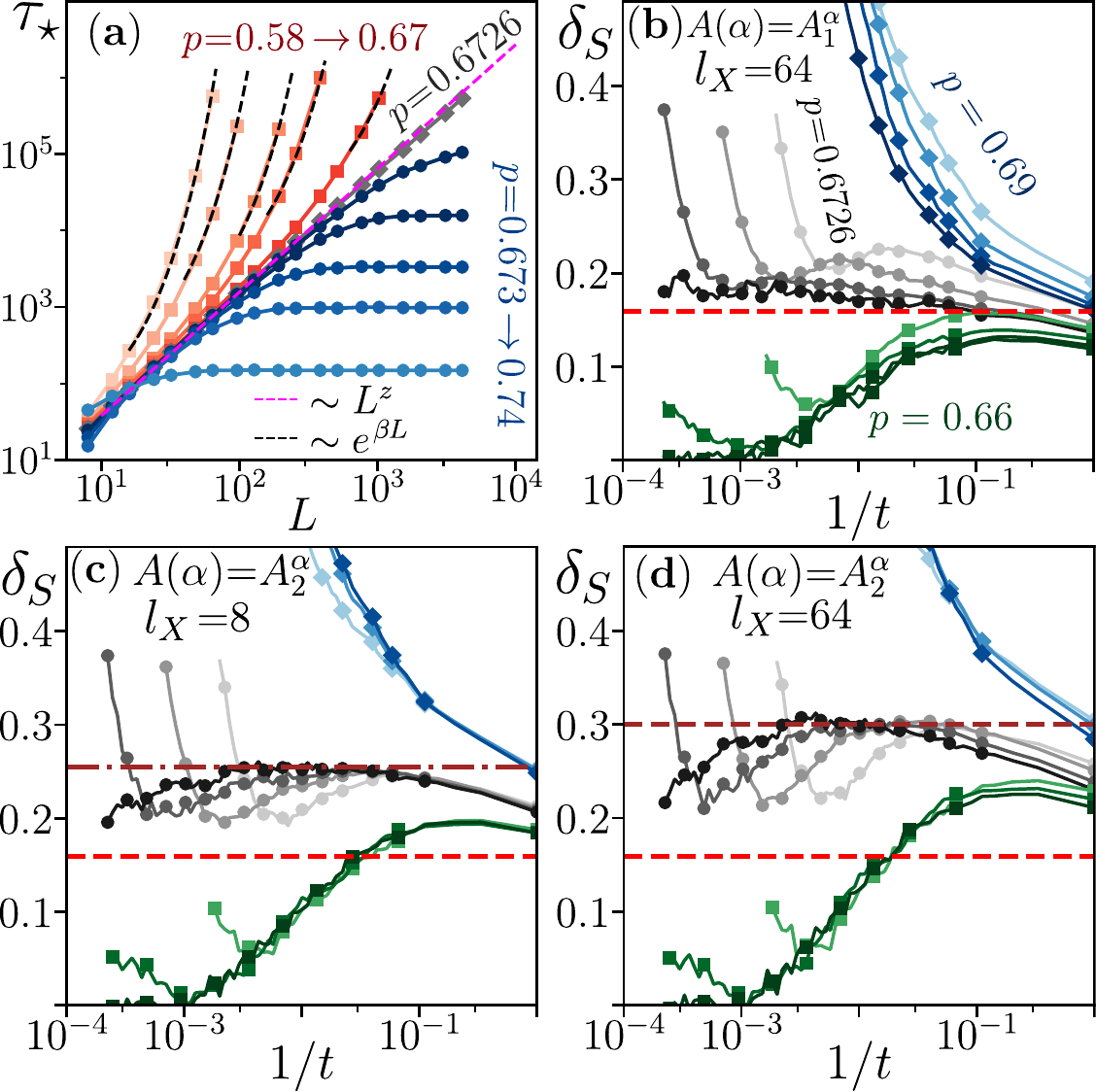}
    \caption{(\textbf{a}) Times $\tau_\star$ 
    of the start of decay to the absorbing state at system size $L$. For $p<p_c^\mathrm{APT}$,  $\tau_\star$ is exponentially diverging with $L$, whereas for $p>p_c^\mathrm{APT}$ it saturates with $L$. (\textbf{b}) Exponent $\delta_S$ for a subsystem size $l_X=64$, cf. Fig.~\ref{fig2}(c). In (\textbf{c,d}), we present $\delta_S$ for $l_X=8,64$ for a different protocol $A_2^\alpha$, see Text; $\delta_S$ saturates to value increasing with $l_X$, denoted by the dashed brown lines:  $\delta=0.255$ in (\textbf{c}) and $\delta=0.3$ in (\textbf{d}). 
    This behavior is distinct from  the $A_1^\alpha$ case. 
    In (\textbf{b,c,d}), darker colors refer to larger system sizes in the range $L=128\div 1024$, and the dashed red line is the DP reference $\delta=0.1595$. \label{fig:3}
}
\end{figure}

To understand in what sense the system undergoes an entanglement transition, we consider a timescale $\tau_\star$ at which $n_{\mathrm{def}}(t)$ and $S(t)$ decay to the absorbing state values at finite system size $L$  \footnote{For $p<p^{\mathrm{APT}}_c$ we find $\tau_\star$ as time at which $n_{\mathrm{def}}(t)$ decays to $1/2$ of the saturation value observed at times $\propto L$; for $p\approx p^{ \mathrm{APT} }_c$, $\tau_\star$ is the time at which $n_{\mathrm{def}}(t)$ is equal to $c t^{\delta}/2$, where $c t^{\delta}$ is a power-law fitted at times $\propto L$; for $p>p^{\mathrm{APT}}_c$, $n_{\mathrm{def}}(t)$ is equal to $n_{\mathrm{def}}(100)/2$ at $\tau_\star$.}. 
For $p<p^{\mathrm{APT}}_c$, we observe an exponential divergence of $\tau_\star$ with $L$, Fig.~\ref{fig:3}(a). At the APT, $\tau_\star \sim L^z$ with exponent $z=1.62(10)$ consistent with the DP class. Finally, at $p>p^{\mathrm{APT}}_c$, the timescale $\tau_\star$ approaches a constant independent of $L$. This behavior of $\tau_\star$ implies that at times proportional to system size $L$, we observe a transition between phases with different entanglement structures: for long-range interactions, $\alpha < \alpha_c \approx 1.3$, there is a MIPT between volume-law phase and a zero-law  phase (with vanishing $S(t)$ at $t\propto L$) at $p=p^{ \mathrm{APT} }_c$~\footnote{
Without feedback, the MIPT is expected to shift toward $p_c^\mathrm{MIPT}\to 1$ while $\alpha\to 1$. Feedback constrains the critical point $p_c^\mathrm{MIPT}$ to reach most $p_c^\mathrm{APT}$. Hence the existence of a (model dependent) $\alpha_c\simeq 1.3$.
}. For short-range interactions $\alpha>2$, there is a transition between an area-law phase and a zero-law phase; for arbitrary $\alpha$ results, see \cite{supmat}.
At finite $L$, we observe deviations from DP behavior when the subsystem size $l_X$ increases, cf. Fig~\ref{fig:3}(b) and Fig~\ref{fig2}(c). Nevertheless, even for the larger $l_X$, the $\delta_S(t)$ approaches $\delta_{DP}$ with increasing $L$, so we expect that the universal behavior of $S(t)$ appears at all values of $l_X$ for $L\to \infty$.

The results reported so far were obtained for the control operation $A(\alpha)\equiv A^{\alpha}_1$, that consists of $N_{\mathrm{flag}}$ two-body gates $U_{i,j}$. Analogous results hold for $A(\alpha)$ that contains more two-body gates (see Ref.~\cite{supmat}, where we also consider a global operation). This behavior may, however, change for other choices of $A(\alpha)$. As an example, we consider $A(\alpha)= A^{\alpha}_2$ in which the 
set $I_\alpha$ contains $ N_{\mathrm{flag}}^2/L$ gates, distributed according to a power-law with exponent $\alpha$. While this choice is sufficient to stabilize the volume-law scaling of $S(t)$ at $p < p^{ \mathrm{APT} }_c$, $A^{\alpha}_2$ is \textit{not} sufficiently strongly entangling to couple the fluctuation of $n_{\mathrm{def}}$ to the behavior of $S(t)$ at the APT (note that $ N_{\mathrm{flag}} \ll L$ at late times at $p = p^{ \mathrm{APT} }_c$). In consequence, we observe that $S(t)$ decays faster than $n_{\mathrm{def}}(t)$, according to a power-law with exponent $\delta > \delta_{DP}$, that increases with the subsystem size $l_X$, see Fig.~\ref{fig:3}(c,d).

\paragraph{Discussion and conclusion.}
We analyzed the effect of variable-range control operations in monitored quantum circuits with an absorbing state. Similarly to the purification transition~\cite{gullans2020dynamicalpurificationphase,fidkowski2021howdynamicalquantum}, MIPT and APT are reflected in the dynamical behavior of  entanglement entropy, and density of defects before the system finally reaches the absorbing state. 
Although the arguments 
in this manuscript are general, we introduced and employed 
a new methodological framework (based on stabilizer circuits supplemented with a classical label) to support our arguments. These flagged stabilizer circuits allow for numerical simulations of much larger systems than in~\cite{odea2022entanglementandabsorbing,ravindranath2022entanglementsteeringin}.
For short-range feedback operations, the system presents separated MIPT and APT (analogously to MIPT and ordering transitions in \cite{sierant2022dissipativefloquet,barratt2022fieldtheoryof,agrawal2022entanglmentandchargesharpening,oshima2022chargefluctuationand}), whereas for long-range interaction the critical points coincide, see also ~\cite{iadecola2022dynamicalentanglementtransition}. 
In the former scenario, the APT is manifested in a transition of the entanglement entropy between two sub-extensive scaling laws. 
Conversely, in the latter case, the critical behavior depends on the control operation properties. 
If the feedback operations are sufficiently entangling, the order parameter fluctuations at APT are fully coupled to the entanglement entropy yielding the same universal dynamics. 
Otherwise, inequivalent universality classes are expected. 
Varying the type of feedback-control operations, we observed both cases.
We expect that our arguments also hold for many-body Floquet circuits and in higher dimensions~\cite{turkeshi2020measurementinducedcriticality,lunt2021measurementinducedcriticality,sierant2022measurementinducedphase}.

A key conclusion of our work is that the post-selection problem is generally \textit{not} solved using absorbing-state transitions, as it requires some fine-tuning of the control operations. In particular, even when the APT and MIPT coincide, it is not a priori clear whether their critical properties are the same or not.
At the same time, we do not exclude that a systematic solution to the post-selection problem may exist in special systems, for instance, free fermions~\cite{buchhold2022revealingmeasurementinduced}, or employing more complex feedback-control operations (e.g., not based on an absorbing state).
In any case, it would be interesting to explore the APT in long-range systems experimentally, for instance, in trapped ion experiments~\cite{pagano1,pagano2,pagano3,pagano4,pagano5,pagano6}. 
Lastly, feedback-control using absorbing states can be employed for state-preparation protocols, for instance, for topological~\cite{lavasani2021measurementinducedtopological,lavasani2021topologicalorderand,zhu2022nishimoriscat,sang2021measurementprotectedquantum,lee2022decodingmeasurementpreparedquantum,klocke2022topologicalorderand} or metrologically relevant absorbing states~\cite{smerzi1,silvia1,silvia,doley,doley1,Dooley2018}.
A more general exploration of how feedback operations affect many-body monitored systems is an interesting direction for further studies.

\textit{Note added: During the completion of this work, we became aware of a closely related work on Clifford circuits by  L. Piroli, Y. Li, R. Vasseur, and A. Nahum appearing in the same arXiv posting~\cite{yaodong2023}. }

\begin{acknowledgments}
\textit{Adcknowledgments.} 
We thank M. Dalmonte, R. Fazio, M. Lewenstein, G. Pagano, S. Sharma, and M. Schirò for discussions and collaborations on related topics. 
We thank T. Iadecola, N. O'Dea, S. Gopalakrishnan, and Z. Yang for insightful discussions on related topics. 
XT thanks S. Pappalardi for the discussions on metrology.
XT acknowledges support from the ANR grant “NonEQuMat.”
(ANR-19-CE47-0001) and computational resources on the Coll\'ge de France IPH cluster. 
PS acknowledges support from: ERC AdG NOQIA; Ministerio de Ciencia y
Innovation Agencia Estatal de Investigaciones (PGC2018-097027-B-
I00/10.13039/501100011033,  CEX2019-000910-S/10.13039/501100011033, Plan National
FIDEUA PID2019-106901GB-I00, FPI, QUANTERA MAQS PCI2019-111828-2,
QUANTERA DYNAMITE PCI2022-132919, Proyectos de I+D+I “Retos Colaboración”
QUSPIN RTC2019-007196-7); MICIIN with funding from European Union
NextGenerationEU(PRTR-C17.I1) and by Generalitat de Catalunya; Fundació Cellex;
Fundació Mir-Puig; Generalitat de Catalunya (European Social Fund FEDER and CERCA
program, AGAUR Grant No. 2017 SGR 134, QuantumCAT \ U16-011424, co-funded by
ERDF Operational Program of Catalonia 2014-2020); EU Horizon 2020 FET-OPEN OPTOlogic (Grant No 899794); EU Horizon Europe Program (Grant Agreement 101080086 — NeQST), National Science Centre, Poland (Symfonia Grant No. 2016/20/W/ST4/00314); ICFO Internal
“QuantumGaudi” project; European Union’s Horizon 2020 research and innovation program
under the Marie-Skłodowska-Curie grant agreement No 101029393 (STREDCH) and No
847648 (“La Caixa” Junior Leaders fellowships ID100010434: LCF/BQ/PI19/11690013,
LCF/BQ/PI20/11760031, LCF/BQ/PR20/11770012, LCF/BQ/PR21/11840013). Views and
opinions expressed in this work are, however, those of the author(s) only and do not
necessarily reflect those of the European Union, European Climate, Infrastructure and
Environment Executive Agency (CINEA), nor any other granting authority. Neither the
European Union nor any granting authority can be held responsible for them.
\end{acknowledgments}


\begin{thebibliography}{158}%
\makeatletter
\providecommand \@ifxundefined [1]{%
 \@ifx{#1\undefined}
}%
\providecommand \@ifnum [1]{%
 \ifnum #1\expandafter \@firstoftwo
 \else \expandafter \@secondoftwo
 \fi
}%
\providecommand \@ifx [1]{%
 \ifx #1\expandafter \@firstoftwo
 \else \expandafter \@secondoftwo
 \fi
}%
\providecommand \natexlab [1]{#1}%
\providecommand \enquote  [1]{``#1''}%
\providecommand \bibnamefont  [1]{#1}%
\providecommand \bibfnamefont [1]{#1}%
\providecommand \citenamefont [1]{#1}%
\providecommand \href@noop [0]{\@secondoftwo}%
\providecommand \href [0]{\begingroup \@sanitize@url \@href}%
\providecommand \@href[1]{\@@startlink{#1}\@@href}%
\providecommand \@@href[1]{\endgroup#1\@@endlink}%
\providecommand \@sanitize@url [0]{\catcode `\\12\catcode `\$12\catcode
  `\&12\catcode `\#12\catcode `\^12\catcode `\_12\catcode `\%12\relax}%
\providecommand \@@startlink[1]{}%
\providecommand \@@endlink[0]{}%
\providecommand \url  [0]{\begingroup\@sanitize@url \@url }%
\providecommand \@url [1]{\endgroup\@href {#1}{\urlprefix }}%
\providecommand \urlprefix  [0]{URL }%
\providecommand \Eprint [0]{\href }%
\providecommand \doibase [0]{https://doi.org/}%
\providecommand \selectlanguage [0]{\@gobble}%
\providecommand \bibinfo  [0]{\@secondoftwo}%
\providecommand \bibfield  [0]{\@secondoftwo}%
\providecommand \translation [1]{[#1]}%
\providecommand \BibitemOpen [0]{}%
\providecommand \bibitemStop [0]{}%
\providecommand \bibitemNoStop [0]{.\EOS\space}%
\providecommand \EOS [0]{\spacefactor3000\relax}%
\providecommand \BibitemShut  [1]{\csname bibitem#1\endcsname}%
\let\auto@bib@innerbib\@empty
\bibitem [{\citenamefont {Fraxanet}\ \emph {et~al.}(2022)\citenamefont
  {Fraxanet}, \citenamefont {Salamon},\ and\ \citenamefont
  {Lewenstein}}]{Fraxanet22}%
  \BibitemOpen
  \bibfield  {author} {\bibinfo {author} {\bibfnamefont {J.}~\bibnamefont
  {Fraxanet}}, \bibinfo {author} {\bibfnamefont {T.}~\bibnamefont {Salamon}},\
  and\ \bibinfo {author} {\bibfnamefont {M.}~\bibnamefont {Lewenstein}},\
  }\href@noop {} {} (\bibinfo {year} {2022}),\ \Eprint
  {https://arxiv.org/abs/2204.08905} {arXiv:2204.08905} \BibitemShut {NoStop}%
\bibitem [{\citenamefont {Preskill}(2018)}]{preskill2018quantumcomputingin}%
  \BibitemOpen
  \bibfield  {author} {\bibinfo {author} {\bibfnamefont {J.}~\bibnamefont
  {Preskill}},\ }\href {https://doi.org/10.22331/q-2018-08-06-79} {\bibfield
  {journal} {\bibinfo  {journal} {{Quantum}}\ }\textbf {\bibinfo {volume}
  {2}},\ \bibinfo {pages} {79} (\bibinfo {year} {2018})}\BibitemShut {NoStop}%
\bibitem [{\citenamefont {Ferris}\ \emph {et~al.}()\citenamefont {Ferris},
  \citenamefont {Rasmusson}, \citenamefont {Bronn},\ and\ \citenamefont
  {Lanes}}]{ferris2022quantumsimulationon}%
  \BibitemOpen
  \bibfield  {author} {\bibinfo {author} {\bibfnamefont {K.~J.}\ \bibnamefont
  {Ferris}}, \bibinfo {author} {\bibfnamefont {A.~J.}\ \bibnamefont
  {Rasmusson}}, \bibinfo {author} {\bibfnamefont {N.~T.}\ \bibnamefont
  {Bronn}},\ and\ \bibinfo {author} {\bibfnamefont {O.}~\bibnamefont {Lanes}},\
  }\href@noop {} {}\Eprint {https://arxiv.org/abs/2209.02795}
  {arXiv:2209.02795} \BibitemShut {NoStop}%
\bibitem [{\citenamefont {Ashida}\ \emph {et~al.}(2020)\citenamefont {Ashida},
  \citenamefont {Gong},\ and\ \citenamefont
  {Ueda}}]{ashida2020nonhermitianphysics}%
  \BibitemOpen
  \bibfield  {author} {\bibinfo {author} {\bibfnamefont {Y.}~\bibnamefont
  {Ashida}}, \bibinfo {author} {\bibfnamefont {Z.}~\bibnamefont {Gong}},\ and\
  \bibinfo {author} {\bibfnamefont {M.}~\bibnamefont {Ueda}},\ }\href
  {https://doi.org/10.1080/00018732.2021.1876991} {\bibfield  {journal}
  {\bibinfo  {journal} {Adv. Phys.}\ }\textbf {\bibinfo {volume} {69}},\
  \bibinfo {pages} {249} (\bibinfo {year} {2020})}\BibitemShut {NoStop}%
\bibitem [{\citenamefont {Fisher}\ \emph {et~al.}()\citenamefont {Fisher},
  \citenamefont {Khemani}, \citenamefont {Nahum},\ and\ \citenamefont
  {Vijay}}]{fisher2022randomquantumcircuits}%
  \BibitemOpen
  \bibfield  {author} {\bibinfo {author} {\bibfnamefont {M.~P.~A.}\
  \bibnamefont {Fisher}}, \bibinfo {author} {\bibfnamefont {V.}~\bibnamefont
  {Khemani}}, \bibinfo {author} {\bibfnamefont {A.}~\bibnamefont {Nahum}},\
  and\ \bibinfo {author} {\bibfnamefont {S.}~\bibnamefont {Vijay}},\
  }\href@noop {} {}\Eprint {https://arxiv.org/abs/2207.14280}
  {arXiv:2207.14280} \BibitemShut {NoStop}%
\bibitem [{\citenamefont {Potter}\ and\ \citenamefont
  {Vasseur}(2022)}]{potter2022entanglementdynamicsin}%
  \BibitemOpen
  \bibfield  {author} {\bibinfo {author} {\bibfnamefont {A.~C.}\ \bibnamefont
  {Potter}}\ and\ \bibinfo {author} {\bibfnamefont {R.}~\bibnamefont
  {Vasseur}},\ }\bibinfo {title} {Entanglement dynamics in hybrid quantum
  circuits},\ in\ \href {https://doi.org/10.1007/978-3-031-03998-0_9} {\emph
  {\bibinfo {booktitle} {Entanglement in Spin Chains: From Theory to Quantum
  Technology Applications}}},\ \bibinfo {editor} {edited by\ \bibinfo {editor}
  {\bibfnamefont {A.}~\bibnamefont {Bayat}}, \bibinfo {editor} {\bibfnamefont
  {S.}~\bibnamefont {Bose}},\ and\ \bibinfo {editor} {\bibfnamefont
  {H.}~\bibnamefont {Johannesson}}}\ (\bibinfo  {publisher} {Springer},\
  \bibinfo {address} {Cham},\ \bibinfo {year} {2022})\ pp.\ \bibinfo {pages}
  {211--249}\BibitemShut {NoStop}%
\bibitem [{\citenamefont {Lunt}\ \emph {et~al.}(2022)\citenamefont {Lunt},
  \citenamefont {Richter},\ and\ \citenamefont
  {Pal}}]{lunt2022quantumsimulationusing}%
  \BibitemOpen
  \bibfield  {author} {\bibinfo {author} {\bibfnamefont {O.}~\bibnamefont
  {Lunt}}, \bibinfo {author} {\bibfnamefont {J.}~\bibnamefont {Richter}},\ and\
  \bibinfo {author} {\bibfnamefont {A.}~\bibnamefont {Pal}},\ }\bibinfo {title}
  {Quantum simulation using noisy unitary circuits and measurements},\ in\
  \href {https://doi.org/10.1007/978-3-031-03998-0_10} {\emph {\bibinfo
  {booktitle} {Entanglement in Spin Chains: From Theory to Quantum Technology
  Applications}}},\ \bibinfo {editor} {edited by\ \bibinfo {editor}
  {\bibfnamefont {A.}~\bibnamefont {Bayat}}, \bibinfo {editor} {\bibfnamefont
  {S.}~\bibnamefont {Bose}},\ and\ \bibinfo {editor} {\bibfnamefont
  {H.}~\bibnamefont {Johannesson}}}\ (\bibinfo  {publisher} {Springer},\
  \bibinfo {address} {Cham},\ \bibinfo {year} {2022})\ pp.\ \bibinfo {pages}
  {251--284}\BibitemShut {NoStop}%
\bibitem [{\citenamefont {Rossini}\ and\ \citenamefont
  {Vicari}(2021)}]{rossini2021coherentanddissipative}%
  \BibitemOpen
  \bibfield  {author} {\bibinfo {author} {\bibfnamefont {D.}~\bibnamefont
  {Rossini}}\ and\ \bibinfo {author} {\bibfnamefont {E.}~\bibnamefont
  {Vicari}},\ }\href
  {https://doi.org/https://doi.org/10.1016/j.physrep.2021.08.003} {\bibfield
  {journal} {\bibinfo  {journal} {Physics Reports}\ }\textbf {\bibinfo {volume}
  {936}},\ \bibinfo {pages} {1} (\bibinfo {year} {2021})},\ \bibinfo {note}
  {coherent and dissipative dynamics at quantum phase transitions}\BibitemShut
  {NoStop}%
\bibitem [{\citenamefont {Chen}\ \emph {et~al.}(2020)\citenamefont {Chen},
  \citenamefont {Li}, \citenamefont {Fisher},\ and\ \citenamefont
  {Lucas}}]{chen2020emergentconformalsymmetry}%
  \BibitemOpen
  \bibfield  {author} {\bibinfo {author} {\bibfnamefont {X.}~\bibnamefont
  {Chen}}, \bibinfo {author} {\bibfnamefont {Y.}~\bibnamefont {Li}}, \bibinfo
  {author} {\bibfnamefont {M.~P.~A.}\ \bibnamefont {Fisher}},\ and\ \bibinfo
  {author} {\bibfnamefont {A.}~\bibnamefont {Lucas}},\ }\href
  {https://doi.org/10.1103/PhysRevResearch.2.033017} {\bibfield  {journal}
  {\bibinfo  {journal} {Phys. Rev. Research}\ }\textbf {\bibinfo {volume}
  {2}},\ \bibinfo {pages} {033017} (\bibinfo {year} {2020})}\BibitemShut
  {NoStop}%
\bibitem [{\citenamefont {Chen}()}]{chen2021nonunitaryfree}%
  \BibitemOpen
  \bibfield  {author} {\bibinfo {author} {\bibfnamefont {X.}~\bibnamefont
  {Chen}},\ }\href@noop {} {}\Eprint {https://arxiv.org/abs/2110.12230}
  {arXiv:2110.12230} \BibitemShut {NoStop}%
\bibitem [{\citenamefont {Cao}\ \emph {et~al.}(2019)\citenamefont {Cao},
  \citenamefont {Tilloy},\ and\ \citenamefont {Luca}}]{cao2019entanglementina}%
  \BibitemOpen
  \bibfield  {author} {\bibinfo {author} {\bibfnamefont {X.}~\bibnamefont
  {Cao}}, \bibinfo {author} {\bibfnamefont {A.}~\bibnamefont {Tilloy}},\ and\
  \bibinfo {author} {\bibfnamefont {A.~D.}\ \bibnamefont {Luca}},\ }\href
  {https://doi.org/10.21468/SciPostPhys.7.2.024} {\bibfield  {journal}
  {\bibinfo  {journal} {SciPost Phys.}\ }\textbf {\bibinfo {volume} {7}},\
  \bibinfo {pages} {024} (\bibinfo {year} {2019})}\BibitemShut {NoStop}%
\bibitem [{\citenamefont {Skinner}\ \emph {et~al.}(2019)\citenamefont
  {Skinner}, \citenamefont {Ruhman},\ and\ \citenamefont
  {Nahum}}]{skinner2019measurementinducedphase}%
  \BibitemOpen
  \bibfield  {author} {\bibinfo {author} {\bibfnamefont {B.}~\bibnamefont
  {Skinner}}, \bibinfo {author} {\bibfnamefont {J.}~\bibnamefont {Ruhman}},\
  and\ \bibinfo {author} {\bibfnamefont {A.}~\bibnamefont {Nahum}},\ }\href
  {https://doi.org/10.1103/PhysRevX.9.031009} {\bibfield  {journal} {\bibinfo
  {journal} {Phys. Rev. X}\ }\textbf {\bibinfo {volume} {9}},\ \bibinfo {pages}
  {031009} (\bibinfo {year} {2019})}\BibitemShut {NoStop}%
\bibitem [{\citenamefont {Li}\ \emph {et~al.}(2018)\citenamefont {Li},
  \citenamefont {Chen},\ and\ \citenamefont
  {Fisher}}]{li2018quantumzenoeffect}%
  \BibitemOpen
  \bibfield  {author} {\bibinfo {author} {\bibfnamefont {Y.}~\bibnamefont
  {Li}}, \bibinfo {author} {\bibfnamefont {X.}~\bibnamefont {Chen}},\ and\
  \bibinfo {author} {\bibfnamefont {M.~P.~A.}\ \bibnamefont {Fisher}},\ }\href
  {https://doi.org/10.1103/PhysRevB.98.205136} {\bibfield  {journal} {\bibinfo
  {journal} {Phys. Rev. B}\ }\textbf {\bibinfo {volume} {98}},\ \bibinfo
  {pages} {205136} (\bibinfo {year} {2018})}\BibitemShut {NoStop}%
\bibitem [{\citenamefont {Li}\ \emph {et~al.}(2019)\citenamefont {Li},
  \citenamefont {Chen},\ and\ \citenamefont
  {Fisher}}]{li2019measurementdrivenentanglement}%
  \BibitemOpen
  \bibfield  {author} {\bibinfo {author} {\bibfnamefont {Y.}~\bibnamefont
  {Li}}, \bibinfo {author} {\bibfnamefont {X.}~\bibnamefont {Chen}},\ and\
  \bibinfo {author} {\bibfnamefont {M.~P.~A.}\ \bibnamefont {Fisher}},\ }\href
  {https://doi.org/10.1103/PhysRevB.100.134306} {\bibfield  {journal} {\bibinfo
   {journal} {Phys. Rev. B}\ }\textbf {\bibinfo {volume} {100}},\ \bibinfo
  {pages} {134306} (\bibinfo {year} {2019})}\BibitemShut {NoStop}%
\bibitem [{\citenamefont {Chan}\ \emph {et~al.}(2019)\citenamefont {Chan},
  \citenamefont {Nandkishore}, \citenamefont {Pretko},\ and\ \citenamefont
  {Smith}}]{chan2019unitaryprojective}%
  \BibitemOpen
  \bibfield  {author} {\bibinfo {author} {\bibfnamefont {A.}~\bibnamefont
  {Chan}}, \bibinfo {author} {\bibfnamefont {R.~M.}\ \bibnamefont
  {Nandkishore}}, \bibinfo {author} {\bibfnamefont {M.}~\bibnamefont
  {Pretko}},\ and\ \bibinfo {author} {\bibfnamefont {G.}~\bibnamefont
  {Smith}},\ }\href {https://doi.org/10.1103/PhysRevB.99.224307} {\bibfield
  {journal} {\bibinfo  {journal} {Phys. Rev. B}\ }\textbf {\bibinfo {volume}
  {99}},\ \bibinfo {pages} {224307} (\bibinfo {year} {2019})}\BibitemShut
  {NoStop}%
\bibitem [{\citenamefont {Vasseur}\ \emph {et~al.}(2019)\citenamefont
  {Vasseur}, \citenamefont {Potter}, \citenamefont {You},\ and\ \citenamefont
  {Ludwig}}]{vasseur2019entanglementtransitionsfrom}%
  \BibitemOpen
  \bibfield  {author} {\bibinfo {author} {\bibfnamefont {R.}~\bibnamefont
  {Vasseur}}, \bibinfo {author} {\bibfnamefont {A.~C.}\ \bibnamefont {Potter}},
  \bibinfo {author} {\bibfnamefont {Y.-Z.}\ \bibnamefont {You}},\ and\ \bibinfo
  {author} {\bibfnamefont {A.~W.~W.}\ \bibnamefont {Ludwig}},\ }\href
  {https://doi.org/10.1103/PhysRevB.100.134203} {\bibfield  {journal} {\bibinfo
   {journal} {Phys. Rev. B}\ }\textbf {\bibinfo {volume} {100}},\ \bibinfo
  {pages} {134203} (\bibinfo {year} {2019})}\BibitemShut {NoStop}%
\bibitem [{\citenamefont {Jian}\ \emph {et~al.}(2020)\citenamefont {Jian},
  \citenamefont {You}, \citenamefont {Vasseur},\ and\ \citenamefont
  {Ludwig}}]{jian2020measurementinducedcriticality}%
  \BibitemOpen
  \bibfield  {author} {\bibinfo {author} {\bibfnamefont {C.-M.}\ \bibnamefont
  {Jian}}, \bibinfo {author} {\bibfnamefont {Y.-Z.}\ \bibnamefont {You}},
  \bibinfo {author} {\bibfnamefont {R.}~\bibnamefont {Vasseur}},\ and\ \bibinfo
  {author} {\bibfnamefont {A.~W.~W.}\ \bibnamefont {Ludwig}},\ }\href
  {https://doi.org/10.1103/PhysRevB.101.104302} {\bibfield  {journal} {\bibinfo
   {journal} {Phys. Rev. B}\ }\textbf {\bibinfo {volume} {101}},\ \bibinfo
  {pages} {104302} (\bibinfo {year} {2020})}\BibitemShut {NoStop}%
\bibitem [{\citenamefont {Nahum}\ \emph {et~al.}(2021)\citenamefont {Nahum},
  \citenamefont {Roy}, \citenamefont {Skinner},\ and\ \citenamefont
  {Ruhman}}]{nahum2921measurementandentanglement}%
  \BibitemOpen
  \bibfield  {author} {\bibinfo {author} {\bibfnamefont {A.}~\bibnamefont
  {Nahum}}, \bibinfo {author} {\bibfnamefont {S.}~\bibnamefont {Roy}}, \bibinfo
  {author} {\bibfnamefont {B.}~\bibnamefont {Skinner}},\ and\ \bibinfo {author}
  {\bibfnamefont {J.}~\bibnamefont {Ruhman}},\ }\href
  {https://doi.org/10.1103/PRXQuantum.2.010352} {\bibfield  {journal} {\bibinfo
   {journal} {PRX Quantum}\ }\textbf {\bibinfo {volume} {2}},\ \bibinfo {pages}
  {010352} (\bibinfo {year} {2021})}\BibitemShut {NoStop}%
\bibitem [{\citenamefont {Bao}\ \emph {et~al.}(2020)\citenamefont {Bao},
  \citenamefont {Choi},\ and\ \citenamefont {Altman}}]{bao2020theoryofthe}%
  \BibitemOpen
  \bibfield  {author} {\bibinfo {author} {\bibfnamefont {Y.}~\bibnamefont
  {Bao}}, \bibinfo {author} {\bibfnamefont {S.}~\bibnamefont {Choi}},\ and\
  \bibinfo {author} {\bibfnamefont {E.}~\bibnamefont {Altman}},\ }\href
  {https://doi.org/10.1103/PhysRevB.101.104301} {\bibfield  {journal} {\bibinfo
   {journal} {Phys. Rev. B}\ }\textbf {\bibinfo {volume} {101}},\ \bibinfo
  {pages} {104301} (\bibinfo {year} {2020})}\BibitemShut {NoStop}%
\bibitem [{\citenamefont {Choi}\ \emph {et~al.}(2020)\citenamefont {Choi},
  \citenamefont {Bao}, \citenamefont {Qi},\ and\ \citenamefont
  {Altman}}]{choi2020quantumerrorcorrection}%
  \BibitemOpen
  \bibfield  {author} {\bibinfo {author} {\bibfnamefont {S.}~\bibnamefont
  {Choi}}, \bibinfo {author} {\bibfnamefont {Y.}~\bibnamefont {Bao}}, \bibinfo
  {author} {\bibfnamefont {X.-L.}\ \bibnamefont {Qi}},\ and\ \bibinfo {author}
  {\bibfnamefont {E.}~\bibnamefont {Altman}},\ }\href
  {https://doi.org/10.1103/PhysRevLett.125.030505} {\bibfield  {journal}
  {\bibinfo  {journal} {Phys. Rev. Lett.}\ }\textbf {\bibinfo {volume} {125}},\
  \bibinfo {pages} {030505} (\bibinfo {year} {2020})}\BibitemShut {NoStop}%
\bibitem [{\citenamefont {Gullans}\ and\ \citenamefont
  {Huse}(2020{\natexlab{a}})}]{gullans2020dynamicalpurificationphase}%
  \BibitemOpen
  \bibfield  {author} {\bibinfo {author} {\bibfnamefont {M.~J.}\ \bibnamefont
  {Gullans}}\ and\ \bibinfo {author} {\bibfnamefont {D.~A.}\ \bibnamefont
  {Huse}},\ }\href {https://doi.org/10.1103/PhysRevX.10.041020} {\bibfield
  {journal} {\bibinfo  {journal} {Phys. Rev. X}\ }\textbf {\bibinfo {volume}
  {10}},\ \bibinfo {pages} {041020} (\bibinfo {year}
  {2020}{\natexlab{a}})}\BibitemShut {NoStop}%
\bibitem [{\citenamefont {Gullans}\ and\ \citenamefont
  {Huse}(2020{\natexlab{b}})}]{gullans2020scalableprobesof}%
  \BibitemOpen
  \bibfield  {author} {\bibinfo {author} {\bibfnamefont {M.~J.}\ \bibnamefont
  {Gullans}}\ and\ \bibinfo {author} {\bibfnamefont {D.~A.}\ \bibnamefont
  {Huse}},\ }\href {https://doi.org/10.1103/PhysRevLett.125.070606} {\bibfield
  {journal} {\bibinfo  {journal} {Phys. Rev. Lett.}\ }\textbf {\bibinfo
  {volume} {125}},\ \bibinfo {pages} {070606} (\bibinfo {year}
  {2020}{\natexlab{b}})}\BibitemShut {NoStop}%
\bibitem [{\citenamefont {Czischek}\ \emph {et~al.}(2021)\citenamefont
  {Czischek}, \citenamefont {Torlai}, \citenamefont {Ray}, \citenamefont
  {Islam},\ and\ \citenamefont {Melko}}]{czischek2021simulating}%
  \BibitemOpen
  \bibfield  {author} {\bibinfo {author} {\bibfnamefont {S.}~\bibnamefont
  {Czischek}}, \bibinfo {author} {\bibfnamefont {G.}~\bibnamefont {Torlai}},
  \bibinfo {author} {\bibfnamefont {S.}~\bibnamefont {Ray}}, \bibinfo {author}
  {\bibfnamefont {R.}~\bibnamefont {Islam}},\ and\ \bibinfo {author}
  {\bibfnamefont {R.~G.}\ \bibnamefont {Melko}},\ }\href
  {https://doi.org/10.1103/PhysRevA.104.062405} {\bibfield  {journal} {\bibinfo
   {journal} {Phys. Rev. A}\ }\textbf {\bibinfo {volume} {104}},\ \bibinfo
  {pages} {062405} (\bibinfo {year} {2021})}\BibitemShut {NoStop}%
\bibitem [{\citenamefont {Han}\ and\ \citenamefont
  {Chen}()}]{han2022entanglementstructure}%
  \BibitemOpen
  \bibfield  {author} {\bibinfo {author} {\bibfnamefont {Y.}~\bibnamefont
  {Han}}\ and\ \bibinfo {author} {\bibfnamefont {X.}~\bibnamefont {Chen}},\
  }\href@noop {} {}\Eprint {https://arxiv.org/abs/2207.02165}
  {arXiv:2207.02165} \BibitemShut {NoStop}%
\bibitem [{\citenamefont {Fidkowski}\ \emph {et~al.}(2021)\citenamefont
  {Fidkowski}, \citenamefont {Haah},\ and\ \citenamefont
  {Hastings}}]{fidkowski2021howdynamicalquantum}%
  \BibitemOpen
  \bibfield  {author} {\bibinfo {author} {\bibfnamefont {L.}~\bibnamefont
  {Fidkowski}}, \bibinfo {author} {\bibfnamefont {J.}~\bibnamefont {Haah}},\
  and\ \bibinfo {author} {\bibfnamefont {M.~B.}\ \bibnamefont {Hastings}},\
  }\href {https://doi.org/10.22331/q-2021-01-17-382} {\bibfield  {journal}
  {\bibinfo  {journal} {{Quantum}}\ }\textbf {\bibinfo {volume} {5}},\ \bibinfo
  {pages} {382} (\bibinfo {year} {2021})}\BibitemShut {NoStop}%
\bibitem [{\citenamefont {Altland}\ \emph {et~al.}(2022)\citenamefont
  {Altland}, \citenamefont {Buchhold}, \citenamefont {Diehl},\ and\
  \citenamefont {Micklitz}}]{altland2022dynamicsofmeasured}%
  \BibitemOpen
  \bibfield  {author} {\bibinfo {author} {\bibfnamefont {A.}~\bibnamefont
  {Altland}}, \bibinfo {author} {\bibfnamefont {M.}~\bibnamefont {Buchhold}},
  \bibinfo {author} {\bibfnamefont {S.}~\bibnamefont {Diehl}},\ and\ \bibinfo
  {author} {\bibfnamefont {T.}~\bibnamefont {Micklitz}},\ }\href
  {https://doi.org/10.1103/PhysRevResearch.4.L022066} {\bibfield  {journal}
  {\bibinfo  {journal} {Phys. Rev. Research}\ }\textbf {\bibinfo {volume}
  {4}},\ \bibinfo {pages} {L022066} (\bibinfo {year} {2022})}\BibitemShut
  {NoStop}%
\bibitem [{\citenamefont {Fuji}\ and\ \citenamefont
  {Ashida}(2020)}]{fuji2020measurementinducedquantum}%
  \BibitemOpen
  \bibfield  {author} {\bibinfo {author} {\bibfnamefont {Y.}~\bibnamefont
  {Fuji}}\ and\ \bibinfo {author} {\bibfnamefont {Y.}~\bibnamefont {Ashida}},\
  }\href {https://doi.org/10.1103/PhysRevB.102.054302} {\bibfield  {journal}
  {\bibinfo  {journal} {Phys. Rev. B}\ }\textbf {\bibinfo {volume} {102}},\
  \bibinfo {pages} {054302} (\bibinfo {year} {2020})}\BibitemShut {NoStop}%
\bibitem [{\citenamefont {Biella}\ and\ \citenamefont
  {Schir{\'{o}}}(2021)}]{biella2021manybodyquantumzeno}%
  \BibitemOpen
  \bibfield  {author} {\bibinfo {author} {\bibfnamefont {A.}~\bibnamefont
  {Biella}}\ and\ \bibinfo {author} {\bibfnamefont {M.}~\bibnamefont
  {Schir{\'{o}}}},\ }\href {https://doi.org/10.22331/q-2021-08-19-528}
  {\bibfield  {journal} {\bibinfo  {journal} {{Quantum}}\ }\textbf {\bibinfo
  {volume} {5}},\ \bibinfo {pages} {528} (\bibinfo {year} {2021})}\BibitemShut
  {NoStop}%
\bibitem [{\citenamefont {Gopalakrishnan}\ and\ \citenamefont
  {Gullans}(2021)}]{gopalakrishnan2021entanglementandpurification}%
  \BibitemOpen
  \bibfield  {author} {\bibinfo {author} {\bibfnamefont {S.}~\bibnamefont
  {Gopalakrishnan}}\ and\ \bibinfo {author} {\bibfnamefont {M.~J.}\
  \bibnamefont {Gullans}},\ }\href
  {https://doi.org/10.1103/PhysRevLett.126.170503} {\bibfield  {journal}
  {\bibinfo  {journal} {Phys. Rev. Lett.}\ }\textbf {\bibinfo {volume} {126}},\
  \bibinfo {pages} {170503} (\bibinfo {year} {2021})}\BibitemShut {NoStop}%
\bibitem [{\citenamefont {Jian}\ \emph
  {et~al.}(2021{\natexlab{a}})\citenamefont {Jian}, \citenamefont {Yang},
  \citenamefont {Bi},\ and\ \citenamefont {Chen}}]{jian2021yangleeedge}%
  \BibitemOpen
  \bibfield  {author} {\bibinfo {author} {\bibfnamefont {S.-K.}\ \bibnamefont
  {Jian}}, \bibinfo {author} {\bibfnamefont {Z.-C.}\ \bibnamefont {Yang}},
  \bibinfo {author} {\bibfnamefont {Z.}~\bibnamefont {Bi}},\ and\ \bibinfo
  {author} {\bibfnamefont {X.}~\bibnamefont {Chen}},\ }\href
  {https://doi.org/10.1103/PhysRevB.104.L161107} {\bibfield  {journal}
  {\bibinfo  {journal} {Phys. Rev. B}\ }\textbf {\bibinfo {volume} {104}},\
  \bibinfo {pages} {L161107} (\bibinfo {year}
  {2021}{\natexlab{a}})}\BibitemShut {NoStop}%
\bibitem [{\citenamefont {Ippoliti}\ \emph {et~al.}(2021)\citenamefont
  {Ippoliti}, \citenamefont {Gullans}, \citenamefont {Gopalakrishnan},
  \citenamefont {Huse},\ and\ \citenamefont
  {Khemani}}]{ippoliti2021entanglementphasetransitions}%
  \BibitemOpen
  \bibfield  {author} {\bibinfo {author} {\bibfnamefont {M.}~\bibnamefont
  {Ippoliti}}, \bibinfo {author} {\bibfnamefont {M.~J.}\ \bibnamefont
  {Gullans}}, \bibinfo {author} {\bibfnamefont {S.}~\bibnamefont
  {Gopalakrishnan}}, \bibinfo {author} {\bibfnamefont {D.~A.}\ \bibnamefont
  {Huse}},\ and\ \bibinfo {author} {\bibfnamefont {V.}~\bibnamefont
  {Khemani}},\ }\href {https://doi.org/10.1103/PhysRevX.11.011030} {\bibfield
  {journal} {\bibinfo  {journal} {Phys. Rev. X}\ }\textbf {\bibinfo {volume}
  {11}},\ \bibinfo {pages} {011030} (\bibinfo {year} {2021})}\BibitemShut
  {NoStop}%
\bibitem [{\citenamefont {Ippoliti}\ \emph {et~al.}(2022)\citenamefont
  {Ippoliti}, \citenamefont {Rakovszky},\ and\ \citenamefont
  {Khemani}}]{ippoliti2022fractallogarithmicand}%
  \BibitemOpen
  \bibfield  {author} {\bibinfo {author} {\bibfnamefont {M.}~\bibnamefont
  {Ippoliti}}, \bibinfo {author} {\bibfnamefont {T.}~\bibnamefont
  {Rakovszky}},\ and\ \bibinfo {author} {\bibfnamefont {V.}~\bibnamefont
  {Khemani}},\ }\href {https://doi.org/10.1103/PhysRevX.12.011045} {\bibfield
  {journal} {\bibinfo  {journal} {Phys. Rev. X}\ }\textbf {\bibinfo {volume}
  {12}},\ \bibinfo {pages} {011045} (\bibinfo {year} {2022})}\BibitemShut
  {NoStop}%
\bibitem [{\citenamefont {Lang}\ and\ \citenamefont
  {B\"uchler}(2020)}]{lang2020entanglementtransitionin}%
  \BibitemOpen
  \bibfield  {author} {\bibinfo {author} {\bibfnamefont {N.}~\bibnamefont
  {Lang}}\ and\ \bibinfo {author} {\bibfnamefont {H.~P.}\ \bibnamefont
  {B\"uchler}},\ }\href {https://doi.org/10.1103/PhysRevB.102.094204}
  {\bibfield  {journal} {\bibinfo  {journal} {Phys. Rev. B}\ }\textbf {\bibinfo
  {volume} {102}},\ \bibinfo {pages} {094204} (\bibinfo {year}
  {2020})}\BibitemShut {NoStop}%
\bibitem [{\citenamefont {Li}\ and\ \citenamefont
  {Fisher}()}]{li2021robustdecodingin}%
  \BibitemOpen
  \bibfield  {author} {\bibinfo {author} {\bibfnamefont {Y.}~\bibnamefont
  {Li}}\ and\ \bibinfo {author} {\bibfnamefont {M.~P.~A.}\ \bibnamefont
  {Fisher}},\ }\href@noop {} {}\Eprint {https://arxiv.org/abs/2108.04274}
  {arXiv:2108.04274} \BibitemShut {NoStop}%
\bibitem [{\citenamefont {Li}\ \emph {et~al.}({\natexlab{a}})\citenamefont
  {Li}, \citenamefont {Vasseur}, \citenamefont {Fisher},\ and\ \citenamefont
  {Ludwig}}]{li2021statisticalmechanicsmodel}%
  \BibitemOpen
  \bibfield  {author} {\bibinfo {author} {\bibfnamefont {Y.}~\bibnamefont
  {Li}}, \bibinfo {author} {\bibfnamefont {R.}~\bibnamefont {Vasseur}},
  \bibinfo {author} {\bibfnamefont {M.~P.~A.}\ \bibnamefont {Fisher}},\ and\
  \bibinfo {author} {\bibfnamefont {A.~W.~W.}\ \bibnamefont {Ludwig}},\
  }\href@noop {} {} ({\natexlab{a}}),\ \Eprint
  {https://arxiv.org/abs/2110.02988} {arXiv:2110.02988} \BibitemShut {NoStop}%
\bibitem [{\citenamefont {Li}\ \emph {et~al.}({\natexlab{b}})\citenamefont
  {Li}, \citenamefont {Vijay},\ and\ \citenamefont
  {Fisher}}]{li2021entanglementdomainwalls}%
  \BibitemOpen
  \bibfield  {author} {\bibinfo {author} {\bibfnamefont {Y.}~\bibnamefont
  {Li}}, \bibinfo {author} {\bibfnamefont {S.}~\bibnamefont {Vijay}},\ and\
  \bibinfo {author} {\bibfnamefont {M.~P.~A.}\ \bibnamefont {Fisher}},\
  }\href@noop {} {} ({\natexlab{b}}),\ \Eprint
  {https://arxiv.org/abs/2105.13352} {arXiv:2105.13352} \BibitemShut {NoStop}%
\bibitem [{\citenamefont {Li}\ and\ \citenamefont
  {Fisher}(2021)}]{li2021statisticalmechanicsof}%
  \BibitemOpen
  \bibfield  {author} {\bibinfo {author} {\bibfnamefont {Y.}~\bibnamefont
  {Li}}\ and\ \bibinfo {author} {\bibfnamefont {M.~P.~A.}\ \bibnamefont
  {Fisher}},\ }\href {https://doi.org/10.1103/PhysRevB.103.104306} {\bibfield
  {journal} {\bibinfo  {journal} {Phys. Rev. B}\ }\textbf {\bibinfo {volume}
  {103}},\ \bibinfo {pages} {104306} (\bibinfo {year} {2021})}\BibitemShut
  {NoStop}%
\bibitem [{\citenamefont {Jian}\ \emph
  {et~al.}(2021{\natexlab{b}})\citenamefont {Jian}, \citenamefont {Liu},
  \citenamefont {Chen}, \citenamefont {Swingle},\ and\ \citenamefont
  {Zhang}}]{jian2021measurementinducedphase}%
  \BibitemOpen
  \bibfield  {author} {\bibinfo {author} {\bibfnamefont {S.-K.}\ \bibnamefont
  {Jian}}, \bibinfo {author} {\bibfnamefont {C.}~\bibnamefont {Liu}}, \bibinfo
  {author} {\bibfnamefont {X.}~\bibnamefont {Chen}}, \bibinfo {author}
  {\bibfnamefont {B.}~\bibnamefont {Swingle}},\ and\ \bibinfo {author}
  {\bibfnamefont {P.}~\bibnamefont {Zhang}},\ }\href
  {https://doi.org/10.1103/PhysRevLett.127.140601} {\bibfield  {journal}
  {\bibinfo  {journal} {Phys. Rev. Lett.}\ }\textbf {\bibinfo {volume} {127}},\
  \bibinfo {pages} {140601} (\bibinfo {year} {2021}{\natexlab{b}})}\BibitemShut
  {NoStop}%
\bibitem [{\citenamefont {Lopez-Piqueres}\ \emph {et~al.}(2020)\citenamefont
  {Lopez-Piqueres}, \citenamefont {Ware},\ and\ \citenamefont
  {Vasseur}}]{lopezpiqueres2020meanfieldentanglement}%
  \BibitemOpen
  \bibfield  {author} {\bibinfo {author} {\bibfnamefont {J.}~\bibnamefont
  {Lopez-Piqueres}}, \bibinfo {author} {\bibfnamefont {B.}~\bibnamefont
  {Ware}},\ and\ \bibinfo {author} {\bibfnamefont {R.}~\bibnamefont
  {Vasseur}},\ }\href {https://doi.org/10.1103/PhysRevB.102.064202} {\bibfield
  {journal} {\bibinfo  {journal} {Phys. Rev. B}\ }\textbf {\bibinfo {volume}
  {102}},\ \bibinfo {pages} {064202} (\bibinfo {year} {2020})}\BibitemShut
  {NoStop}%
\bibitem [{\citenamefont {Jin}\ and\ \citenamefont {Martin}()}]{jin2022KPZ}%
  \BibitemOpen
  \bibfield  {author} {\bibinfo {author} {\bibfnamefont {T.}~\bibnamefont
  {Jin}}\ and\ \bibinfo {author} {\bibfnamefont {D.~G.}\ \bibnamefont
  {Martin}},\ }\href@noop {} {}\Eprint {https://arxiv.org/abs/2204.00070}
  {arXiv:2204.00070} \BibitemShut {NoStop}%
\bibitem [{\citenamefont {Willsher}\ \emph {et~al.}(2022)\citenamefont
  {Willsher}, \citenamefont {Liu}, \citenamefont {Moessner},\ and\
  \citenamefont {Knolle}}]{willsher2022measurementinducedphase}%
  \BibitemOpen
  \bibfield  {author} {\bibinfo {author} {\bibfnamefont {J.}~\bibnamefont
  {Willsher}}, \bibinfo {author} {\bibfnamefont {S.-W.}\ \bibnamefont {Liu}},
  \bibinfo {author} {\bibfnamefont {R.}~\bibnamefont {Moessner}},\ and\
  \bibinfo {author} {\bibfnamefont {J.}~\bibnamefont {Knolle}},\ }\href
  {https://doi.org/10.1103/PhysRevB.106.024305} {\bibfield  {journal} {\bibinfo
   {journal} {Phys. Rev. B}\ }\textbf {\bibinfo {volume} {106}},\ \bibinfo
  {pages} {024305} (\bibinfo {year} {2022})}\BibitemShut {NoStop}%
\bibitem [{\citenamefont {Pizzi}\ \emph {et~al.}()\citenamefont {Pizzi},
  \citenamefont {Malz}, \citenamefont {Nunnenkamp},\ and\ \citenamefont
  {Knolle}}]{pizzi2022bridgingthegap}%
  \BibitemOpen
  \bibfield  {author} {\bibinfo {author} {\bibfnamefont {A.}~\bibnamefont
  {Pizzi}}, \bibinfo {author} {\bibfnamefont {D.}~\bibnamefont {Malz}},
  \bibinfo {author} {\bibfnamefont {A.}~\bibnamefont {Nunnenkamp}},\ and\
  \bibinfo {author} {\bibfnamefont {J.}~\bibnamefont {Knolle}},\ }\href@noop {}
  {}\Eprint {https://arxiv.org/abs/2204.03016} {arXiv:2204.03016} \BibitemShut
  {NoStop}%
\bibitem [{\citenamefont {Lyons}\ \emph {et~al.}()\citenamefont {Lyons},
  \citenamefont {Choi},\ and\ \citenamefont
  {Altman}}]{lyons2022auniversalcrossover}%
  \BibitemOpen
  \bibfield  {author} {\bibinfo {author} {\bibfnamefont {A.}~\bibnamefont
  {Lyons}}, \bibinfo {author} {\bibfnamefont {S.}~\bibnamefont {Choi}},\ and\
  \bibinfo {author} {\bibfnamefont {E.}~\bibnamefont {Altman}},\ }\href@noop {}
  {}\Eprint {https://arxiv.org/abs/2208.02217} {arXiv:2208.02217} \BibitemShut
  {NoStop}%
\bibitem [{\citenamefont {Zhang}\ \emph {et~al.}(2020)\citenamefont {Zhang},
  \citenamefont {Reyes}, \citenamefont {Kourtis}, \citenamefont {Chamon},
  \citenamefont {Mucciolo},\ and\ \citenamefont
  {Ruckenstein}}]{zhang2020nonuniversalentanglementlevel}%
  \BibitemOpen
  \bibfield  {author} {\bibinfo {author} {\bibfnamefont {L.}~\bibnamefont
  {Zhang}}, \bibinfo {author} {\bibfnamefont {J.~A.}\ \bibnamefont {Reyes}},
  \bibinfo {author} {\bibfnamefont {S.}~\bibnamefont {Kourtis}}, \bibinfo
  {author} {\bibfnamefont {C.}~\bibnamefont {Chamon}}, \bibinfo {author}
  {\bibfnamefont {E.~R.}\ \bibnamefont {Mucciolo}},\ and\ \bibinfo {author}
  {\bibfnamefont {A.~E.}\ \bibnamefont {Ruckenstein}},\ }\href
  {https://doi.org/10.1103/PhysRevB.101.235104} {\bibfield  {journal} {\bibinfo
   {journal} {Phys. Rev. B}\ }\textbf {\bibinfo {volume} {101}},\ \bibinfo
  {pages} {235104} (\bibinfo {year} {2020})}\BibitemShut {NoStop}%
\bibitem [{\citenamefont {Zhang}\ \emph {et~al.}(2021)\citenamefont {Zhang},
  \citenamefont {Jian}, \citenamefont {Liu},\ and\ \citenamefont
  {Chen}}]{zhang2021emergentreplica}%
  \BibitemOpen
  \bibfield  {author} {\bibinfo {author} {\bibfnamefont {P.}~\bibnamefont
  {Zhang}}, \bibinfo {author} {\bibfnamefont {S.-K.}\ \bibnamefont {Jian}},
  \bibinfo {author} {\bibfnamefont {C.}~\bibnamefont {Liu}},\ and\ \bibinfo
  {author} {\bibfnamefont {X.}~\bibnamefont {Chen}},\ }\href
  {https://doi.org/10.22331/q-2021-11-16-579} {\bibfield  {journal} {\bibinfo
  {journal} {{Quantum}}\ }\textbf {\bibinfo {volume} {5}},\ \bibinfo {pages}
  {579} (\bibinfo {year} {2021})}\BibitemShut {NoStop}%
\bibitem [{\citenamefont {Zhang}\ \emph {et~al.}(2022)\citenamefont {Zhang},
  \citenamefont {Liu}, \citenamefont {Jian},\ and\ \citenamefont
  {Chen}}]{zhang2022universalentanglementtransitions}%
  \BibitemOpen
  \bibfield  {author} {\bibinfo {author} {\bibfnamefont {P.}~\bibnamefont
  {Zhang}}, \bibinfo {author} {\bibfnamefont {C.}~\bibnamefont {Liu}}, \bibinfo
  {author} {\bibfnamefont {S.-K.}\ \bibnamefont {Jian}},\ and\ \bibinfo
  {author} {\bibfnamefont {X.}~\bibnamefont {Chen}},\ }\href
  {https://doi.org/10.22331/q-2022-05-27-723} {\bibfield  {journal} {\bibinfo
  {journal} {{Quantum}}\ }\textbf {\bibinfo {volume} {6}},\ \bibinfo {pages}
  {723} (\bibinfo {year} {2022})}\BibitemShut {NoStop}%
\bibitem [{\citenamefont {Zhou}\ and\ \citenamefont
  {Chen}(2021)}]{zhou2021nonunitaryentanglementdynamics}%
  \BibitemOpen
  \bibfield  {author} {\bibinfo {author} {\bibfnamefont {T.}~\bibnamefont
  {Zhou}}\ and\ \bibinfo {author} {\bibfnamefont {X.}~\bibnamefont {Chen}},\
  }\href {https://doi.org/10.1103/PhysRevB.104.L180301} {\bibfield  {journal}
  {\bibinfo  {journal} {Phys. Rev. B}\ }\textbf {\bibinfo {volume} {104}},\
  \bibinfo {pages} {L180301} (\bibinfo {year} {2021})}\BibitemShut {NoStop}%
\bibitem [{\citenamefont {Bentsen}\ \emph {et~al.}(2021)\citenamefont
  {Bentsen}, \citenamefont {Sahu},\ and\ \citenamefont
  {Swingle}}]{bentsen2021measurementinducedpurification}%
  \BibitemOpen
  \bibfield  {author} {\bibinfo {author} {\bibfnamefont {G.~S.}\ \bibnamefont
  {Bentsen}}, \bibinfo {author} {\bibfnamefont {S.}~\bibnamefont {Sahu}},\ and\
  \bibinfo {author} {\bibfnamefont {B.}~\bibnamefont {Swingle}},\ }\href
  {https://doi.org/10.1103/PhysRevB.104.094304} {\bibfield  {journal} {\bibinfo
   {journal} {Phys. Rev. B}\ }\textbf {\bibinfo {volume} {104}},\ \bibinfo
  {pages} {094304} (\bibinfo {year} {2021})}\BibitemShut {NoStop}%
\bibitem [{\citenamefont {Yang}\ \emph {et~al.}(2022)\citenamefont {Yang},
  \citenamefont {Li}, \citenamefont {Fisher},\ and\ \citenamefont
  {Chen}}]{yang2022entanglementphasetransitions}%
  \BibitemOpen
  \bibfield  {author} {\bibinfo {author} {\bibfnamefont {Z.-C.}\ \bibnamefont
  {Yang}}, \bibinfo {author} {\bibfnamefont {Y.}~\bibnamefont {Li}}, \bibinfo
  {author} {\bibfnamefont {M.~P.~A.}\ \bibnamefont {Fisher}},\ and\ \bibinfo
  {author} {\bibfnamefont {X.}~\bibnamefont {Chen}},\ }\href
  {https://doi.org/10.1103/PhysRevB.105.104306} {\bibfield  {journal} {\bibinfo
   {journal} {Phys. Rev. B}\ }\textbf {\bibinfo {volume} {105}},\ \bibinfo
  {pages} {104306} (\bibinfo {year} {2022})}\BibitemShut {NoStop}%
\bibitem [{\citenamefont {Rossini}\ and\ \citenamefont
  {Vicari}(2020)}]{rossini2020measurementinduceddynamics}%
  \BibitemOpen
  \bibfield  {author} {\bibinfo {author} {\bibfnamefont {D.}~\bibnamefont
  {Rossini}}\ and\ \bibinfo {author} {\bibfnamefont {E.}~\bibnamefont
  {Vicari}},\ }\href {https://doi.org/10.1103/PhysRevB.102.035119} {\bibfield
  {journal} {\bibinfo  {journal} {Phys. Rev. B}\ }\textbf {\bibinfo {volume}
  {102}},\ \bibinfo {pages} {035119} (\bibinfo {year} {2020})}\BibitemShut
  {NoStop}%
\bibitem [{\citenamefont {Medina}\ \emph {et~al.}(2021)\citenamefont {Medina},
  \citenamefont {Vasseur},\ and\ \citenamefont
  {Serbyn}}]{medina2021entanglementtransitionsfrom}%
  \BibitemOpen
  \bibfield  {author} {\bibinfo {author} {\bibfnamefont {R.}~\bibnamefont
  {Medina}}, \bibinfo {author} {\bibfnamefont {R.}~\bibnamefont {Vasseur}},\
  and\ \bibinfo {author} {\bibfnamefont {M.}~\bibnamefont {Serbyn}},\ }\href
  {https://doi.org/10.1103/PhysRevB.104.104205} {\bibfield  {journal} {\bibinfo
   {journal} {Phys. Rev. B}\ }\textbf {\bibinfo {volume} {104}},\ \bibinfo
  {pages} {104205} (\bibinfo {year} {2021})}\BibitemShut {NoStop}%
\bibitem [{\citenamefont {Lunt}\ and\ \citenamefont
  {Pal}(2020)}]{lunt2020measurementinducedentanglement}%
  \BibitemOpen
  \bibfield  {author} {\bibinfo {author} {\bibfnamefont {O.}~\bibnamefont
  {Lunt}}\ and\ \bibinfo {author} {\bibfnamefont {A.}~\bibnamefont {Pal}},\
  }\href {https://doi.org/10.1103/PhysRevResearch.2.043072} {\bibfield
  {journal} {\bibinfo  {journal} {Phys. Rev. Research}\ }\textbf {\bibinfo
  {volume} {2}},\ \bibinfo {pages} {043072} (\bibinfo {year}
  {2020})}\BibitemShut {NoStop}%
\bibitem [{\citenamefont {Nahum}\ and\ \citenamefont
  {Skinner}(2020)}]{nahum2020entanglementanddynamics}%
  \BibitemOpen
  \bibfield  {author} {\bibinfo {author} {\bibfnamefont {A.}~\bibnamefont
  {Nahum}}\ and\ \bibinfo {author} {\bibfnamefont {B.}~\bibnamefont
  {Skinner}},\ }\href {https://doi.org/10.1103/PhysRevResearch.2.023288}
  {\bibfield  {journal} {\bibinfo  {journal} {Phys. Rev. Research}\ }\textbf
  {\bibinfo {volume} {2}},\ \bibinfo {pages} {023288} (\bibinfo {year}
  {2020})}\BibitemShut {NoStop}%
\bibitem [{\citenamefont {Kelly}\ \emph {et~al.}()\citenamefont {Kelly},
  \citenamefont {Poschinger}, \citenamefont {Schmidt-Kaler}, \citenamefont
  {Fisher},\ and\ \citenamefont {Marino}}]{kelly2022coherencerequirementsfor}%
  \BibitemOpen
  \bibfield  {author} {\bibinfo {author} {\bibfnamefont {S.~P.}\ \bibnamefont
  {Kelly}}, \bibinfo {author} {\bibfnamefont {U.}~\bibnamefont {Poschinger}},
  \bibinfo {author} {\bibfnamefont {F.}~\bibnamefont {Schmidt-Kaler}}, \bibinfo
  {author} {\bibfnamefont {M.~P.~A.}\ \bibnamefont {Fisher}},\ and\ \bibinfo
  {author} {\bibfnamefont {J.}~\bibnamefont {Marino}},\ }\href@noop {}
  {}\Eprint {https://arxiv.org/abs/2210.11547} {arXiv:2210.11547} \BibitemShut
  {NoStop}%
\bibitem [{\citenamefont {Szyniszewski}\ \emph {et~al.}(2019)\citenamefont
  {Szyniszewski}, \citenamefont {Romito},\ and\ \citenamefont
  {Schomerus}}]{szyniszewski2019entanglementtransitionfrom}%
  \BibitemOpen
  \bibfield  {author} {\bibinfo {author} {\bibfnamefont {M.}~\bibnamefont
  {Szyniszewski}}, \bibinfo {author} {\bibfnamefont {A.}~\bibnamefont
  {Romito}},\ and\ \bibinfo {author} {\bibfnamefont {H.}~\bibnamefont
  {Schomerus}},\ }\href {https://doi.org/10.1103/PhysRevB.100.064204}
  {\bibfield  {journal} {\bibinfo  {journal} {Phys. Rev. B}\ }\textbf {\bibinfo
  {volume} {100}},\ \bibinfo {pages} {064204} (\bibinfo {year}
  {2019})}\BibitemShut {NoStop}%
\bibitem [{\citenamefont {Szyniszewski}\ \emph {et~al.}(2020)\citenamefont
  {Szyniszewski}, \citenamefont {Romito},\ and\ \citenamefont
  {Schomerus}}]{szyniszewski2020universalityofentanglement}%
  \BibitemOpen
  \bibfield  {author} {\bibinfo {author} {\bibfnamefont {M.}~\bibnamefont
  {Szyniszewski}}, \bibinfo {author} {\bibfnamefont {A.}~\bibnamefont
  {Romito}},\ and\ \bibinfo {author} {\bibfnamefont {H.}~\bibnamefont
  {Schomerus}},\ }\href {https://doi.org/10.1103/PhysRevLett.125.210602}
  {\bibfield  {journal} {\bibinfo  {journal} {Phys. Rev. Lett.}\ }\textbf
  {\bibinfo {volume} {125}},\ \bibinfo {pages} {210602} (\bibinfo {year}
  {2020})}\BibitemShut {NoStop}%
\bibitem [{\citenamefont {Kumar}\ \emph {et~al.}(2020)\citenamefont {Kumar},
  \citenamefont {Romito},\ and\ \citenamefont
  {Snizhko}}]{parveen2020quantumzenoeffect}%
  \BibitemOpen
  \bibfield  {author} {\bibinfo {author} {\bibfnamefont {P.}~\bibnamefont
  {Kumar}}, \bibinfo {author} {\bibfnamefont {A.}~\bibnamefont {Romito}},\ and\
  \bibinfo {author} {\bibfnamefont {K.}~\bibnamefont {Snizhko}},\ }\href
  {https://doi.org/10.1103/PhysRevResearch.2.043420} {\bibfield  {journal}
  {\bibinfo  {journal} {Phys. Rev. Research}\ }\textbf {\bibinfo {volume}
  {2}},\ \bibinfo {pages} {043420} (\bibinfo {year} {2020})}\BibitemShut
  {NoStop}%
\bibitem [{\citenamefont {Tang}\ and\ \citenamefont
  {Zhu}(2020)}]{tang2020measurementinducedphase}%
  \BibitemOpen
  \bibfield  {author} {\bibinfo {author} {\bibfnamefont {Q.}~\bibnamefont
  {Tang}}\ and\ \bibinfo {author} {\bibfnamefont {W.}~\bibnamefont {Zhu}},\
  }\href {https://doi.org/10.1103/PhysRevResearch.2.013022} {\bibfield
  {journal} {\bibinfo  {journal} {Phys. Rev. Research}\ }\textbf {\bibinfo
  {volume} {2}},\ \bibinfo {pages} {013022} (\bibinfo {year}
  {2020})}\BibitemShut {NoStop}%
\bibitem [{\citenamefont {Tang}\ \emph {et~al.}(2021)\citenamefont {Tang},
  \citenamefont {Chen},\ and\ \citenamefont
  {Zhu}}]{tang2021quantumcriticalityin}%
  \BibitemOpen
  \bibfield  {author} {\bibinfo {author} {\bibfnamefont {Q.}~\bibnamefont
  {Tang}}, \bibinfo {author} {\bibfnamefont {X.}~\bibnamefont {Chen}},\ and\
  \bibinfo {author} {\bibfnamefont {W.}~\bibnamefont {Zhu}},\ }\href
  {https://doi.org/10.1103/PhysRevB.103.174303} {\bibfield  {journal} {\bibinfo
   {journal} {Phys. Rev. B}\ }\textbf {\bibinfo {volume} {103}},\ \bibinfo
  {pages} {174303} (\bibinfo {year} {2021})}\BibitemShut {NoStop}%
\bibitem [{\citenamefont {Shtanko}\ \emph {et~al.}()\citenamefont {Shtanko},
  \citenamefont {Kharkov}, \citenamefont {García-Pintos},\ and\ \citenamefont
  {Gorshkov}}]{shtanko2020classicalmodelsof}%
  \BibitemOpen
  \bibfield  {author} {\bibinfo {author} {\bibfnamefont {O.}~\bibnamefont
  {Shtanko}}, \bibinfo {author} {\bibfnamefont {Y.~A.}\ \bibnamefont
  {Kharkov}}, \bibinfo {author} {\bibfnamefont {L.~P.}\ \bibnamefont
  {García-Pintos}},\ and\ \bibinfo {author} {\bibfnamefont {A.~V.}\
  \bibnamefont {Gorshkov}},\ }\href@noop {} {}\Eprint
  {https://arxiv.org/abs/2004.06736} {arXiv:2004.06736} \BibitemShut {NoStop}%
\bibitem [{\citenamefont {Van~Regemortel}\ \emph {et~al.}(2021)\citenamefont
  {Van~Regemortel}, \citenamefont {Cian}, \citenamefont {Seif}, \citenamefont
  {Dehghani},\ and\ \citenamefont
  {Hafezi}}]{van2021entanglemententropyscaling}%
  \BibitemOpen
  \bibfield  {author} {\bibinfo {author} {\bibfnamefont {M.}~\bibnamefont
  {Van~Regemortel}}, \bibinfo {author} {\bibfnamefont {Z.-P.}\ \bibnamefont
  {Cian}}, \bibinfo {author} {\bibfnamefont {A.}~\bibnamefont {Seif}}, \bibinfo
  {author} {\bibfnamefont {H.}~\bibnamefont {Dehghani}},\ and\ \bibinfo
  {author} {\bibfnamefont {M.}~\bibnamefont {Hafezi}},\ }\href
  {https://doi.org/10.1103/PhysRevLett.126.123604} {\bibfield  {journal}
  {\bibinfo  {journal} {Phys. Rev. Lett.}\ }\textbf {\bibinfo {volume} {126}},\
  \bibinfo {pages} {123604} (\bibinfo {year} {2021})}\BibitemShut {NoStop}%
\bibitem [{\citenamefont {Vijay}()}]{vijay2020measurementdrivenphase}%
  \BibitemOpen
  \bibfield  {author} {\bibinfo {author} {\bibfnamefont {S.}~\bibnamefont
  {Vijay}},\ }\href@noop {} {}\Eprint {https://arxiv.org/abs/2005.03052}
  {arXiv:2005.03052} \BibitemShut {NoStop}%
\bibitem [{\citenamefont {Zabalo}\ \emph {et~al.}(2020)\citenamefont {Zabalo},
  \citenamefont {Gullans}, \citenamefont {Wilson}, \citenamefont
  {Gopalakrishnan}, \citenamefont {Huse},\ and\ \citenamefont
  {Pixley}}]{zabalo2020criticalpropertiesof}%
  \BibitemOpen
  \bibfield  {author} {\bibinfo {author} {\bibfnamefont {A.}~\bibnamefont
  {Zabalo}}, \bibinfo {author} {\bibfnamefont {M.~J.}\ \bibnamefont {Gullans}},
  \bibinfo {author} {\bibfnamefont {J.~H.}\ \bibnamefont {Wilson}}, \bibinfo
  {author} {\bibfnamefont {S.}~\bibnamefont {Gopalakrishnan}}, \bibinfo
  {author} {\bibfnamefont {D.~A.}\ \bibnamefont {Huse}},\ and\ \bibinfo
  {author} {\bibfnamefont {J.~H.}\ \bibnamefont {Pixley}},\ }\href
  {https://doi.org/10.1103/PhysRevB.101.060301} {\bibfield  {journal} {\bibinfo
   {journal} {Phys. Rev. B}\ }\textbf {\bibinfo {volume} {101}},\ \bibinfo
  {pages} {060301} (\bibinfo {year} {2020})}\BibitemShut {NoStop}%
\bibitem [{\citenamefont {Fan}\ \emph {et~al.}(2021)\citenamefont {Fan},
  \citenamefont {Vijay}, \citenamefont {Vishwanath},\ and\ \citenamefont
  {You}}]{fan2021selforganizederror}%
  \BibitemOpen
  \bibfield  {author} {\bibinfo {author} {\bibfnamefont {R.}~\bibnamefont
  {Fan}}, \bibinfo {author} {\bibfnamefont {S.}~\bibnamefont {Vijay}}, \bibinfo
  {author} {\bibfnamefont {A.}~\bibnamefont {Vishwanath}},\ and\ \bibinfo
  {author} {\bibfnamefont {Y.-Z.}\ \bibnamefont {You}},\ }\href
  {https://doi.org/10.1103/PhysRevB.103.174309} {\bibfield  {journal} {\bibinfo
   {journal} {Phys. Rev. B}\ }\textbf {\bibinfo {volume} {103}},\ \bibinfo
  {pages} {174309} (\bibinfo {year} {2021})}\BibitemShut {NoStop}%
\bibitem [{\citenamefont {Sang}\ \emph {et~al.}(2021)\citenamefont {Sang},
  \citenamefont {Li}, \citenamefont {Zhou}, \citenamefont {Chen}, \citenamefont
  {Hsieh},\ and\ \citenamefont {Fisher}}]{sang2021entanglementnegativityat}%
  \BibitemOpen
  \bibfield  {author} {\bibinfo {author} {\bibfnamefont {S.}~\bibnamefont
  {Sang}}, \bibinfo {author} {\bibfnamefont {Y.}~\bibnamefont {Li}}, \bibinfo
  {author} {\bibfnamefont {T.}~\bibnamefont {Zhou}}, \bibinfo {author}
  {\bibfnamefont {X.}~\bibnamefont {Chen}}, \bibinfo {author} {\bibfnamefont
  {T.~H.}\ \bibnamefont {Hsieh}},\ and\ \bibinfo {author} {\bibfnamefont
  {M.~P.}\ \bibnamefont {Fisher}},\ }\href
  {https://doi.org/10.1103/PRXQuantum.2.030313} {\bibfield  {journal} {\bibinfo
   {journal} {PRX Quantum}\ }\textbf {\bibinfo {volume} {2}},\ \bibinfo {pages}
  {030313} (\bibinfo {year} {2021})}\BibitemShut {NoStop}%
\bibitem [{\citenamefont {Shi}\ \emph {et~al.}()\citenamefont {Shi},
  \citenamefont {Dai},\ and\ \citenamefont
  {Lu}}]{shi2020entanglementnegativityat}%
  \BibitemOpen
  \bibfield  {author} {\bibinfo {author} {\bibfnamefont {B.}~\bibnamefont
  {Shi}}, \bibinfo {author} {\bibfnamefont {X.}~\bibnamefont {Dai}},\ and\
  \bibinfo {author} {\bibfnamefont {Y.-M.}\ \bibnamefont {Lu}},\ }\href@noop {}
  {}\Eprint {https://arxiv.org/abs/2012.00040} {arXiv:2012.00040} \BibitemShut
  {NoStop}%
\bibitem [{\citenamefont {Weinstein}\ \emph {et~al.}(2022)\citenamefont
  {Weinstein}, \citenamefont {Bao},\ and\ \citenamefont
  {Altman}}]{weinstein2022measurementinducedpower}%
  \BibitemOpen
  \bibfield  {author} {\bibinfo {author} {\bibfnamefont {Z.}~\bibnamefont
  {Weinstein}}, \bibinfo {author} {\bibfnamefont {Y.}~\bibnamefont {Bao}},\
  and\ \bibinfo {author} {\bibfnamefont {E.}~\bibnamefont {Altman}},\ }\href
  {https://doi.org/10.1103/PhysRevLett.129.080501} {\bibfield  {journal}
  {\bibinfo  {journal} {Phys. Rev. Lett.}\ }\textbf {\bibinfo {volume} {129}},\
  \bibinfo {pages} {080501} (\bibinfo {year} {2022})}\BibitemShut {NoStop}%
\bibitem [{\citenamefont {Li}\ \emph {et~al.}(2021)\citenamefont {Li},
  \citenamefont {Chen}, \citenamefont {Ludwig},\ and\ \citenamefont
  {Fisher}}]{li2021conformal}%
  \BibitemOpen
  \bibfield  {author} {\bibinfo {author} {\bibfnamefont {Y.}~\bibnamefont
  {Li}}, \bibinfo {author} {\bibfnamefont {X.}~\bibnamefont {Chen}}, \bibinfo
  {author} {\bibfnamefont {A.~W.~W.}\ \bibnamefont {Ludwig}},\ and\ \bibinfo
  {author} {\bibfnamefont {M.~P.~A.}\ \bibnamefont {Fisher}},\ }\href
  {https://doi.org/10.1103/PhysRevB.104.104305} {\bibfield  {journal} {\bibinfo
   {journal} {Phys. Rev. B}\ }\textbf {\bibinfo {volume} {104}},\ \bibinfo
  {pages} {104305} (\bibinfo {year} {2021})}\BibitemShut {NoStop}%
\bibitem [{\citenamefont {Liu}\ \emph {et~al.}(2022)\citenamefont {Liu},
  \citenamefont {Zhou},\ and\ \citenamefont
  {Chen}}]{liu2022measurementinducedentanglement}%
  \BibitemOpen
  \bibfield  {author} {\bibinfo {author} {\bibfnamefont {H.}~\bibnamefont
  {Liu}}, \bibinfo {author} {\bibfnamefont {T.}~\bibnamefont {Zhou}},\ and\
  \bibinfo {author} {\bibfnamefont {X.}~\bibnamefont {Chen}},\ }\href
  {https://doi.org/10.1103/PhysRevB.106.144311} {\bibfield  {journal} {\bibinfo
   {journal} {Phys. Rev. B}\ }\textbf {\bibinfo {volume} {106}},\ \bibinfo
  {pages} {144311} (\bibinfo {year} {2022})}\BibitemShut {NoStop}%
\bibitem [{\citenamefont {Bao}\ \emph {et~al.}()\citenamefont {Bao},
  \citenamefont {Block},\ and\ \citenamefont
  {Altman}}]{bao2021finitetimeteleportation}%
  \BibitemOpen
  \bibfield  {author} {\bibinfo {author} {\bibfnamefont {Y.}~\bibnamefont
  {Bao}}, \bibinfo {author} {\bibfnamefont {M.}~\bibnamefont {Block}},\ and\
  \bibinfo {author} {\bibfnamefont {E.}~\bibnamefont {Altman}},\ }\href@noop {}
  {}\Eprint {https://arxiv.org/abs/2110.06963} {arXiv:2110.06963} \BibitemShut
  {NoStop}%
\bibitem [{\citenamefont {Agarwal}\ \emph {et~al.}()\citenamefont {Agarwal},
  \citenamefont {Langlett},\ and\ \citenamefont
  {Xu}}]{agrawal2022longrangebell}%
  \BibitemOpen
  \bibfield  {author} {\bibinfo {author} {\bibfnamefont {L.}~\bibnamefont
  {Agarwal}}, \bibinfo {author} {\bibfnamefont {C.~M.}\ \bibnamefont
  {Langlett}},\ and\ \bibinfo {author} {\bibfnamefont {S.}~\bibnamefont {Xu}},\
  }\href@noop {} {}\Eprint {https://arxiv.org/abs/2205.02782}
  {arXiv:2205.02782} \BibitemShut {NoStop}%
\bibitem [{\citenamefont {Garratt}\ \emph {et~al.}()\citenamefont {Garratt},
  \citenamefont {Weinstein},\ and\ \citenamefont
  {Altman}}]{garratt2022measurementsconspirenonlocally}%
  \BibitemOpen
  \bibfield  {author} {\bibinfo {author} {\bibfnamefont {S.~J.}\ \bibnamefont
  {Garratt}}, \bibinfo {author} {\bibfnamefont {Z.}~\bibnamefont {Weinstein}},\
  and\ \bibinfo {author} {\bibfnamefont {E.}~\bibnamefont {Altman}},\
  }\href@noop {} {}\Eprint {https://arxiv.org/abs/2207.09476}
  {arXiv:2207.09476} \BibitemShut {NoStop}%
\bibitem [{\citenamefont
  {Turkeshi}(2022)}]{turkeshi2022measurementinducedcriticality}%
  \BibitemOpen
  \bibfield  {author} {\bibinfo {author} {\bibfnamefont {X.}~\bibnamefont
  {Turkeshi}},\ }\href {https://doi.org/10.1103/PhysRevB.106.144313} {\bibfield
   {journal} {\bibinfo  {journal} {Phys. Rev. B}\ }\textbf {\bibinfo {volume}
  {106}},\ \bibinfo {pages} {144313} (\bibinfo {year} {2022})}\BibitemShut
  {NoStop}%
\bibitem [{\citenamefont {Barratt}\ \emph
  {et~al.}(2022{\natexlab{a}})\citenamefont {Barratt}, \citenamefont {Agrawal},
  \citenamefont {Potter}, \citenamefont {Gopalakrishnan},\ and\ \citenamefont
  {Vasseur}}]{barratt2022transitions}%
  \BibitemOpen
  \bibfield  {author} {\bibinfo {author} {\bibfnamefont {F.}~\bibnamefont
  {Barratt}}, \bibinfo {author} {\bibfnamefont {U.}~\bibnamefont {Agrawal}},
  \bibinfo {author} {\bibfnamefont {A.~C.}\ \bibnamefont {Potter}}, \bibinfo
  {author} {\bibfnamefont {S.}~\bibnamefont {Gopalakrishnan}},\ and\ \bibinfo
  {author} {\bibfnamefont {R.}~\bibnamefont {Vasseur}},\ }\href
  {https://doi.org/10.1103/PhysRevLett.129.200602} {\bibfield  {journal}
  {\bibinfo  {journal} {Phys. Rev. Lett.}\ }\textbf {\bibinfo {volume} {129}},\
  \bibinfo {pages} {200602} (\bibinfo {year} {2022}{\natexlab{a}})}\BibitemShut
  {NoStop}%
\bibitem [{\citenamefont {Dehghani}\ \emph {et~al.}()\citenamefont {Dehghani},
  \citenamefont {Lavasani}, \citenamefont {Hafezi},\ and\ \citenamefont
  {Gullans}}]{dehgani2022neuralnetworkdecoders}%
  \BibitemOpen
  \bibfield  {author} {\bibinfo {author} {\bibfnamefont {H.}~\bibnamefont
  {Dehghani}}, \bibinfo {author} {\bibfnamefont {A.}~\bibnamefont {Lavasani}},
  \bibinfo {author} {\bibfnamefont {M.}~\bibnamefont {Hafezi}},\ and\ \bibinfo
  {author} {\bibfnamefont {M.~J.}\ \bibnamefont {Gullans}},\ }\href@noop {}
  {}\Eprint {https://arxiv.org/abs/2204.10904} {arXiv:2204.10904} \BibitemShut
  {NoStop}%
\bibitem [{\citenamefont {Iaconis}\ \emph {et~al.}(2020)\citenamefont
  {Iaconis}, \citenamefont {Lucas},\ and\ \citenamefont
  {Chen}}]{iaconis2020measurementinducedphase}%
  \BibitemOpen
  \bibfield  {author} {\bibinfo {author} {\bibfnamefont {J.}~\bibnamefont
  {Iaconis}}, \bibinfo {author} {\bibfnamefont {A.}~\bibnamefont {Lucas}},\
  and\ \bibinfo {author} {\bibfnamefont {X.}~\bibnamefont {Chen}},\ }\href
  {https://doi.org/10.1103/PhysRevB.102.224311} {\bibfield  {journal} {\bibinfo
   {journal} {Phys. Rev. B}\ }\textbf {\bibinfo {volume} {102}},\ \bibinfo
  {pages} {224311} (\bibinfo {year} {2020})}\BibitemShut {NoStop}%
\bibitem [{\citenamefont {Han}\ and\ \citenamefont
  {Chen}(2022)}]{han2022measurementinducedcriticality}%
  \BibitemOpen
  \bibfield  {author} {\bibinfo {author} {\bibfnamefont {Y.}~\bibnamefont
  {Han}}\ and\ \bibinfo {author} {\bibfnamefont {X.}~\bibnamefont {Chen}},\
  }\href {https://doi.org/10.1103/PhysRevB.105.064306} {\bibfield  {journal}
  {\bibinfo  {journal} {Phys. Rev. B}\ }\textbf {\bibinfo {volume} {105}},\
  \bibinfo {pages} {064306} (\bibinfo {year} {2022})}\BibitemShut {NoStop}%
\bibitem [{\citenamefont {Sierant}\ and\ \citenamefont
  {Turkeshi}(2022)}]{sierant2022universalbehaviorbeyond}%
  \BibitemOpen
  \bibfield  {author} {\bibinfo {author} {\bibfnamefont {P.}~\bibnamefont
  {Sierant}}\ and\ \bibinfo {author} {\bibfnamefont {X.}~\bibnamefont
  {Turkeshi}},\ }\href {https://doi.org/10.1103/PhysRevLett.128.130605}
  {\bibfield  {journal} {\bibinfo  {journal} {Phys. Rev. Lett.}\ }\textbf
  {\bibinfo {volume} {128}},\ \bibinfo {pages} {130605} (\bibinfo {year}
  {2022})}\BibitemShut {NoStop}%
\bibitem [{\citenamefont {Zabalo}\ \emph {et~al.}(2022)\citenamefont {Zabalo},
  \citenamefont {Gullans}, \citenamefont {Wilson}, \citenamefont {Vasseur},
  \citenamefont {Ludwig}, \citenamefont {Gopalakrishnan}, \citenamefont
  {Huse},\ and\ \citenamefont {Pixley}}]{zabalo2022operatorscalingdimensions}%
  \BibitemOpen
  \bibfield  {author} {\bibinfo {author} {\bibfnamefont {A.}~\bibnamefont
  {Zabalo}}, \bibinfo {author} {\bibfnamefont {M.~J.}\ \bibnamefont {Gullans}},
  \bibinfo {author} {\bibfnamefont {J.~H.}\ \bibnamefont {Wilson}}, \bibinfo
  {author} {\bibfnamefont {R.}~\bibnamefont {Vasseur}}, \bibinfo {author}
  {\bibfnamefont {A.~W.~W.}\ \bibnamefont {Ludwig}}, \bibinfo {author}
  {\bibfnamefont {S.}~\bibnamefont {Gopalakrishnan}}, \bibinfo {author}
  {\bibfnamefont {D.~A.}\ \bibnamefont {Huse}},\ and\ \bibinfo {author}
  {\bibfnamefont {J.~H.}\ \bibnamefont {Pixley}},\ }\href
  {https://doi.org/10.1103/PhysRevLett.128.050602} {\bibfield  {journal}
  {\bibinfo  {journal} {Phys. Rev. Lett.}\ }\textbf {\bibinfo {volume} {128}},\
  \bibinfo {pages} {050602} (\bibinfo {year} {2022})}\BibitemShut {NoStop}%
\bibitem [{\citenamefont {Iaconis}\ and\ \citenamefont
  {Chen}(2021)}]{iaconis2021multifractalityinnonunitary}%
  \BibitemOpen
  \bibfield  {author} {\bibinfo {author} {\bibfnamefont {J.}~\bibnamefont
  {Iaconis}}\ and\ \bibinfo {author} {\bibfnamefont {X.}~\bibnamefont {Chen}},\
  }\href {https://doi.org/10.1103/PhysRevB.104.214307} {\bibfield  {journal}
  {\bibinfo  {journal} {Phys. Rev. B}\ }\textbf {\bibinfo {volume} {104}},\
  \bibinfo {pages} {214307} (\bibinfo {year} {2021})}\BibitemShut {NoStop}%
\bibitem [{\citenamefont {Kalsi}\ \emph {et~al.}(2022)\citenamefont {Kalsi},
  \citenamefont {Romito},\ and\ \citenamefont
  {Schomerus}}]{kalsi2022threefoldway}%
  \BibitemOpen
  \bibfield  {author} {\bibinfo {author} {\bibfnamefont {T.}~\bibnamefont
  {Kalsi}}, \bibinfo {author} {\bibfnamefont {A.}~\bibnamefont {Romito}},\ and\
  \bibinfo {author} {\bibfnamefont {H.}~\bibnamefont {Schomerus}},\ }\href
  {https://doi.org/10.1088/1751-8121/ac71e8} {\bibfield  {journal} {\bibinfo
  {journal} {J. Phys. A: Math. Theor.}\ }\textbf {\bibinfo {volume} {55}},\
  \bibinfo {pages} {264009} (\bibinfo {year} {2022})}\BibitemShut {NoStop}%
\bibitem [{\citenamefont {Zabalo}\ \emph {et~al.}()\citenamefont {Zabalo},
  \citenamefont {Wilson}, \citenamefont {Gullans}, \citenamefont {Vasseur},
  \citenamefont {Gopalakrishnan}, \citenamefont {Huse},\ and\ \citenamefont
  {Pixley}}]{zabalo2022infiniterandomnesscriticality}%
  \BibitemOpen
  \bibfield  {author} {\bibinfo {author} {\bibfnamefont {A.}~\bibnamefont
  {Zabalo}}, \bibinfo {author} {\bibfnamefont {J.~H.}\ \bibnamefont {Wilson}},
  \bibinfo {author} {\bibfnamefont {M.~J.}\ \bibnamefont {Gullans}}, \bibinfo
  {author} {\bibfnamefont {R.}~\bibnamefont {Vasseur}}, \bibinfo {author}
  {\bibfnamefont {S.}~\bibnamefont {Gopalakrishnan}}, \bibinfo {author}
  {\bibfnamefont {D.~A.}\ \bibnamefont {Huse}},\ and\ \bibinfo {author}
  {\bibfnamefont {J.~H.}\ \bibnamefont {Pixley}},\ }\href@noop {} {}\Eprint
  {https://arxiv.org/abs/2205.14002} {arXiv:2205.14002} \BibitemShut {NoStop}%
\bibitem [{\citenamefont {Weinstein}\ \emph {et~al.}()\citenamefont
  {Weinstein}, \citenamefont {Kelly}, \citenamefont {Marino},\ and\
  \citenamefont {Altman}}]{weinstein2022scramblngtransitionin}%
  \BibitemOpen
  \bibfield  {author} {\bibinfo {author} {\bibfnamefont {Z.}~\bibnamefont
  {Weinstein}}, \bibinfo {author} {\bibfnamefont {S.~P.}\ \bibnamefont
  {Kelly}}, \bibinfo {author} {\bibfnamefont {J.}~\bibnamefont {Marino}},\ and\
  \bibinfo {author} {\bibfnamefont {E.}~\bibnamefont {Altman}},\ }\href@noop {}
  {}\Eprint {https://arxiv.org/abs/2210.14242} {arXiv:2210.14242} \BibitemShut
  {NoStop}%
\bibitem [{\citenamefont {Alberton}\ \emph {et~al.}(2021)\citenamefont
  {Alberton}, \citenamefont {Buchhold},\ and\ \citenamefont
  {Diehl}}]{alberton2021entanglementtransitionin}%
  \BibitemOpen
  \bibfield  {author} {\bibinfo {author} {\bibfnamefont {O.}~\bibnamefont
  {Alberton}}, \bibinfo {author} {\bibfnamefont {M.}~\bibnamefont {Buchhold}},\
  and\ \bibinfo {author} {\bibfnamefont {S.}~\bibnamefont {Diehl}},\ }\href
  {https://doi.org/10.1103/PhysRevLett.126.170602} {\bibfield  {journal}
  {\bibinfo  {journal} {Phys. Rev. Lett.}\ }\textbf {\bibinfo {volume} {126}},\
  \bibinfo {pages} {170602} (\bibinfo {year} {2021})}\BibitemShut {NoStop}%
\bibitem [{\citenamefont {Buchhold}\ \emph {et~al.}(2021)\citenamefont
  {Buchhold}, \citenamefont {Minoguchi}, \citenamefont {Altland},\ and\
  \citenamefont {Diehl}}]{buchhold2021effectivetheoryfor}%
  \BibitemOpen
  \bibfield  {author} {\bibinfo {author} {\bibfnamefont {M.}~\bibnamefont
  {Buchhold}}, \bibinfo {author} {\bibfnamefont {Y.}~\bibnamefont {Minoguchi}},
  \bibinfo {author} {\bibfnamefont {A.}~\bibnamefont {Altland}},\ and\ \bibinfo
  {author} {\bibfnamefont {S.}~\bibnamefont {Diehl}},\ }\href
  {https://doi.org/10.1103/PhysRevX.11.041004} {\bibfield  {journal} {\bibinfo
  {journal} {Phys. Rev. X}\ }\textbf {\bibinfo {volume} {11}},\ \bibinfo
  {pages} {041004} (\bibinfo {year} {2021})}\BibitemShut {NoStop}%
\bibitem [{\citenamefont {Turkeshi}\ \emph {et~al.}(2021)\citenamefont
  {Turkeshi}, \citenamefont {Biella}, \citenamefont {Fazio}, \citenamefont
  {Dalmonte},\ and\ \citenamefont
  {Schir\'o}}]{turkeshi2021measurementinducedentanglement}%
  \BibitemOpen
  \bibfield  {author} {\bibinfo {author} {\bibfnamefont {X.}~\bibnamefont
  {Turkeshi}}, \bibinfo {author} {\bibfnamefont {A.}~\bibnamefont {Biella}},
  \bibinfo {author} {\bibfnamefont {R.}~\bibnamefont {Fazio}}, \bibinfo
  {author} {\bibfnamefont {M.}~\bibnamefont {Dalmonte}},\ and\ \bibinfo
  {author} {\bibfnamefont {M.}~\bibnamefont {Schir\'o}},\ }\href
  {https://doi.org/10.1103/PhysRevB.103.224210} {\bibfield  {journal} {\bibinfo
   {journal} {Phys. Rev. B}\ }\textbf {\bibinfo {volume} {103}},\ \bibinfo
  {pages} {224210} (\bibinfo {year} {2021})}\BibitemShut {NoStop}%
\bibitem [{\citenamefont {Turkeshi}\ \emph
  {et~al.}(2022{\natexlab{a}})\citenamefont {Turkeshi}, \citenamefont
  {Dalmonte}, \citenamefont {Fazio},\ and\ \citenamefont
  {Schir\`o}}]{turkeshi2022entanglementtransitionsfrom}%
  \BibitemOpen
  \bibfield  {author} {\bibinfo {author} {\bibfnamefont {X.}~\bibnamefont
  {Turkeshi}}, \bibinfo {author} {\bibfnamefont {M.}~\bibnamefont {Dalmonte}},
  \bibinfo {author} {\bibfnamefont {R.}~\bibnamefont {Fazio}},\ and\ \bibinfo
  {author} {\bibfnamefont {M.}~\bibnamefont {Schir\`o}},\ }\href
  {https://doi.org/10.1103/PhysRevB.105.L241114} {\bibfield  {journal}
  {\bibinfo  {journal} {Phys. Rev. B}\ }\textbf {\bibinfo {volume} {105}},\
  \bibinfo {pages} {L241114} (\bibinfo {year}
  {2022}{\natexlab{a}})}\BibitemShut {NoStop}%
\bibitem [{\citenamefont {Turkeshi}\ and\ \citenamefont
  {Schiró}()}]{turkeshi2022entanglementandcorrelation}%
  \BibitemOpen
  \bibfield  {author} {\bibinfo {author} {\bibfnamefont {X.}~\bibnamefont
  {Turkeshi}}\ and\ \bibinfo {author} {\bibfnamefont {M.}~\bibnamefont
  {Schiró}},\ }\href@noop {} {}\Eprint {https://arxiv.org/abs/2201.09895}
  {arXiv:2201.09895} \BibitemShut {NoStop}%
\bibitem [{\citenamefont {Gal}\ \emph {et~al.}()\citenamefont {Gal},
  \citenamefont {Turkeshi},\ and\ \citenamefont
  {Schirò}}]{gal2022volumetoarea}%
  \BibitemOpen
  \bibfield  {author} {\bibinfo {author} {\bibfnamefont {Y.~L.}\ \bibnamefont
  {Gal}}, \bibinfo {author} {\bibfnamefont {X.}~\bibnamefont {Turkeshi}},\ and\
  \bibinfo {author} {\bibfnamefont {M.}~\bibnamefont {Schirò}},\ }\href@noop
  {} {}\Eprint {https://arxiv.org/abs/2210.11937} {arXiv:2210.11937}
  \BibitemShut {NoStop}%
\bibitem [{\citenamefont {Turkeshi}\ \emph
  {et~al.}(2022{\natexlab{b}})\citenamefont {Turkeshi}, \citenamefont
  {Piroli},\ and\ \citenamefont
  {Schir\'o}}]{turkeshi2022enhancedentanglementnegativity}%
  \BibitemOpen
  \bibfield  {author} {\bibinfo {author} {\bibfnamefont {X.}~\bibnamefont
  {Turkeshi}}, \bibinfo {author} {\bibfnamefont {L.}~\bibnamefont {Piroli}},\
  and\ \bibinfo {author} {\bibfnamefont {M.}~\bibnamefont {Schir\'o}},\ }\href
  {https://doi.org/10.1103/PhysRevB.106.024304} {\bibfield  {journal} {\bibinfo
   {journal} {Phys. Rev. B}\ }\textbf {\bibinfo {volume} {106}},\ \bibinfo
  {pages} {024304} (\bibinfo {year} {2022}{\natexlab{b}})}\BibitemShut
  {NoStop}%
\bibitem [{\citenamefont {Ladewig}\ \emph {et~al.}(2022)\citenamefont
  {Ladewig}, \citenamefont {Diehl},\ and\ \citenamefont
  {Buchhold}}]{ladewig2022monitoredopenfermion}%
  \BibitemOpen
  \bibfield  {author} {\bibinfo {author} {\bibfnamefont {B.}~\bibnamefont
  {Ladewig}}, \bibinfo {author} {\bibfnamefont {S.}~\bibnamefont {Diehl}},\
  and\ \bibinfo {author} {\bibfnamefont {M.}~\bibnamefont {Buchhold}},\ }\href
  {https://doi.org/10.1103/PhysRevResearch.4.033001} {\bibfield  {journal}
  {\bibinfo  {journal} {Phys. Rev. Research}\ }\textbf {\bibinfo {volume}
  {4}},\ \bibinfo {pages} {033001} (\bibinfo {year} {2022})}\BibitemShut
  {NoStop}%
\bibitem [{\citenamefont {Coppola}\ \emph {et~al.}(2022)\citenamefont
  {Coppola}, \citenamefont {Tirrito}, \citenamefont {Karevski},\ and\
  \citenamefont {Collura}}]{coppola2022growth}%
  \BibitemOpen
  \bibfield  {author} {\bibinfo {author} {\bibfnamefont {M.}~\bibnamefont
  {Coppola}}, \bibinfo {author} {\bibfnamefont {E.}~\bibnamefont {Tirrito}},
  \bibinfo {author} {\bibfnamefont {D.}~\bibnamefont {Karevski}},\ and\
  \bibinfo {author} {\bibfnamefont {M.}~\bibnamefont {Collura}},\ }\href
  {https://doi.org/10.1103/PhysRevB.105.094303} {\bibfield  {journal} {\bibinfo
   {journal} {Phys. Rev. B}\ }\textbf {\bibinfo {volume} {105}},\ \bibinfo
  {pages} {094303} (\bibinfo {year} {2022})}\BibitemShut {NoStop}%
\bibitem [{\citenamefont {Boorman}\ \emph {et~al.}(2022)\citenamefont
  {Boorman}, \citenamefont {Szyniszewski}, \citenamefont {Schomerus},\ and\
  \citenamefont {Romito}}]{boorman2022diagonisticsofentanglement}%
  \BibitemOpen
  \bibfield  {author} {\bibinfo {author} {\bibfnamefont {T.}~\bibnamefont
  {Boorman}}, \bibinfo {author} {\bibfnamefont {M.}~\bibnamefont
  {Szyniszewski}}, \bibinfo {author} {\bibfnamefont {H.}~\bibnamefont
  {Schomerus}},\ and\ \bibinfo {author} {\bibfnamefont {A.}~\bibnamefont
  {Romito}},\ }\href {https://doi.org/10.1103/PhysRevB.105.144202} {\bibfield
  {journal} {\bibinfo  {journal} {Phys. Rev. B}\ }\textbf {\bibinfo {volume}
  {105}},\ \bibinfo {pages} {144202} (\bibinfo {year} {2022})}\BibitemShut
  {NoStop}%
\bibitem [{\citenamefont {Botzung}\ \emph {et~al.}(2021)\citenamefont
  {Botzung}, \citenamefont {Diehl},\ and\ \citenamefont
  {M\"uller}}]{botzung2021engineereddissipationinduced}%
  \BibitemOpen
  \bibfield  {author} {\bibinfo {author} {\bibfnamefont {T.}~\bibnamefont
  {Botzung}}, \bibinfo {author} {\bibfnamefont {S.}~\bibnamefont {Diehl}},\
  and\ \bibinfo {author} {\bibfnamefont {M.}~\bibnamefont {M\"uller}},\ }\href
  {https://doi.org/10.1103/PhysRevB.104.184422} {\bibfield  {journal} {\bibinfo
   {journal} {Phys. Rev. B}\ }\textbf {\bibinfo {volume} {104}},\ \bibinfo
  {pages} {184422} (\bibinfo {year} {2021})}\BibitemShut {NoStop}%
\bibitem [{\citenamefont {Fleckenstein}\ \emph {et~al.}(2022)\citenamefont
  {Fleckenstein}, \citenamefont {Zorzato}, \citenamefont {Varjas},
  \citenamefont {Bergholtz}, \citenamefont {Bardarson},\ and\ \citenamefont
  {Tiwari}}]{fleckenstein2022nonhermitiantopology}%
  \BibitemOpen
  \bibfield  {author} {\bibinfo {author} {\bibfnamefont {C.}~\bibnamefont
  {Fleckenstein}}, \bibinfo {author} {\bibfnamefont {A.}~\bibnamefont
  {Zorzato}}, \bibinfo {author} {\bibfnamefont {D.}~\bibnamefont {Varjas}},
  \bibinfo {author} {\bibfnamefont {E.~J.}\ \bibnamefont {Bergholtz}}, \bibinfo
  {author} {\bibfnamefont {J.~H.}\ \bibnamefont {Bardarson}},\ and\ \bibinfo
  {author} {\bibfnamefont {A.}~\bibnamefont {Tiwari}},\ }\href
  {https://doi.org/10.1103/PhysRevResearch.4.L032026} {\bibfield  {journal}
  {\bibinfo  {journal} {Phys. Rev. Research}\ }\textbf {\bibinfo {volume}
  {4}},\ \bibinfo {pages} {L032026} (\bibinfo {year} {2022})}\BibitemShut
  {NoStop}%
\bibitem [{\citenamefont {Kells}\ \emph {et~al.}()\citenamefont {Kells},
  \citenamefont {Meidan},\ and\ \citenamefont
  {Romito}}]{kells2021topologicaltransitionswith}%
  \BibitemOpen
  \bibfield  {author} {\bibinfo {author} {\bibfnamefont {G.}~\bibnamefont
  {Kells}}, \bibinfo {author} {\bibfnamefont {D.}~\bibnamefont {Meidan}},\ and\
  \bibinfo {author} {\bibfnamefont {A.}~\bibnamefont {Romito}},\ }\href@noop {}
  {}\Eprint {https://arxiv.org/abs/2112.09787} {arXiv:2112.09787} \BibitemShut
  {NoStop}%
\bibitem [{\citenamefont {Szyniszewski}\ \emph {et~al.}()\citenamefont
  {Szyniszewski}, \citenamefont {Lunt},\ and\ \citenamefont
  {Pal}}]{szyniszewski2022disorderedmonitoredfree}%
  \BibitemOpen
  \bibfield  {author} {\bibinfo {author} {\bibfnamefont {M.}~\bibnamefont
  {Szyniszewski}}, \bibinfo {author} {\bibfnamefont {O.}~\bibnamefont {Lunt}},\
  and\ \bibinfo {author} {\bibfnamefont {A.}~\bibnamefont {Pal}},\ }\href@noop
  {} {}\Eprint {https://arxiv.org/abs/2211.02534} {arXiv:2211.02534}
  \BibitemShut {NoStop}%
\bibitem [{\citenamefont {Piccitto}\ \emph {et~al.}(2022)\citenamefont
  {Piccitto}, \citenamefont {Russomanno},\ and\ \citenamefont
  {Rossini}}]{piccitto2022entanglementtransitionsin}%
  \BibitemOpen
  \bibfield  {author} {\bibinfo {author} {\bibfnamefont {G.}~\bibnamefont
  {Piccitto}}, \bibinfo {author} {\bibfnamefont {A.}~\bibnamefont
  {Russomanno}},\ and\ \bibinfo {author} {\bibfnamefont {D.}~\bibnamefont
  {Rossini}},\ }\href {https://doi.org/10.1103/PhysRevB.105.064305} {\bibfield
  {journal} {\bibinfo  {journal} {Phys. Rev. B}\ }\textbf {\bibinfo {volume}
  {105}},\ \bibinfo {pages} {064305} (\bibinfo {year} {2022})}\BibitemShut
  {NoStop}%
\bibitem [{\citenamefont {Kawabata}\ \emph {et~al.}()\citenamefont {Kawabata},
  \citenamefont {Numasawa},\ and\ \citenamefont
  {Ryu}}]{kawabata2022entanglementphasetransition}%
  \BibitemOpen
  \bibfield  {author} {\bibinfo {author} {\bibfnamefont {K.}~\bibnamefont
  {Kawabata}}, \bibinfo {author} {\bibfnamefont {T.}~\bibnamefont {Numasawa}},\
  and\ \bibinfo {author} {\bibfnamefont {S.}~\bibnamefont {Ryu}},\ }\href@noop
  {} {}\Eprint {https://arxiv.org/abs/2206.05384} {arXiv:2206.05384}
  \BibitemShut {NoStop}%
\bibitem [{\citenamefont {Minoguchi}\ \emph {et~al.}(2022)\citenamefont
  {Minoguchi}, \citenamefont {Rabl},\ and\ \citenamefont
  {Buchhold}}]{minoguchi2022continuousgaussianmeasurements}%
  \BibitemOpen
  \bibfield  {author} {\bibinfo {author} {\bibfnamefont {Y.}~\bibnamefont
  {Minoguchi}}, \bibinfo {author} {\bibfnamefont {P.}~\bibnamefont {Rabl}},\
  and\ \bibinfo {author} {\bibfnamefont {M.}~\bibnamefont {Buchhold}},\ }\href
  {https://doi.org/10.21468/SciPostPhys.12.1.009} {\bibfield  {journal}
  {\bibinfo  {journal} {SciPost Phys.}\ }\textbf {\bibinfo {volume} {12}},\
  \bibinfo {pages} {009} (\bibinfo {year} {2022})}\BibitemShut {NoStop}%
\bibitem [{\citenamefont {M\"uller}\ \emph {et~al.}(2022)\citenamefont
  {M\"uller}, \citenamefont {Diehl},\ and\ \citenamefont
  {Buchhold}}]{muller2022measurementinduceddark}%
  \BibitemOpen
  \bibfield  {author} {\bibinfo {author} {\bibfnamefont {T.}~\bibnamefont
  {M\"uller}}, \bibinfo {author} {\bibfnamefont {S.}~\bibnamefont {Diehl}},\
  and\ \bibinfo {author} {\bibfnamefont {M.}~\bibnamefont {Buchhold}},\ }\href
  {https://doi.org/10.1103/PhysRevLett.128.010605} {\bibfield  {journal}
  {\bibinfo  {journal} {Phys. Rev. Lett.}\ }\textbf {\bibinfo {volume} {128}},\
  \bibinfo {pages} {010605} (\bibinfo {year} {2022})}\BibitemShut {NoStop}%
\bibitem [{\citenamefont {Koh}\ \emph {et~al.}()\citenamefont {Koh},
  \citenamefont {Sun}, \citenamefont {Motta},\ and\ \citenamefont
  {Minnich}}]{koh2022experimentalrealizationof}%
  \BibitemOpen
  \bibfield  {author} {\bibinfo {author} {\bibfnamefont {J.~M.}\ \bibnamefont
  {Koh}}, \bibinfo {author} {\bibfnamefont {S.-N.}\ \bibnamefont {Sun}},
  \bibinfo {author} {\bibfnamefont {M.}~\bibnamefont {Motta}},\ and\ \bibinfo
  {author} {\bibfnamefont {A.~J.}\ \bibnamefont {Minnich}},\ }\href@noop {}
  {}\Eprint {https://arxiv.org/abs/2203.04338} {arXiv:2203.04338} \BibitemShut
  {NoStop}%
\bibitem [{\citenamefont {Ippoliti}\ and\ \citenamefont
  {Khemani}(2021)}]{ippoliti2021postselectionfreeentanglement}%
  \BibitemOpen
  \bibfield  {author} {\bibinfo {author} {\bibfnamefont {M.}~\bibnamefont
  {Ippoliti}}\ and\ \bibinfo {author} {\bibfnamefont {V.}~\bibnamefont
  {Khemani}},\ }\href {https://doi.org/10.1103/PhysRevLett.126.060501}
  {\bibfield  {journal} {\bibinfo  {journal} {Phys. Rev. Lett.}\ }\textbf
  {\bibinfo {volume} {126}},\ \bibinfo {pages} {060501} (\bibinfo {year}
  {2021})}\BibitemShut {NoStop}%
\bibitem [{\citenamefont {Lu}\ and\ \citenamefont
  {Grover}(2021)}]{lu2021spacetimeduality}%
  \BibitemOpen
  \bibfield  {author} {\bibinfo {author} {\bibfnamefont {T.-C.}\ \bibnamefont
  {Lu}}\ and\ \bibinfo {author} {\bibfnamefont {T.}~\bibnamefont {Grover}},\
  }\href {https://doi.org/10.1103/PRXQuantum.2.040319} {\bibfield  {journal}
  {\bibinfo  {journal} {PRX Quantum}\ }\textbf {\bibinfo {volume} {2}},\
  \bibinfo {pages} {040319} (\bibinfo {year} {2021})}\BibitemShut {NoStop}%
\bibitem [{\citenamefont {Noel}\ \emph {et~al.}(2022)\citenamefont {Noel},
  \citenamefont {Niroula}, \citenamefont {Zhu}, \citenamefont {Risinger},
  \citenamefont {Egan}, \citenamefont {Biswas}, \citenamefont {Cetina},
  \citenamefont {Gorshkov}, \citenamefont {Gullans}, \citenamefont {Huse},\
  and\ \citenamefont {Monroe}}]{noel2022measurementinducedquantum}%
  \BibitemOpen
  \bibfield  {author} {\bibinfo {author} {\bibfnamefont {C.}~\bibnamefont
  {Noel}}, \bibinfo {author} {\bibfnamefont {P.}~\bibnamefont {Niroula}},
  \bibinfo {author} {\bibfnamefont {D.}~\bibnamefont {Zhu}}, \bibinfo {author}
  {\bibfnamefont {A.}~\bibnamefont {Risinger}}, \bibinfo {author}
  {\bibfnamefont {L.}~\bibnamefont {Egan}}, \bibinfo {author} {\bibfnamefont
  {D.}~\bibnamefont {Biswas}}, \bibinfo {author} {\bibfnamefont
  {M.}~\bibnamefont {Cetina}}, \bibinfo {author} {\bibfnamefont {A.~V.}\
  \bibnamefont {Gorshkov}}, \bibinfo {author} {\bibfnamefont {M.~J.}\
  \bibnamefont {Gullans}}, \bibinfo {author} {\bibfnamefont {D.~A.}\
  \bibnamefont {Huse}},\ and\ \bibinfo {author} {\bibfnamefont
  {C.}~\bibnamefont {Monroe}},\ }\href
  {https://doi.org/10.1038/s41567-022-01619-7} {\bibfield  {journal} {\bibinfo
  {journal} {Nature Physics}\ }\textbf {\bibinfo {volume} {18}},\ \bibinfo
  {pages} {760} (\bibinfo {year} {2022})}\BibitemShut {NoStop}%
\bibitem [{\citenamefont {Claeys}\ \emph {et~al.}()\citenamefont {Claeys},
  \citenamefont {Henry}, \citenamefont {Vicary},\ and\ \citenamefont
  {Lamacraft}}]{claeys2022exactdynamicsin}%
  \BibitemOpen
  \bibfield  {author} {\bibinfo {author} {\bibfnamefont {P.~W.}\ \bibnamefont
  {Claeys}}, \bibinfo {author} {\bibfnamefont {M.}~\bibnamefont {Henry}},
  \bibinfo {author} {\bibfnamefont {J.}~\bibnamefont {Vicary}},\ and\ \bibinfo
  {author} {\bibfnamefont {A.}~\bibnamefont {Lamacraft}},\ }\href@noop {}
  {}\Eprint {https://arxiv.org/abs/2207.00025} {arXiv:2207.00025} \BibitemShut
  {NoStop}%
\bibitem [{\citenamefont {Li}\ \emph {et~al.}({\natexlab{c}})\citenamefont
  {Li}, \citenamefont {Zou}, \citenamefont {Glorioso}, \citenamefont {Altman},\
  and\ \citenamefont {Fisher}}]{li2022crossentropybenchmark}%
  \BibitemOpen
  \bibfield  {author} {\bibinfo {author} {\bibfnamefont {Y.}~\bibnamefont
  {Li}}, \bibinfo {author} {\bibfnamefont {Y.}~\bibnamefont {Zou}}, \bibinfo
  {author} {\bibfnamefont {P.}~\bibnamefont {Glorioso}}, \bibinfo {author}
  {\bibfnamefont {E.}~\bibnamefont {Altman}},\ and\ \bibinfo {author}
  {\bibfnamefont {M.~P.~A.}\ \bibnamefont {Fisher}},\ }\href@noop {} {}
  ({\natexlab{c}}),\ \Eprint {https://arxiv.org/abs/2209.00609}
  {arXiv:2209.00609} \BibitemShut {NoStop}%
\bibitem [{\citenamefont {Feng}\ \emph {et~al.}()\citenamefont {Feng},
  \citenamefont {Skinner},\ and\ \citenamefont
  {Nahum}}]{feng2022measurementinducedphase}%
  \BibitemOpen
  \bibfield  {author} {\bibinfo {author} {\bibfnamefont {X.}~\bibnamefont
  {Feng}}, \bibinfo {author} {\bibfnamefont {B.}~\bibnamefont {Skinner}},\ and\
  \bibinfo {author} {\bibfnamefont {A.}~\bibnamefont {Nahum}},\ }\href@noop {}
  {}\Eprint {https://arxiv.org/abs/2210.07264} {arXiv:2210.07264} \BibitemShut
  {NoStop}%
\bibitem [{\citenamefont {Iadecola}\ \emph {et~al.}()\citenamefont {Iadecola},
  \citenamefont {Ganeshan}, \citenamefont {Pixley},\ and\ \citenamefont
  {Wilson}}]{iadecola2022dynamicalentanglementtransition}%
  \BibitemOpen
  \bibfield  {author} {\bibinfo {author} {\bibfnamefont {T.}~\bibnamefont
  {Iadecola}}, \bibinfo {author} {\bibfnamefont {S.}~\bibnamefont {Ganeshan}},
  \bibinfo {author} {\bibfnamefont {J.~H.}\ \bibnamefont {Pixley}},\ and\
  \bibinfo {author} {\bibfnamefont {J.~H.}\ \bibnamefont {Wilson}},\
  }\href@noop {} {}\Eprint {https://arxiv.org/abs/2207.12415}
  {arXiv:2207.12415} \BibitemShut {NoStop}%
\bibitem [{\citenamefont {Buchhold}\ \emph {et~al.}()\citenamefont {Buchhold},
  \citenamefont {Müller},\ and\ \citenamefont
  {Diehl}}]{buchhold2022revealingmeasurementinduced}%
  \BibitemOpen
  \bibfield  {author} {\bibinfo {author} {\bibfnamefont {M.}~\bibnamefont
  {Buchhold}}, \bibinfo {author} {\bibfnamefont {T.}~\bibnamefont {Müller}},\
  and\ \bibinfo {author} {\bibfnamefont {S.}~\bibnamefont {Diehl}},\
  }\href@noop {} {}\Eprint {https://arxiv.org/abs/2208.10506}
  {arXiv:2208.10506} \BibitemShut {NoStop}%
\bibitem [{\citenamefont {Wang}\ \emph {et~al.}()\citenamefont {Wang},
  \citenamefont {Fang},\ and\ \citenamefont
  {Ren}}]{wang2022absenceofentanglement}%
  \BibitemOpen
  \bibfield  {author} {\bibinfo {author} {\bibfnamefont {Y.-p.}\ \bibnamefont
  {Wang}}, \bibinfo {author} {\bibfnamefont {C.}~\bibnamefont {Fang}},\ and\
  \bibinfo {author} {\bibfnamefont {J.}~\bibnamefont {Ren}},\ }\href@noop {}
  {}\Eprint {https://arxiv.org/abs/2209.11241} {arXiv:2209.11241} \BibitemShut
  {NoStop}%
\bibitem [{\citenamefont {Ravindranath}\ \emph {et~al.}()\citenamefont
  {Ravindranath}, \citenamefont {Han}, \citenamefont {Yang},\ and\
  \citenamefont {Chen}}]{ravindranath2022entanglementsteeringin}%
  \BibitemOpen
  \bibfield  {author} {\bibinfo {author} {\bibfnamefont {V.}~\bibnamefont
  {Ravindranath}}, \bibinfo {author} {\bibfnamefont {Y.}~\bibnamefont {Han}},
  \bibinfo {author} {\bibfnamefont {Z.-C.}\ \bibnamefont {Yang}},\ and\
  \bibinfo {author} {\bibfnamefont {X.}~\bibnamefont {Chen}},\ }\href@noop {}
  {}\Eprint {https://arxiv.org/abs/2211.05162} {arXiv:2211.05162} \BibitemShut
  {NoStop}%
\bibitem [{\citenamefont {O'Dea}\ \emph {et~al.}()\citenamefont {O'Dea},
  \citenamefont {Morningstar}, \citenamefont {Gopalakrishnan},\ and\
  \citenamefont {Khemani}}]{odea2022entanglementandabsorbing}%
  \BibitemOpen
  \bibfield  {author} {\bibinfo {author} {\bibfnamefont {N.}~\bibnamefont
  {O'Dea}}, \bibinfo {author} {\bibfnamefont {A.}~\bibnamefont {Morningstar}},
  \bibinfo {author} {\bibfnamefont {S.}~\bibnamefont {Gopalakrishnan}},\ and\
  \bibinfo {author} {\bibfnamefont {V.}~\bibnamefont {Khemani}},\ }\href@noop
  {} {}\Eprint {https://arxiv.org/abs/2211.12526} {arXiv:2211.12526}
  \BibitemShut {NoStop}%
\bibitem [{\citenamefont {Friedman}\ \emph {et~al.}()\citenamefont {Friedman},
  \citenamefont {Hart},\ and\ \citenamefont
  {Nandkishore}}]{friedman2022measurementinducedphases}%
  \BibitemOpen
  \bibfield  {author} {\bibinfo {author} {\bibfnamefont {A.~J.}\ \bibnamefont
  {Friedman}}, \bibinfo {author} {\bibfnamefont {O.}~\bibnamefont {Hart}},\
  and\ \bibinfo {author} {\bibfnamefont {R.}~\bibnamefont {Nandkishore}},\
  }\href@noop {} {}\Eprint {https://arxiv.org/abs/2210.07256}
  {arXiv:2210.07256} \BibitemShut {NoStop}%
\bibitem [{sup()}]{supmat}%
  \BibitemOpen
  \href@noop {} {\bibinfo {title} {See the supplementary material, including
  ref.~\cite{Kawashima1993}.}}\BibitemShut {Stop}%
\bibitem [{\citenamefont {Aaronson}\ and\ \citenamefont
  {Gottesman}(2004)}]{aaronson2004improvedsimulationof}%
  \BibitemOpen
  \bibfield  {author} {\bibinfo {author} {\bibfnamefont {S.}~\bibnamefont
  {Aaronson}}\ and\ \bibinfo {author} {\bibfnamefont {D.}~\bibnamefont
  {Gottesman}},\ }\href {https://doi.org/10.1103/PhysRevA.70.052328} {\bibfield
   {journal} {\bibinfo  {journal} {Phys. Rev. A}\ }\textbf {\bibinfo {volume}
  {70}},\ \bibinfo {pages} {052328} (\bibinfo {year} {2004})}\BibitemShut
  {NoStop}%
\bibitem [{\citenamefont {Gidney}(2021)}]{gidney2021stimfaststabilizer}%
  \BibitemOpen
  \bibfield  {author} {\bibinfo {author} {\bibfnamefont {C.}~\bibnamefont
  {Gidney}},\ }\href {https://doi.org/10.22331/q-2021-07-06-497} {\bibfield
  {journal} {\bibinfo  {journal} {{Quantum}}\ }\textbf {\bibinfo {volume}
  {5}},\ \bibinfo {pages} {497} (\bibinfo {year} {2021})}\BibitemShut {NoStop}%
\bibitem [{\citenamefont {Nielsen}\ and\ \citenamefont
  {Chuang}(2000)}]{nielsen00}%
  \BibitemOpen
  \bibfield  {author} {\bibinfo {author} {\bibfnamefont {M.~A.}\ \bibnamefont
  {Nielsen}}\ and\ \bibinfo {author} {\bibfnamefont {I.~L.}\ \bibnamefont
  {Chuang}},\ }\href@noop {} {\emph {\bibinfo {title} {Quantum Computation and
  Quantum Information}}}\ (\bibinfo  {publisher} {Cambridge University Press},\
  \bibinfo {year} {2000})\BibitemShut {NoStop}%
\bibitem [{\citenamefont {Gross}\ \emph {et~al.}(2021)\citenamefont {Gross},
  \citenamefont {Nezami},\ and\ \citenamefont {Walter}}]{Gross2021}%
  \BibitemOpen
  \bibfield  {author} {\bibinfo {author} {\bibfnamefont {D.}~\bibnamefont
  {Gross}}, \bibinfo {author} {\bibfnamefont {S.}~\bibnamefont {Nezami}},\ and\
  \bibinfo {author} {\bibfnamefont {M.}~\bibnamefont {Walter}},\ }\href
  {https://doi.org/10.1007/s00220-021-04118-7} {\bibfield  {journal} {\bibinfo
  {journal} {Communications in Mathematical Physics}\ }\textbf {\bibinfo
  {volume} {385}},\ \bibinfo {pages} {1325} (\bibinfo {year}
  {2021})}\BibitemShut {NoStop}%
\bibitem [{Note1()}]{Note1}%
  \BibitemOpen
  \bibinfo {note} {We denote the Pauli matrices by $X_i$, $Y_i$, $Z_i$;
  $\mathinner {|{\uparrow _i}\rangle }$ and $\mathinner {|{\downarrow
  _i}\rangle }$ are the +1 and -1 eigenvectors of $Z_i$.}\BibitemShut {Stop}%
\bibitem [{\citenamefont {Hamma}\ \emph
  {et~al.}(2005{\natexlab{a}})\citenamefont {Hamma}, \citenamefont
  {Ionicioiu},\ and\ \citenamefont {Zanardi}}]{hamma2004}%
  \BibitemOpen
  \bibfield  {author} {\bibinfo {author} {\bibfnamefont {A.}~\bibnamefont
  {Hamma}}, \bibinfo {author} {\bibfnamefont {R.}~\bibnamefont {Ionicioiu}},\
  and\ \bibinfo {author} {\bibfnamefont {P.}~\bibnamefont {Zanardi}},\ }\href
  {https://doi.org/10.1103/PhysRevA.71.022315} {\bibfield  {journal} {\bibinfo
  {journal} {Phys. Rev. A}\ }\textbf {\bibinfo {volume} {71}},\ \bibinfo
  {pages} {022315} (\bibinfo {year} {2005}{\natexlab{a}})}\BibitemShut
  {NoStop}%
\bibitem [{\citenamefont {Hamma}\ \emph
  {et~al.}(2005{\natexlab{b}})\citenamefont {Hamma}, \citenamefont
  {Ionicioiu},\ and\ \citenamefont {Zanardi}}]{Hamma_2005}%
  \BibitemOpen
  \bibfield  {author} {\bibinfo {author} {\bibfnamefont {A.}~\bibnamefont
  {Hamma}}, \bibinfo {author} {\bibfnamefont {R.}~\bibnamefont {Ionicioiu}},\
  and\ \bibinfo {author} {\bibfnamefont {P.}~\bibnamefont {Zanardi}},\ }\href
  {https://doi.org/10.1016/j.physleta.2005.01.060} {\bibfield  {journal}
  {\bibinfo  {journal} {Phys. Lett. A}\ }\textbf {\bibinfo {volume} {337}},\
  \bibinfo {pages} {22} (\bibinfo {year} {2005}{\natexlab{b}})}\BibitemShut
  {NoStop}%
\bibitem [{\citenamefont {Minato}\ \emph {et~al.}(2022)\citenamefont {Minato},
  \citenamefont {Sugimoto}, \citenamefont {Kuwahara},\ and\ \citenamefont
  {Saito}}]{minato2022fateofmeasurementinduced}%
  \BibitemOpen
  \bibfield  {author} {\bibinfo {author} {\bibfnamefont {T.}~\bibnamefont
  {Minato}}, \bibinfo {author} {\bibfnamefont {K.}~\bibnamefont {Sugimoto}},
  \bibinfo {author} {\bibfnamefont {T.}~\bibnamefont {Kuwahara}},\ and\
  \bibinfo {author} {\bibfnamefont {K.}~\bibnamefont {Saito}},\ }\href
  {https://doi.org/10.1103/PhysRevLett.128.010603} {\bibfield  {journal}
  {\bibinfo  {journal} {Phys. Rev. Lett.}\ }\textbf {\bibinfo {volume} {128}},\
  \bibinfo {pages} {010603} (\bibinfo {year} {2022})}\BibitemShut {NoStop}%
\bibitem [{\citenamefont {Block}\ \emph {et~al.}(2022)\citenamefont {Block},
  \citenamefont {Bao}, \citenamefont {Choi}, \citenamefont {Altman},\ and\
  \citenamefont {Yao}}]{block2022measurementinducedtransition}%
  \BibitemOpen
  \bibfield  {author} {\bibinfo {author} {\bibfnamefont {M.}~\bibnamefont
  {Block}}, \bibinfo {author} {\bibfnamefont {Y.}~\bibnamefont {Bao}}, \bibinfo
  {author} {\bibfnamefont {S.}~\bibnamefont {Choi}}, \bibinfo {author}
  {\bibfnamefont {E.}~\bibnamefont {Altman}},\ and\ \bibinfo {author}
  {\bibfnamefont {N.~Y.}\ \bibnamefont {Yao}},\ }\href
  {https://doi.org/10.1103/PhysRevLett.128.010604} {\bibfield  {journal}
  {\bibinfo  {journal} {Phys. Rev. Lett.}\ }\textbf {\bibinfo {volume} {128}},\
  \bibinfo {pages} {010604} (\bibinfo {year} {2022})}\BibitemShut {NoStop}%
\bibitem [{\citenamefont {Sharma}\ \emph {et~al.}(2022)\citenamefont {Sharma},
  \citenamefont {Turkeshi}, \citenamefont {Fazio},\ and\ \citenamefont
  {Dalmonte}}]{sharma2022measurementinducedcriticality}%
  \BibitemOpen
  \bibfield  {author} {\bibinfo {author} {\bibfnamefont {S.}~\bibnamefont
  {Sharma}}, \bibinfo {author} {\bibfnamefont {X.}~\bibnamefont {Turkeshi}},
  \bibinfo {author} {\bibfnamefont {R.}~\bibnamefont {Fazio}},\ and\ \bibinfo
  {author} {\bibfnamefont {M.}~\bibnamefont {Dalmonte}},\ }\href
  {https://doi.org/10.21468/SciPostPhysCore.5.2.023} {\bibfield  {journal}
  {\bibinfo  {journal} {SciPost Phys. Core}\ }\textbf {\bibinfo {volume} {5}},\
  \bibinfo {pages} {023} (\bibinfo {year} {2022})}\BibitemShut {NoStop}%
\bibitem [{\citenamefont {Hashizume}\ \emph {et~al.}(2022)\citenamefont
  {Hashizume}, \citenamefont {Bentsen},\ and\ \citenamefont
  {Daley}}]{hashizum2022measurementinducedphase}%
  \BibitemOpen
  \bibfield  {author} {\bibinfo {author} {\bibfnamefont {T.}~\bibnamefont
  {Hashizume}}, \bibinfo {author} {\bibfnamefont {G.}~\bibnamefont {Bentsen}},\
  and\ \bibinfo {author} {\bibfnamefont {A.~J.}\ \bibnamefont {Daley}},\ }\href
  {https://doi.org/10.1103/PhysRevResearch.4.013174} {\bibfield  {journal}
  {\bibinfo  {journal} {Phys. Rev. Research}\ }\textbf {\bibinfo {volume}
  {4}},\ \bibinfo {pages} {013174} (\bibinfo {year} {2022})}\BibitemShut
  {NoStop}%
\bibitem [{\citenamefont {Wang}\ \emph {et~al.}(2013)\citenamefont {Wang},
  \citenamefont {Zhou}, \citenamefont {Liu}, \citenamefont {Garoni},\ and\
  \citenamefont {Deng}}]{Wang13}%
  \BibitemOpen
  \bibfield  {author} {\bibinfo {author} {\bibfnamefont {J.}~\bibnamefont
  {Wang}}, \bibinfo {author} {\bibfnamefont {Z.}~\bibnamefont {Zhou}}, \bibinfo
  {author} {\bibfnamefont {Q.}~\bibnamefont {Liu}}, \bibinfo {author}
  {\bibfnamefont {T.~M.}\ \bibnamefont {Garoni}},\ and\ \bibinfo {author}
  {\bibfnamefont {Y.}~\bibnamefont {Deng}},\ }\href
  {https://doi.org/10.1103/PhysRevE.88.042102} {\bibfield  {journal} {\bibinfo
  {journal} {Phys. Rev. E}\ }\textbf {\bibinfo {volume} {88}},\ \bibinfo
  {pages} {042102} (\bibinfo {year} {2013})}\BibitemShut {NoStop}%
\bibitem [{\citenamefont {Mendon\c{c}a}(2011)}]{Mendon11}%
  \BibitemOpen
  \bibfield  {author} {\bibinfo {author} {\bibfnamefont {J.~R.~G.}\
  \bibnamefont {Mendon\c{c}a}},\ }\href
  {https://doi.org/10.1103/PhysRevE.83.012102} {\bibfield  {journal} {\bibinfo
  {journal} {Phys. Rev. E}\ }\textbf {\bibinfo {volume} {83}},\ \bibinfo
  {pages} {012102} (\bibinfo {year} {2011})}\BibitemShut {NoStop}%
\bibitem [{Note2()}]{Note2}%
  \BibitemOpen
  \bibinfo {note} {For $p<p^{\protect \mathrm {APT}}_c$ we find $\tau _\star $
  as time at which $n_{\protect \mathrm {def}}(t)$ decays to $1/2$ of the
  saturation value observed at times $\propto L$; for $p\approx p^{ \protect
  \mathrm {APT} }_c$, $\tau _\star $ is the time at which $n_{\protect \mathrm
  {def}}(t)$ is equal to $c t^{\delta }/2$, where $c t^{\delta }$ is a
  power-law fitted at times $\propto L$; for $p>p^{\protect \mathrm {APT}}_c$,
  $n_{\protect \mathrm {def}}(t)$ is equal to $n_{\protect \mathrm
  {def}}(100)/2$ at $\tau _\star $.}\BibitemShut {Stop}%
\bibitem [{Note3()}]{Note3}%
  \BibitemOpen
  \bibinfo {note} {Without feedback, the MIPT is expected to shift toward
  $p_c^\protect \mathrm {MIPT}\to 1$ while $\alpha \to 1$. Feedback constrains
  the critical point $p_c^\protect \mathrm {MIPT}$ to reach most $p_c^\protect
  \mathrm {APT}$. Hence the existence of a (model dependent) $\alpha _c\simeq
  1.3$.}\BibitemShut {Stop}%
\bibitem [{\citenamefont {Sierant}\ \emph
  {et~al.}(2022{\natexlab{a}})\citenamefont {Sierant}, \citenamefont
  {Chiriac{\`{o}}}, \citenamefont {Surace}, \citenamefont {Sharma},
  \citenamefont {Turkeshi}, \citenamefont {Dalmonte}, \citenamefont {Fazio},\
  and\ \citenamefont {Pagano}}]{sierant2022dissipativefloquet}%
  \BibitemOpen
  \bibfield  {author} {\bibinfo {author} {\bibfnamefont {P.}~\bibnamefont
  {Sierant}}, \bibinfo {author} {\bibfnamefont {G.}~\bibnamefont
  {Chiriac{\`{o}}}}, \bibinfo {author} {\bibfnamefont {F.~M.}\ \bibnamefont
  {Surace}}, \bibinfo {author} {\bibfnamefont {S.}~\bibnamefont {Sharma}},
  \bibinfo {author} {\bibfnamefont {X.}~\bibnamefont {Turkeshi}}, \bibinfo
  {author} {\bibfnamefont {M.}~\bibnamefont {Dalmonte}}, \bibinfo {author}
  {\bibfnamefont {R.}~\bibnamefont {Fazio}},\ and\ \bibinfo {author}
  {\bibfnamefont {G.}~\bibnamefont {Pagano}},\ }\href
  {https://doi.org/10.22331/q-2022-02-02-638} {\bibfield  {journal} {\bibinfo
  {journal} {{Quantum}}\ }\textbf {\bibinfo {volume} {6}},\ \bibinfo {pages}
  {638} (\bibinfo {year} {2022}{\natexlab{a}})}\BibitemShut {NoStop}%
\bibitem [{\citenamefont {Barratt}\ \emph
  {et~al.}(2022{\natexlab{b}})\citenamefont {Barratt}, \citenamefont {Agrawal},
  \citenamefont {Gopalakrishnan}, \citenamefont {Huse}, \citenamefont
  {Vasseur},\ and\ \citenamefont {Potter}}]{barratt2022fieldtheoryof}%
  \BibitemOpen
  \bibfield  {author} {\bibinfo {author} {\bibfnamefont {F.}~\bibnamefont
  {Barratt}}, \bibinfo {author} {\bibfnamefont {U.}~\bibnamefont {Agrawal}},
  \bibinfo {author} {\bibfnamefont {S.}~\bibnamefont {Gopalakrishnan}},
  \bibinfo {author} {\bibfnamefont {D.~A.}\ \bibnamefont {Huse}}, \bibinfo
  {author} {\bibfnamefont {R.}~\bibnamefont {Vasseur}},\ and\ \bibinfo {author}
  {\bibfnamefont {A.~C.}\ \bibnamefont {Potter}},\ }\href
  {https://doi.org/10.1103/PhysRevLett.129.120604} {\bibfield  {journal}
  {\bibinfo  {journal} {Phys. Rev. Lett.}\ }\textbf {\bibinfo {volume} {129}},\
  \bibinfo {pages} {120604} (\bibinfo {year} {2022}{\natexlab{b}})}\BibitemShut
  {NoStop}%
\bibitem [{\citenamefont {Agrawal}\ \emph {et~al.}(2022)\citenamefont
  {Agrawal}, \citenamefont {Zabalo}, \citenamefont {Chen}, \citenamefont
  {Wilson}, \citenamefont {Potter}, \citenamefont {Pixley}, \citenamefont
  {Gopalakrishnan},\ and\ \citenamefont
  {Vasseur}}]{agrawal2022entanglmentandchargesharpening}%
  \BibitemOpen
  \bibfield  {author} {\bibinfo {author} {\bibfnamefont {U.}~\bibnamefont
  {Agrawal}}, \bibinfo {author} {\bibfnamefont {A.}~\bibnamefont {Zabalo}},
  \bibinfo {author} {\bibfnamefont {K.}~\bibnamefont {Chen}}, \bibinfo {author}
  {\bibfnamefont {J.~H.}\ \bibnamefont {Wilson}}, \bibinfo {author}
  {\bibfnamefont {A.~C.}\ \bibnamefont {Potter}}, \bibinfo {author}
  {\bibfnamefont {J.~H.}\ \bibnamefont {Pixley}}, \bibinfo {author}
  {\bibfnamefont {S.}~\bibnamefont {Gopalakrishnan}},\ and\ \bibinfo {author}
  {\bibfnamefont {R.}~\bibnamefont {Vasseur}},\ }\href
  {https://doi.org/10.1103/PhysRevX.12.041002} {\bibfield  {journal} {\bibinfo
  {journal} {Phys. Rev. X}\ }\textbf {\bibinfo {volume} {12}},\ \bibinfo
  {pages} {041002} (\bibinfo {year} {2022})}\BibitemShut {NoStop}%
\bibitem [{\citenamefont {Oshima}\ and\ \citenamefont
  {Fuji}()}]{oshima2022chargefluctuationand}%
  \BibitemOpen
  \bibfield  {author} {\bibinfo {author} {\bibfnamefont {H.}~\bibnamefont
  {Oshima}}\ and\ \bibinfo {author} {\bibfnamefont {Y.}~\bibnamefont {Fuji}},\
  }\href@noop {} {}\Eprint {https://arxiv.org/abs/2210.16009}
  {arXiv:2210.16009} \BibitemShut {NoStop}%
\bibitem [{\citenamefont {Turkeshi}\ \emph {et~al.}(2020)\citenamefont
  {Turkeshi}, \citenamefont {Fazio},\ and\ \citenamefont
  {Dalmonte}}]{turkeshi2020measurementinducedcriticality}%
  \BibitemOpen
  \bibfield  {author} {\bibinfo {author} {\bibfnamefont {X.}~\bibnamefont
  {Turkeshi}}, \bibinfo {author} {\bibfnamefont {R.}~\bibnamefont {Fazio}},\
  and\ \bibinfo {author} {\bibfnamefont {M.}~\bibnamefont {Dalmonte}},\ }\href
  {https://doi.org/10.1103/PhysRevB.102.014315} {\bibfield  {journal} {\bibinfo
   {journal} {Phys. Rev. B}\ }\textbf {\bibinfo {volume} {102}},\ \bibinfo
  {pages} {014315} (\bibinfo {year} {2020})}\BibitemShut {NoStop}%
\bibitem [{\citenamefont {Lunt}\ \emph {et~al.}(2021)\citenamefont {Lunt},
  \citenamefont {Szyniszewski},\ and\ \citenamefont
  {Pal}}]{lunt2021measurementinducedcriticality}%
  \BibitemOpen
  \bibfield  {author} {\bibinfo {author} {\bibfnamefont {O.}~\bibnamefont
  {Lunt}}, \bibinfo {author} {\bibfnamefont {M.}~\bibnamefont {Szyniszewski}},\
  and\ \bibinfo {author} {\bibfnamefont {A.}~\bibnamefont {Pal}},\ }\href
  {https://doi.org/10.1103/PhysRevB.104.155111} {\bibfield  {journal} {\bibinfo
   {journal} {Phys. Rev. B}\ }\textbf {\bibinfo {volume} {104}},\ \bibinfo
  {pages} {155111} (\bibinfo {year} {2021})}\BibitemShut {NoStop}%
\bibitem [{\citenamefont {Sierant}\ \emph
  {et~al.}(2022{\natexlab{b}})\citenamefont {Sierant}, \citenamefont
  {Schir\`o}, \citenamefont {Lewenstein},\ and\ \citenamefont
  {Turkeshi}}]{sierant2022measurementinducedphase}%
  \BibitemOpen
  \bibfield  {author} {\bibinfo {author} {\bibfnamefont {P.}~\bibnamefont
  {Sierant}}, \bibinfo {author} {\bibfnamefont {M.}~\bibnamefont {Schir\`o}},
  \bibinfo {author} {\bibfnamefont {M.}~\bibnamefont {Lewenstein}},\ and\
  \bibinfo {author} {\bibfnamefont {X.}~\bibnamefont {Turkeshi}},\ }\href
  {https://doi.org/10.1103/PhysRevB.106.214316} {\bibfield  {journal} {\bibinfo
   {journal} {Phys. Rev. B}\ }\textbf {\bibinfo {volume} {106}},\ \bibinfo
  {pages} {214316} (\bibinfo {year} {2022}{\natexlab{b}})}\BibitemShut
  {NoStop}%
\bibitem [{\citenamefont {Monroe}\ \emph {et~al.}(2021)\citenamefont {Monroe},
  \citenamefont {Campbell}, \citenamefont {Duan}, \citenamefont {Gong},
  \citenamefont {Gorshkov}, \citenamefont {Hess}, \citenamefont {Islam},
  \citenamefont {Kim}, \citenamefont {Linke}, \citenamefont {Pagano},
  \citenamefont {Richerme}, \citenamefont {Senko},\ and\ \citenamefont
  {Yao}}]{pagano1}%
  \BibitemOpen
  \bibfield  {author} {\bibinfo {author} {\bibfnamefont {C.}~\bibnamefont
  {Monroe}}, \bibinfo {author} {\bibfnamefont {W.~C.}\ \bibnamefont
  {Campbell}}, \bibinfo {author} {\bibfnamefont {L.-M.}\ \bibnamefont {Duan}},
  \bibinfo {author} {\bibfnamefont {Z.-X.}\ \bibnamefont {Gong}}, \bibinfo
  {author} {\bibfnamefont {A.~V.}\ \bibnamefont {Gorshkov}}, \bibinfo {author}
  {\bibfnamefont {P.~W.}\ \bibnamefont {Hess}}, \bibinfo {author}
  {\bibfnamefont {R.}~\bibnamefont {Islam}}, \bibinfo {author} {\bibfnamefont
  {K.}~\bibnamefont {Kim}}, \bibinfo {author} {\bibfnamefont {N.~M.}\
  \bibnamefont {Linke}}, \bibinfo {author} {\bibfnamefont {G.}~\bibnamefont
  {Pagano}}, \bibinfo {author} {\bibfnamefont {P.}~\bibnamefont {Richerme}},
  \bibinfo {author} {\bibfnamefont {C.}~\bibnamefont {Senko}},\ and\ \bibinfo
  {author} {\bibfnamefont {N.~Y.}\ \bibnamefont {Yao}},\ }\href
  {https://doi.org/10.1103/RevModPhys.93.025001} {\bibfield  {journal}
  {\bibinfo  {journal} {Rev. Mod. Phys.}\ }\textbf {\bibinfo {volume} {93}},\
  \bibinfo {pages} {025001} (\bibinfo {year} {2021})}\BibitemShut {NoStop}%
\bibitem [{\citenamefont {Zhang}\ \emph {et~al.}(2017)\citenamefont {Zhang},
  \citenamefont {Pagano}, \citenamefont {Hess}, \citenamefont {Kyprianidis},
  \citenamefont {Becker}, \citenamefont {Kaplan}, \citenamefont {Gorshkov},
  \citenamefont {Gong},\ and\ \citenamefont {Monroe}}]{pagano2}%
  \BibitemOpen
  \bibfield  {author} {\bibinfo {author} {\bibfnamefont {J.}~\bibnamefont
  {Zhang}}, \bibinfo {author} {\bibfnamefont {G.}~\bibnamefont {Pagano}},
  \bibinfo {author} {\bibfnamefont {P.~W.}\ \bibnamefont {Hess}}, \bibinfo
  {author} {\bibfnamefont {A.}~\bibnamefont {Kyprianidis}}, \bibinfo {author}
  {\bibfnamefont {P.}~\bibnamefont {Becker}}, \bibinfo {author} {\bibfnamefont
  {H.}~\bibnamefont {Kaplan}}, \bibinfo {author} {\bibfnamefont {A.~V.}\
  \bibnamefont {Gorshkov}}, \bibinfo {author} {\bibfnamefont {Z.-X.}\
  \bibnamefont {Gong}},\ and\ \bibinfo {author} {\bibfnamefont
  {C.}~\bibnamefont {Monroe}},\ }\href {https://doi.org/10.1038/nature24654}
  {\bibfield  {journal} {\bibinfo  {journal} {Nature}\ }\textbf {\bibinfo
  {volume} {551}},\ \bibinfo {pages} {601} (\bibinfo {year}
  {2017})}\BibitemShut {NoStop}%
\bibitem [{\citenamefont {Tan}\ \emph {et~al.}(2021)\citenamefont {Tan},
  \citenamefont {Becker}, \citenamefont {Liu}, \citenamefont {Pagano},
  \citenamefont {Collins}, \citenamefont {De}, \citenamefont {Feng},
  \citenamefont {Kaplan}, \citenamefont {Kyprianidis}, \citenamefont
  {Lundgren}, \citenamefont {Morong}, \citenamefont {Whitsitt}, \citenamefont
  {Gorshkov},\ and\ \citenamefont {Monroe}}]{pagano3}%
  \BibitemOpen
  \bibfield  {author} {\bibinfo {author} {\bibfnamefont {W.~L.}\ \bibnamefont
  {Tan}}, \bibinfo {author} {\bibfnamefont {P.}~\bibnamefont {Becker}},
  \bibinfo {author} {\bibfnamefont {F.}~\bibnamefont {Liu}}, \bibinfo {author}
  {\bibfnamefont {G.}~\bibnamefont {Pagano}}, \bibinfo {author} {\bibfnamefont
  {K.~S.}\ \bibnamefont {Collins}}, \bibinfo {author} {\bibfnamefont
  {A.}~\bibnamefont {De}}, \bibinfo {author} {\bibfnamefont {L.}~\bibnamefont
  {Feng}}, \bibinfo {author} {\bibfnamefont {H.~B.}\ \bibnamefont {Kaplan}},
  \bibinfo {author} {\bibfnamefont {A.}~\bibnamefont {Kyprianidis}}, \bibinfo
  {author} {\bibfnamefont {R.}~\bibnamefont {Lundgren}}, \bibinfo {author}
  {\bibfnamefont {W.}~\bibnamefont {Morong}}, \bibinfo {author} {\bibfnamefont
  {S.}~\bibnamefont {Whitsitt}}, \bibinfo {author} {\bibfnamefont {A.~V.}\
  \bibnamefont {Gorshkov}},\ and\ \bibinfo {author} {\bibfnamefont
  {C.}~\bibnamefont {Monroe}},\ }\href
  {https://doi.org/10.1038/s41567-021-01194-3} {\bibfield  {journal} {\bibinfo
  {journal} {Nature Physics}\ }\textbf {\bibinfo {volume} {17}},\ \bibinfo
  {pages} {742} (\bibinfo {year} {2021})}\BibitemShut {NoStop}%
\bibitem [{\citenamefont {Kyprianidis}\ \emph {et~al.}(2021)\citenamefont
  {Kyprianidis}, \citenamefont {Machado}, \citenamefont {Morong}, \citenamefont
  {Becker}, \citenamefont {Collins}, \citenamefont {Else}, \citenamefont
  {Feng}, \citenamefont {Hess}, \citenamefont {Nayak}, \citenamefont {Pagano},
  \citenamefont {Yao},\ and\ \citenamefont {Monroe}}]{pagano4}%
  \BibitemOpen
  \bibfield  {author} {\bibinfo {author} {\bibfnamefont {A.}~\bibnamefont
  {Kyprianidis}}, \bibinfo {author} {\bibfnamefont {F.}~\bibnamefont
  {Machado}}, \bibinfo {author} {\bibfnamefont {W.}~\bibnamefont {Morong}},
  \bibinfo {author} {\bibfnamefont {P.}~\bibnamefont {Becker}}, \bibinfo
  {author} {\bibfnamefont {K.~S.}\ \bibnamefont {Collins}}, \bibinfo {author}
  {\bibfnamefont {D.~V.}\ \bibnamefont {Else}}, \bibinfo {author}
  {\bibfnamefont {L.}~\bibnamefont {Feng}}, \bibinfo {author} {\bibfnamefont
  {P.~W.}\ \bibnamefont {Hess}}, \bibinfo {author} {\bibfnamefont
  {C.}~\bibnamefont {Nayak}}, \bibinfo {author} {\bibfnamefont
  {G.}~\bibnamefont {Pagano}}, \bibinfo {author} {\bibfnamefont {N.~Y.}\
  \bibnamefont {Yao}},\ and\ \bibinfo {author} {\bibfnamefont {C.}~\bibnamefont
  {Monroe}},\ }\href {https://doi.org/10.1126/science.abg8102} {\bibfield
  {journal} {\bibinfo  {journal} {Science}\ }\textbf {\bibinfo {volume}
  {372}},\ \bibinfo {pages} {1192} (\bibinfo {year} {2021})}\BibitemShut
  {NoStop}%
\bibitem [{\citenamefont {Pagano}\ \emph {et~al.}(2020)\citenamefont {Pagano},
  \citenamefont {Bapat}, \citenamefont {Becker}, \citenamefont {Collins},
  \citenamefont {De}, \citenamefont {Hess}, \citenamefont {Kaplan},
  \citenamefont {Kyprianidis}, \citenamefont {Tan}, \citenamefont {Baldwin},
  \citenamefont {Brady}, \citenamefont {Deshpande}, \citenamefont {Liu},
  \citenamefont {Jordan}, \citenamefont {Gorshkov},\ and\ \citenamefont
  {Monroe}}]{pagano5}%
  \BibitemOpen
  \bibfield  {author} {\bibinfo {author} {\bibfnamefont {G.}~\bibnamefont
  {Pagano}}, \bibinfo {author} {\bibfnamefont {A.}~\bibnamefont {Bapat}},
  \bibinfo {author} {\bibfnamefont {P.}~\bibnamefont {Becker}}, \bibinfo
  {author} {\bibfnamefont {K.~S.}\ \bibnamefont {Collins}}, \bibinfo {author}
  {\bibfnamefont {A.}~\bibnamefont {De}}, \bibinfo {author} {\bibfnamefont
  {P.~W.}\ \bibnamefont {Hess}}, \bibinfo {author} {\bibfnamefont {H.~B.}\
  \bibnamefont {Kaplan}}, \bibinfo {author} {\bibfnamefont {A.}~\bibnamefont
  {Kyprianidis}}, \bibinfo {author} {\bibfnamefont {W.~L.}\ \bibnamefont
  {Tan}}, \bibinfo {author} {\bibfnamefont {C.}~\bibnamefont {Baldwin}},
  \bibinfo {author} {\bibfnamefont {L.~T.}\ \bibnamefont {Brady}}, \bibinfo
  {author} {\bibfnamefont {A.}~\bibnamefont {Deshpande}}, \bibinfo {author}
  {\bibfnamefont {F.}~\bibnamefont {Liu}}, \bibinfo {author} {\bibfnamefont
  {S.}~\bibnamefont {Jordan}}, \bibinfo {author} {\bibfnamefont {A.~V.}\
  \bibnamefont {Gorshkov}},\ and\ \bibinfo {author} {\bibfnamefont
  {C.}~\bibnamefont {Monroe}},\ }\href
  {https://doi.org/10.1073/pnas.2006373117} {\bibfield  {journal} {\bibinfo
  {journal} {Proceedings of the National Academy of Sciences}\ }\textbf
  {\bibinfo {volume} {117}},\ \bibinfo {pages} {25396} (\bibinfo {year}
  {2020})}\BibitemShut {NoStop}%
\bibitem [{\citenamefont {Kaplan}\ \emph {et~al.}(2020)\citenamefont {Kaplan},
  \citenamefont {Guo}, \citenamefont {Tan}, \citenamefont {De}, \citenamefont
  {Marquardt}, \citenamefont {Pagano},\ and\ \citenamefont {Monroe}}]{pagano6}%
  \BibitemOpen
  \bibfield  {author} {\bibinfo {author} {\bibfnamefont {H.~B.}\ \bibnamefont
  {Kaplan}}, \bibinfo {author} {\bibfnamefont {L.}~\bibnamefont {Guo}},
  \bibinfo {author} {\bibfnamefont {W.~L.}\ \bibnamefont {Tan}}, \bibinfo
  {author} {\bibfnamefont {A.}~\bibnamefont {De}}, \bibinfo {author}
  {\bibfnamefont {F.}~\bibnamefont {Marquardt}}, \bibinfo {author}
  {\bibfnamefont {G.}~\bibnamefont {Pagano}},\ and\ \bibinfo {author}
  {\bibfnamefont {C.}~\bibnamefont {Monroe}},\ }\href
  {https://doi.org/10.1103/PhysRevLett.125.120605} {\bibfield  {journal}
  {\bibinfo  {journal} {Phys. Rev. Lett.}\ }\textbf {\bibinfo {volume} {125}},\
  \bibinfo {pages} {120605} (\bibinfo {year} {2020})}\BibitemShut {NoStop}%
\bibitem [{\citenamefont {Lavasani}\ \emph
  {et~al.}(2021{\natexlab{a}})\citenamefont {Lavasani}, \citenamefont
  {Alavirad},\ and\ \citenamefont
  {Barkeshli}}]{lavasani2021measurementinducedtopological}%
  \BibitemOpen
  \bibfield  {author} {\bibinfo {author} {\bibfnamefont {A.}~\bibnamefont
  {Lavasani}}, \bibinfo {author} {\bibfnamefont {Y.}~\bibnamefont {Alavirad}},\
  and\ \bibinfo {author} {\bibfnamefont {M.}~\bibnamefont {Barkeshli}},\ }\href
  {https://doi.org/10.1038/s41567-020-01112-z} {\bibfield  {journal} {\bibinfo
  {journal} {Nature Phys.}\ }\textbf {\bibinfo {volume} {17}},\ \bibinfo
  {pages} {342} (\bibinfo {year} {2021}{\natexlab{a}})}\BibitemShut {NoStop}%
\bibitem [{\citenamefont {Lavasani}\ \emph
  {et~al.}(2021{\natexlab{b}})\citenamefont {Lavasani}, \citenamefont
  {Alavirad},\ and\ \citenamefont
  {Barkeshli}}]{lavasani2021topologicalorderand}%
  \BibitemOpen
  \bibfield  {author} {\bibinfo {author} {\bibfnamefont {A.}~\bibnamefont
  {Lavasani}}, \bibinfo {author} {\bibfnamefont {Y.}~\bibnamefont {Alavirad}},\
  and\ \bibinfo {author} {\bibfnamefont {M.}~\bibnamefont {Barkeshli}},\ }\href
  {https://doi.org/10.1103/PhysRevLett.127.235701} {\bibfield  {journal}
  {\bibinfo  {journal} {Phys. Rev. Lett.}\ }\textbf {\bibinfo {volume} {127}},\
  \bibinfo {pages} {235701} (\bibinfo {year} {2021}{\natexlab{b}})}\BibitemShut
  {NoStop}%
\bibitem [{\citenamefont {Zhu}\ \emph {et~al.}()\citenamefont {Zhu},
  \citenamefont {Tantivasadakarn}, \citenamefont {Vishwanath}, \citenamefont
  {Trebst},\ and\ \citenamefont {Verresen}}]{zhu2022nishimoriscat}%
  \BibitemOpen
  \bibfield  {author} {\bibinfo {author} {\bibfnamefont {G.-Y.}\ \bibnamefont
  {Zhu}}, \bibinfo {author} {\bibfnamefont {N.}~\bibnamefont
  {Tantivasadakarn}}, \bibinfo {author} {\bibfnamefont {A.}~\bibnamefont
  {Vishwanath}}, \bibinfo {author} {\bibfnamefont {S.}~\bibnamefont {Trebst}},\
  and\ \bibinfo {author} {\bibfnamefont {R.}~\bibnamefont {Verresen}},\
  }\href@noop {} {}\Eprint {https://arxiv.org/abs/2208.11136}
  {arXiv:2208.11136} \BibitemShut {NoStop}%
\bibitem [{\citenamefont {Sang}\ and\ \citenamefont
  {Hsieh}(2021)}]{sang2021measurementprotectedquantum}%
  \BibitemOpen
  \bibfield  {author} {\bibinfo {author} {\bibfnamefont {S.}~\bibnamefont
  {Sang}}\ and\ \bibinfo {author} {\bibfnamefont {T.~H.}\ \bibnamefont
  {Hsieh}},\ }\href {https://doi.org/10.1103/PhysRevResearch.3.023200}
  {\bibfield  {journal} {\bibinfo  {journal} {Phys. Rev. Research}\ }\textbf
  {\bibinfo {volume} {3}},\ \bibinfo {pages} {023200} (\bibinfo {year}
  {2021})}\BibitemShut {NoStop}%
\bibitem [{\citenamefont {Lee}\ \emph {et~al.}()\citenamefont {Lee},
  \citenamefont {Ji}, \citenamefont {Bi},\ and\ \citenamefont
  {Fisher}}]{lee2022decodingmeasurementpreparedquantum}%
  \BibitemOpen
  \bibfield  {author} {\bibinfo {author} {\bibfnamefont {J.~Y.}\ \bibnamefont
  {Lee}}, \bibinfo {author} {\bibfnamefont {W.}~\bibnamefont {Ji}}, \bibinfo
  {author} {\bibfnamefont {Z.}~\bibnamefont {Bi}},\ and\ \bibinfo {author}
  {\bibfnamefont {M.~P.~A.}\ \bibnamefont {Fisher}},\ }\href@noop {} {}\Eprint
  {https://arxiv.org/abs/2208.11699} {arXiv:2208.11699} \BibitemShut {NoStop}%
\bibitem [{\citenamefont {Klocke}\ and\ \citenamefont
  {Buchhold}(2022)}]{klocke2022topologicalorderand}%
  \BibitemOpen
  \bibfield  {author} {\bibinfo {author} {\bibfnamefont {K.}~\bibnamefont
  {Klocke}}\ and\ \bibinfo {author} {\bibfnamefont {M.}~\bibnamefont
  {Buchhold}},\ }\href {https://doi.org/10.1103/PhysRevB.106.104307} {\bibfield
   {journal} {\bibinfo  {journal} {Phys. Rev. B}\ }\textbf {\bibinfo {volume}
  {106}},\ \bibinfo {pages} {104307} (\bibinfo {year} {2022})}\BibitemShut
  {NoStop}%
\bibitem [{\citenamefont {Pezz\`e}\ \emph {et~al.}(2018)\citenamefont
  {Pezz\`e}, \citenamefont {Smerzi}, \citenamefont {Oberthaler}, \citenamefont
  {Schmied},\ and\ \citenamefont {Treutlein}}]{smerzi1}%
  \BibitemOpen
  \bibfield  {author} {\bibinfo {author} {\bibfnamefont {L.}~\bibnamefont
  {Pezz\`e}}, \bibinfo {author} {\bibfnamefont {A.}~\bibnamefont {Smerzi}},
  \bibinfo {author} {\bibfnamefont {M.~K.}\ \bibnamefont {Oberthaler}},
  \bibinfo {author} {\bibfnamefont {R.}~\bibnamefont {Schmied}},\ and\ \bibinfo
  {author} {\bibfnamefont {P.}~\bibnamefont {Treutlein}},\ }\href
  {https://doi.org/10.1103/RevModPhys.90.035005} {\bibfield  {journal}
  {\bibinfo  {journal} {Rev. Mod. Phys.}\ }\textbf {\bibinfo {volume} {90}},\
  \bibinfo {pages} {035005} (\bibinfo {year} {2018})}\BibitemShut {NoStop}%
\bibitem [{\citenamefont {Dooley}\ \emph {et~al.}(2022)\citenamefont {Dooley},
  \citenamefont {Pappalardi},\ and\ \citenamefont {Goold}}]{silvia1}%
  \BibitemOpen
  \bibfield  {author} {\bibinfo {author} {\bibfnamefont {S.}~\bibnamefont
  {Dooley}}, \bibinfo {author} {\bibfnamefont {S.}~\bibnamefont {Pappalardi}},\
  and\ \bibinfo {author} {\bibfnamefont {J.}~\bibnamefont {Goold}},\ }\href
  {https://doi.org/10.48550/ARXIV.2207.13521} {\bibinfo {title} {Entanglement
  enhanced metrology with quantum many-body scars}} (\bibinfo {year}
  {2022})\BibitemShut {NoStop}%
\bibitem [{\citenamefont {Desaules}\ \emph {et~al.}(2022)\citenamefont
  {Desaules}, \citenamefont {Pietracaprina}, \citenamefont {Papi\'{c}},
  \citenamefont {Goold},\ and\ \citenamefont {Pappalardi}}]{silvia}%
  \BibitemOpen
  \bibfield  {author} {\bibinfo {author} {\bibfnamefont {J.-Y.}\ \bibnamefont
  {Desaules}}, \bibinfo {author} {\bibfnamefont {F.}~\bibnamefont
  {Pietracaprina}}, \bibinfo {author} {\bibfnamefont {Z.}~\bibnamefont
  {Papi\'{c}}}, \bibinfo {author} {\bibfnamefont {J.}~\bibnamefont {Goold}},\
  and\ \bibinfo {author} {\bibfnamefont {S.}~\bibnamefont {Pappalardi}},\
  }\href {https://doi.org/10.1103/PhysRevLett.129.020601} {\bibfield  {journal}
  {\bibinfo  {journal} {Phys. Rev. Lett.}\ }\textbf {\bibinfo {volume} {129}},\
  \bibinfo {pages} {020601} (\bibinfo {year} {2022})}\BibitemShut {NoStop}%
\bibitem [{\citenamefont {Dooley}(2021)}]{doley}%
  \BibitemOpen
  \bibfield  {author} {\bibinfo {author} {\bibfnamefont {S.}~\bibnamefont
  {Dooley}},\ }\href {https://doi.org/10.1103/PRXQuantum.2.020330} {\bibfield
  {journal} {\bibinfo  {journal} {PRX Quantum}\ }\textbf {\bibinfo {volume}
  {2}},\ \bibinfo {pages} {020330} (\bibinfo {year} {2021})}\BibitemShut
  {NoStop}%
\bibitem [{\citenamefont {Hayes}\ \emph {et~al.}(2018)\citenamefont {Hayes},
  \citenamefont {Dooley}, \citenamefont {Munro}, \citenamefont {Nemoto},\ and\
  \citenamefont {Dunningham}}]{doley1}%
  \BibitemOpen
  \bibfield  {author} {\bibinfo {author} {\bibfnamefont {A.~J.}\ \bibnamefont
  {Hayes}}, \bibinfo {author} {\bibfnamefont {S.}~\bibnamefont {Dooley}},
  \bibinfo {author} {\bibfnamefont {W.~J.}\ \bibnamefont {Munro}}, \bibinfo
  {author} {\bibfnamefont {K.}~\bibnamefont {Nemoto}},\ and\ \bibinfo {author}
  {\bibfnamefont {J.}~\bibnamefont {Dunningham}},\ }\href
  {https://doi.org/10.48550/ARXIV.1801.03452} {\bibinfo {title} {Making the
  most of time in quantum metrology: concurrent state preparation and sensing}}
  (\bibinfo {year} {2018})\BibitemShut {NoStop}%
\bibitem [{\citenamefont {Dooley}\ \emph {et~al.}(2018)\citenamefont {Dooley},
  \citenamefont {Hanks}, \citenamefont {Nakayama}, \citenamefont {Munro},\ and\
  \citenamefont {Nemoto}}]{Dooley2018}%
  \BibitemOpen
  \bibfield  {author} {\bibinfo {author} {\bibfnamefont {S.}~\bibnamefont
  {Dooley}}, \bibinfo {author} {\bibfnamefont {M.}~\bibnamefont {Hanks}},
  \bibinfo {author} {\bibfnamefont {S.}~\bibnamefont {Nakayama}}, \bibinfo
  {author} {\bibfnamefont {W.~J.}\ \bibnamefont {Munro}},\ and\ \bibinfo
  {author} {\bibfnamefont {K.}~\bibnamefont {Nemoto}},\ }\bibfield  {journal}
  {\bibinfo  {journal} {npj Quantum Information}\ }\textbf {\bibinfo {volume}
  {4}},\ \href {https://doi.org/10.1038/s41534-018-0073-3}
  {10.1038/s41534-018-0073-3} (\bibinfo {year} {2018})\BibitemShut {NoStop}%
\bibitem [{\citenamefont {Piroli}\ \emph {et~al.}(2022)\citenamefont {Piroli},
  \citenamefont {Li}, \citenamefont {Vasseur},\ and\ \citenamefont
  {Nahum}}]{yaodong2023}%
  \BibitemOpen
  \bibfield  {author} {\bibinfo {author} {\bibfnamefont {L.}~\bibnamefont
  {Piroli}}, \bibinfo {author} {\bibfnamefont {Y.}~\bibnamefont {Li}}, \bibinfo
  {author} {\bibfnamefont {R.}~\bibnamefont {Vasseur}},\ and\ \bibinfo {author}
  {\bibfnamefont {A.}~\bibnamefont {Nahum}},\ }\href@noop {} {\bibinfo {title}
  {Appearing parallelly in the same arxiv post.}} (\bibinfo {year} {2022}),\
  \Eprint {https://arxiv.org/abs/2212.xxxxx} {arXiv:2212.xxxxx} \BibitemShut
  {NoStop}%
\bibitem [{Note4()}]{Note4}%
  \BibitemOpen
  \bibinfo {note} {A $k$-design $\protect \mathcal {C}$ is a probabilistic
  ensemble of unitary gates with probability distribution $p_\protect \mathcal
  {C}$ that reproduce $k$-order correlations \begin {equation} \DOTSI \intop
  \ilimits@ _\protect \mathrm {Haar} dU (U\otimes U^\dagger )^k = \DOTSB \sum@
  \slimits@ _{U\in C} p_\protect \mathcal {C}(U) (U\otimes U^\dagger )^k. \end
  {equation} Eq.~\protect \textup {\hbox {\mathsurround \z@ \protect
  \normalfont (\ignorespaces \ref {eq:val}\unskip \@@italiccorr )}} corresponds
  to $k=1$. Therefore Clifford and Haar averages coincide; see Ref.~\cite
  {nielsen00,Gross2021}.}\BibitemShut {Stop}%
\bibitem [{\citenamefont {Kawashima}\ and\ \citenamefont
  {Ito}(1993)}]{Kawashima1993}%
  \BibitemOpen
  \bibfield  {author} {\bibinfo {author} {\bibfnamefont {N.}~\bibnamefont
  {Kawashima}}\ and\ \bibinfo {author} {\bibfnamefont {N.}~\bibnamefont
  {Ito}},\ }\href {https://doi.org/10.1143/jpsj.62.435} {\bibfield  {journal}
  {\bibinfo  {journal} {J. Phys. Soc. Japan}\ }\textbf {\bibinfo {volume}
  {62}},\ \bibinfo {pages} {435} (\bibinfo {year} {1993})}\BibitemShut
  {NoStop}%
\end{thebibliography}
%

\bibliographystyle{apsrev4-2}

\widetext
\clearpage
\begin{center}
\textbf{\large \centering Supplemental Material:\\ Controlling entanglement at absorbing state phase transitions in random circuits}
\end{center}

\setcounter{equation}{0}
\setcounter{figure}{0}
\setcounter{table}{0}
\setcounter{page}{1}
\renewcommand{\theequation}{S\arabic{equation}}
\setcounter{figure}{0}
\renewcommand{\thefigure}{S\arabic{figure}}
\renewcommand{\thepage}{S\arabic{page}}
\renewcommand{\thesection}{S\arabic{section}}
\renewcommand{\thetable}{S\arabic{table}}
\makeatletter

\renewcommand{\thesection}{\arabic{section}}
\renewcommand{\thesubsection}{\thesection.\arabic{subsection}}
\renewcommand{\thesubsubsection}{\thesubsection.\arabic{subsubsection}}

\vspace{1cm}
In this Supplemental Material, we discuss
\begin{enumerate}
    \item The mathematical definition of the Clifford implementation: the flagged Clifford circuits;
    \item The probabilistic cellular automaton arising as the mean-dynamical evolution of the system;
    \item Additional numerical data supporting the claims in the Main Text.
\end{enumerate}

\section{Flagged Clifford Circuits}
\label{supsec:flaggedcliff}
This section defines the Clifford implementation of interest for this paper. We denote the general qubit lattice by $\Lambda$, and its sites by $\vec{i}$. 
As in the Main Text, we fix the absorbing state to be $|\Psi_\mathrm{abs}\rangle = |\uparrow\uparrow\cdots \uparrow\rangle$. 
The elementary building blocks of our circuit are: the unitary two-body gates $U_{\vec{i},\vec{j}}$ drawn uniformly from the Clifford group and the local projective measurements of the operators $Z_{\vec{i}}$. 
We note the measurements stabilize the absorbing state. To make also the unitary component stabilizing the gate, we would require $U_{\vec{i},\vec{j}}$ to leave invariant the state $|\uparrow_{\vec{i}} \uparrow_{\vec{j}}\rangle$.
In principle, such a unitary should be of the form
\begin{equation}
    U_{\vec{i},\vec{j}} = \begin{pmatrix}
        1 & 0 & 0 & 0  \\
        0 & u_{11} & u_{12} & u_{13}\\
         0 & u_{21} & u_{22} & u_{23}\\
          0 & u_{31} & u_{32} & u_{33}\\
    \end{pmatrix}
\end{equation}
where $u$ is a unitary matrix. Nevertheless, a Clifford unitary with this structure would be trivial and not generate genuine entanglement (i.e., cannot generate Bell pairs from product states of eigenstates of $Z_i$ operators and reduces up to phases to a product one-body Clifford gate or a permutation of $\{|\uparrow_{\vec{i}} \downarrow_{\vec{j}} \rangle,|\downarrow_{\vec{i}}  \downarrow_{\vec{j}} \rangle,|\downarrow_{\vec{i}}  \uparrow_{\vec{j}} \rangle\}$).

We consider a flagged Clifford circuit to avoid this problem, extending our state to $|\Psi\rangle\mapsto |\Phi\rangle = |\Psi\rangle\otimes |\mathcal{F}\rangle$. The flag vector $|\mathcal{F}\rangle=\otimes_{\vec{i}\in\Lambda} |f_{\vec{i}}\rangle$ keeps the track of the  post-measurement site polarization. The core idea is that a site $i$ is flagged (i.e. $f_i=1$) if the possible measurements in the layer $\mathcal{M}$ acting on the state of the system $\ket{\Psi_t}$
yield result consistent with the absorbing state, i.e.  $\bra{\Psi_\mathrm{abs}} Z_i  \ket{\Psi_\mathrm{abs}} = 1 = \bra{\Psi_t} Z_i  \ket{\Psi_t}$. 
Conversely, the unflagged sites (i.e., $f_i=0$)  are considered as defects (in the "ordered" absorbing state), for which $\langle \Psi_t|M|\Psi_t\rangle \neq 1$. 
Hence, the absorbing, stationary state of the dynamics may be written as  $|\Phi_\mathrm{abs}\rangle =|\Psi_\mathrm{abs}\rangle\otimes |\uparrow\uparrow\cdots \uparrow\rangle$, which is stabilized by each of the measurement operations.

As in the Main Text, we specialize now to a one-dimensional (1D) lattice of length $L$, denote by $d$ the circuit depth, and split the control unitaries in $\mathcal{A}(\alpha) = A_d(\alpha) U_d$, with 
\begin{equation}
    A_d(\alpha) = \begin{cases}
    \openone, & \text{for } d \text{ odd},\\
    \prod_{(i,j) \in I_\alpha(\mathcal{F}_d)} U_{i,j},
    & \text{for } d \text{ even}.
    \end{cases},\qquad U_d = \begin{cases}
    \prod_{i\ \mathrm{odd}} U^{\mathcal{F}_d}_{i,i+1}, & \text{for } d \text{ odd},\\
    \prod_{i\ \mathrm{even}}U^{\mathcal{F}_d}_{i,i+1},
    & \text{for } d \text{ even}.
    \end{cases}
\end{equation}
These operators include feedback information from the flags. Specifically, $I_\alpha (\mathcal{F}_d)$ is a set of $N_{\mathcal{F}_d}$ pairs drawn from the unflagged sites, i.e., the set of defects $\overline{\mathrm{F}_d}\equiv \{|f_i\rangle \ \text{for } i=1,\dots, L| f_i=0 \}$. The probability distribution of this choice is fixed to $P(r)\propto r^\alpha$, with $r=\min(|i-j|, L-|i-j|)$, with the proportionality factor being the normalization constant. 

In the Main Text, we considered $N_\mathcal{F}=N_\mathrm{flags}$ and $N_\mathcal{F}=N_\mathrm{flags}^2/L$ to demonstrate how various choices of $\mathcal{A}(\alpha)$ change the universal properties of the measurement-induced transition in the long-range limit. (In Sec.~\ref{supsec:numerics}, we present the analysis on additional choices of $A_d(\alpha)$).  
The feedback unitaries $U^{\mathcal{F}_d}_{i,i+1}$ are defined as
\begin{equation}
    U^{\mathcal{F}_d}_{i,i+1}|\Phi_d\rangle = \begin{cases}
        \left(U_{i,j}|\Psi_d\rangle\right)\otimes \left[(\sigma^-_i)^{f_i} (\sigma^-_{i+1})^{f_{i+1}}|\mathcal{F}_d\rangle\right] , & \text{if } f_i f_{i+1} = 0,\\
        |\Phi_d \rangle & \text{otherwise},
    \end{cases}
\end{equation}
with $\sigma^\pm_i$ the raising/lowering operators at position $i$. 
Therefore, the role of the flags is to herald if the two considered sites are both in the ${|\uparrow\rangle}$ polarization.  If $f_{i/j}=1$, the sites are left untouched, otherwise are shuffled through a Clifford random unitary $U_{i,i+1}$, and \emph{unflagged} ($f_{i/j}$ is set to $0$) to reflect the loss of knowledge of the state. Importantly, $f_{i/j}=1$ \textit{is not} equivalent to thesis that the state $\ket{\Psi_t}$ is a product $|\uparrow \uparrow\rangle$ on sites $i$ and $j$ and of a certain state on the rest of the lattice: it may be that prior action of the unitary gates yielded the state $|\uparrow \uparrow\rangle$ on sites $i$ and $j$ \textit{and}, $f_{i/j}$ were set to $0$; such situations are however rare, and dynamics of $N_\mathrm{flags}/L$ and $n_\mathrm{def} \equiv \sum_{i=1}^L \langle \Psi_t| (1-Z_i)|\Psi_t\rangle/(2L)$ have the same quantitative features.
The construction described above allows us to meet the desired condition that both $A_d(\alpha)$ and $U_d$ stabilize the stationary state $|\Phi_\mathrm{abs}\rangle$. 

Lastly, let us define the action of the measurement operations chosen in the Main Text on the doubled space. To enhance the clear separation between measurement-induced and absorbing phase transition for short-range control operations, we chose measurements to act stochastically on pairs of sites
\begin{equation}
    \mathcal{M} = \begin{cases}
    \prod_{i\ \mathrm{odd}} (M^z_i M^z_{i+1})^{r_{i}}, & \text{for } d \text{ odd},\\
    \prod_{i\ \mathrm{even}} (M^z_i M^z_{i+1})^{r_{i}},
    & \text{for } d \text{ even}.
    \end{cases}
\end{equation}
where $r_{i}=0,1$ are randomly picked with probability $1-p,p$ respectively, and $M^z_i$ are the collapsing (according to the Born rule) measurements of the operators $Z_i$. (Recall, periodic boundary conditions are considered, $i+L=i$).  That is, each measurement $M_i^z$ gives $\pm 1$ with probability $p_\pm =\langle \Psi_d|1\pm Z_i|\Psi_d\rangle/2$, and the post-measurement state is 
\begin{equation}
    M^z_i |\Phi_d\rangle =\displaystyle \begin{cases}
        \frac{1}{\sqrt{p_-}}\left(\frac{1-Z_i}{2} |\Psi_d\rangle\right)\otimes \left[(\sigma^-)_i^{f_i}|\mathcal{F}_d\rangle\right],& \text{if the result is }-1,\\
        \frac{1}{\sqrt{p_+}}\left(\frac{1+Z_i}{2} |\Psi_d\rangle\right)\otimes \left[(\sigma^+)_i^{1-f_i}|\mathcal{F}_d\rangle\right],& \text{if the result is }+1.
    \end{cases}  
\end{equation}
The system is in a product state $|\Psi\rangle\otimes |\mathcal{F}\rangle$ at all times. The flags (kept as a vector of $L$ binary entries $f_i$) do not entangle with the qubits' degrees of freedom and serve as a computational tool to reach large system sizes using standard stabilizer circuit simulation techniques.

Our numerical implementation follows the above steps with two caveats. First, the state is initialized in the fully polarized state $|\downarrow\downarrow\dots\downarrow\rangle$, and the initial flags are $|\downarrow\downarrow\dots\downarrow\rangle$.
The above description restores the implementation in Fig.~1 (Main Text), once the time is fixed to $t=d/2$ (i.e., $t=0,1/2,1,3/2,2,\dots$). The Main Text considers the expectation value of observables at integer times.

\section{Average dynamics}
\label{supsec:avedyn}
We are interested in the dynamics of the observable  $n_\mathrm{def} = \sum_i \langle\Psi_t| 1-Z_i|\Psi_t \rangle/{2L}$ under $T$ layers of the measurement-unitary evolution (here we use the notation in the Hilbert space of $L$ qubits). 
We define $\rho_t=|\Psi_t\rangle\langle\Psi_t|$ and $\rho_{\mathcal{F}_t} = |\mathcal{F}_t\rangle\langle \mathcal{F}_t|$, and we are interested in the dynamics of $\overline{\rho_{t}\otimes\rho_{\mathcal{F}_t}} $, that fully capture $n_\mathrm{def}$. 

The key observation is that the average over unitary gates will simplify the problem to a probabilistic cellular automaton (PCA). 
Indeed, the Haar average for every two-body unitary gate is given by
\begin{equation}
    \int dU U_{i,j} |\psi_{i,j}\rangle\langle \psi_{i,j} | = \frac{1}{4} \openone_i\otimes \openone_j\label{eq:val}
\end{equation}
that is fully mixing. Being a 2-design, Clifford gates reproduce Eq.~\eqref{eq:val}~\footnote{A $k$-design $\mathcal{C}$ is a probabilistic ensemble of unitary gates with probability distribution $p_\mathcal{C}$  that reproduce $k$-order correlations
\begin{equation}
    \int_\mathrm{Haar} dU  (U\otimes U^\dagger)^k = \sum_{U\in C} p_\mathcal{C}(U) (U\otimes U^\dagger)^k.
\end{equation}
Eq.~\eqref{eq:val} corresponds to $k=1$. Therefore Clifford and Haar averages coincide; see Ref.~\cite{nielsen00,Gross2021}.
}.
Nevertheless, since  Eq.~\eqref{eq:val} is the only property used to derive the PCA, the following results can be extended to more general Haar circuits (see \cite{odea2022entanglementandabsorbing}).

Introducing the indices $m_i=0,\pm1$ and $\rho_i^{(0)} = \openone_i/2$, $\rho_i^{(\pm1)} = (1\pm Z_i)/2$, it follows the state after action of $A_t(\alpha) U_t \mathcal{M}$ and averaging over the Clifford operation is in the form
\begin{equation}
    R_t\equiv \mathbb{E}_{U\in \mathcal{C}_2}[\rho_t\otimes \rho_{\mathcal{F}_t}]  =\left( \bigotimes_{i=1}^L \rho_i^{(f_i)} \right)\otimes \left(\bigotimes_{i=1}^L \rho_i^{(-1)^{f_i+1}} \right). \label{eq:ziocan}
\end{equation}
In the above equation, we used that the unitary gates fully mix measurement results $-1$ (being unflagged), hence $m_i=f_i$ for the system degrees of freedom. Furthermore, the dynamic becomes a stochastic process driven by the measurement results. Indeed, if the measurements do not polarize and flag two sites to $|\uparrow\rangle$, then the unitary channel will fully mix it.

From Eq.~\eqref{eq:ziocan}, it is clear that the dynamics after the unitary averaging is fully encoded in the flags $\{f_i\}$, therefore mapping to the classical PCA  $R_t$. 
(Indeed, the measurement results fix the probabilistic nature of $R_t$. Taking the average over the measurement results would map $R_t$ to a discrete master equation.)

The dynamics is given by $H=\left(\prod_{i\ \text{even}} H_{i,i+1} \right)\left(\prod_{i\ \text{odd}} H_{i,i+1}\right)$, with the action of $H_{i,i+1}$ specified by the following rules
\begin{itemize}
    \item With probability $p$ and if $f_i+f_{i+1}=0$: $f_i \to 1$ and $f_{i+1}\to 1$ with probability 1/4;
    \item With probability $p$ and if $f_i+f_{i+1}=1$: $f_i \to 1$ and $f_{i+1}\to 1$ with probability 1/2;
    \item With probability $1-p$ and if $f_i+f_{i+1}=2$: $f_i \to 1$ and $f_{i+1}\to 1$;
    \item Otherwise: $f_i \to 0$ and $f_{i+1}\to 0$.
\end{itemize}
We have numerically verified that this PCA reproduces the exact time evolution of the average defect density, $n_\mathrm{def}$, obtained with the numerical simulations of the full quantum dynamics of the stabilizer circuit.

\section{Additional numerical analysis}
\label{supsec:numerics}
This section complements the analysis summarized in the Main Text. We start by analyzing the system's measurement-induced phase transition (MIPT). Then, we present additional numerical results for long-range flagged stabilizer circuits, including complementary choices of the feedback-control operation $A(\alpha)$. 

\subsection{Characterization of the measurement-induced phase transition}
This subsection demonstrates our analysis of the measurement-induced phase transition. Since this phenomenon has been extensively discussed, we report only the salient features and refer to the literature for additional details. We partially used the information presented in this section to build Fig. 1 (Main Text), where the numerically obtained critical measurement rates $p_c^\mathrm{MIPT}$ are included.

We reveal the MIPT using various observables, including the tripartite mutual information and the single-qubit purification time. 
Given a quadripartition $X_1\cup X_2\cup X_3\cup X_4$, the tripartite mutual information is given by 
\begin{equation}
    I_3(X_1:X_2:X_3) = S(X_1) + S(X_2) + S(X_3) +S(X_1\cup X_2\cup X_3) - S(X_1\cup X_2) - S(X_1\cup X_3) - S(X_2\cup X_3),
\end{equation}
where $S(X)=-\mathrm{tr}(\rho_X\log_2\rho_X)$ is the entanglement entropy defined in the Main Text. 
\begin{figure}
    \centering
    \includegraphics[width=\columnwidth]{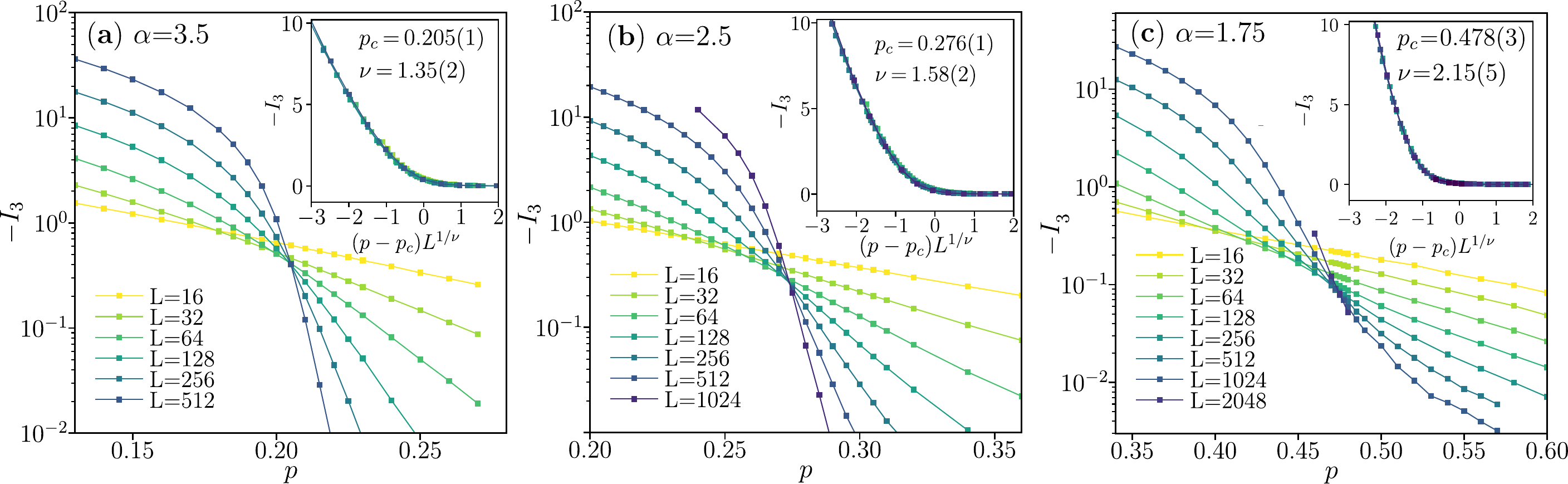}
    \caption{ Measurement-induced phase transition for various exponents $\alpha$ determining the range of feedback-control operations, as signaled by the tripartite quantum mutual information $I_3$. We consider $\alpha=3.5$ (\textbf{a}), $\alpha=2.5$ (\textbf{b}), and $\alpha=1.75$ (\textbf{c}), and extract the critical point and critical exponent $\nu$ as well as the critical measurement rate $p_c$ through finite size scaling of data for various $p$ and $L$, as shown in the insets.\label{supfig:1}
    }
\end{figure}
\begin{figure}
    \centering
    \includegraphics[width=\columnwidth]{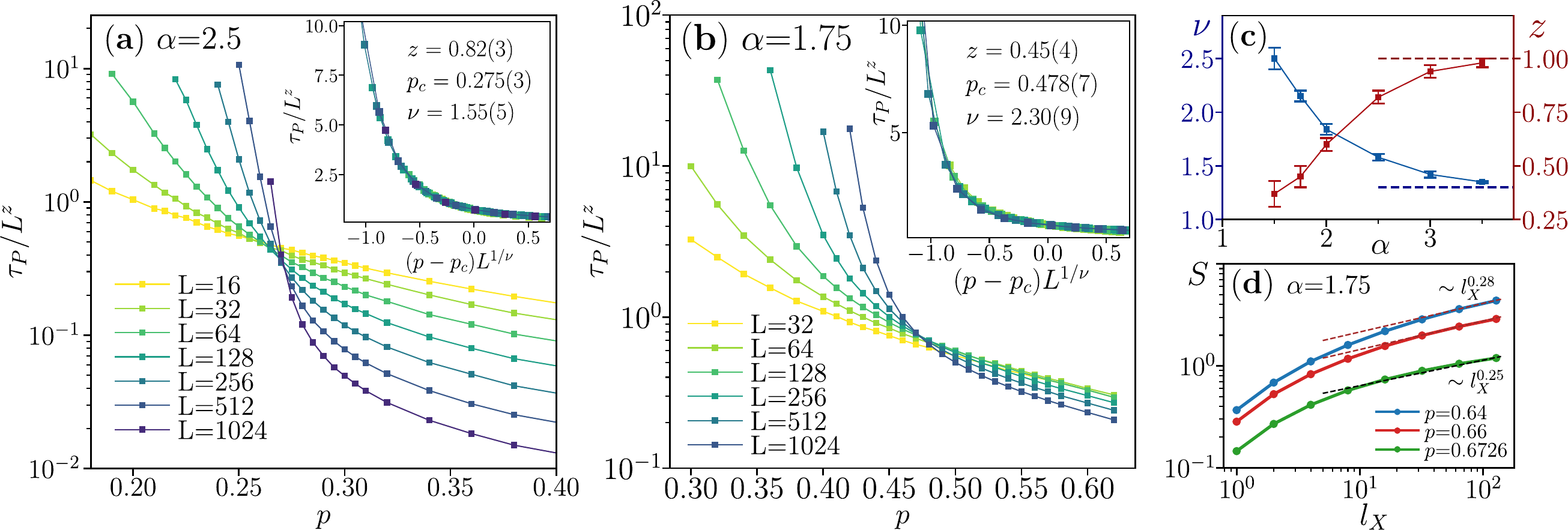}
    \caption{ Measurement-induce phase transition for $\alpha=2.5$ (\textbf{a}) and $\alpha=1.75$ (\textbf{b}) as revealed by the typical purification time $\tau_P$ of an ancilla qubit. The results for the critical parameters $p_c$ and $\nu$ are compatible with the results obtained in the analysis $I_3$, providing a robust cross-check of our findings. (\textbf{c}) The flow of the critical exponents $\nu$ and $z$ at varying $\alpha$. At large $\alpha$, the show the expected conformal behavior, proper of the short-ranged MIPT~\cite{sierant2022measurementinducedphase}. (\textbf{d}) Power-law behavior of the entanglement entropy for $\alpha_c \le\alpha\le 2$, and for $p^\mathrm{MIPT}_c< p \leq p^\mathrm{APT}$ (data for an exemplary value $\alpha=1.75$). Here $L=1024$ is the system size, and subsystems of size $l_X$ are considered. 
    \label{supfig:2}
    }
\end{figure}
To extract the average single qubit purification time $\tau_p$, we entangle the system's initial state with an ancilla qubit, following the prescription introduced in Ref.~\cite{block2022measurementinducedtransition}. Specifically, we consider a global random Clifford gate acting on the chain of $L+1$ sites (system + ancilla) and then let the system evolve through the feedback-monitored dynamics (the gates and measurements are \textit{not} acting on the ancilla qubit).
Eventually, the ancilla qubit will purify. The average time this happens, $\tau_p$, is a probe that allows us to observe the MIPT. 

An important remark is that, at times exponentially large in system size, $\tau_\star \propto e^{\beta L}$, the system will always reach the absorbing state with vanishing entanglement entropy. However, at a fixed time-step, the entanglement, the tripartite mutual information, and the single-qubit purity are non-zero. Moreover, there exists a separation of time scales: when MIPT occurs at $p < p^{\mathrm{APT}}_c$, the quantities characterizing entanglement in the system saturate at times at most proportional to the system size $L$, much smaller than the time of the eventual decay to the absorbing state which commences at $\tau_\star \propto e^{\beta L}$.
This allows us to locate the MIPT and determine its critical exponents by considering $I_3$ at times (circuit depths) proportional to the system size $L$. In the following, we have fixed $t=4L$, but we have checked the stability of our results for choices of different proportionality constants. The same separation of time scales applies to the purification transition witnessed by the single qubit purification time $\tau_p$. Indeed, at MIPT, one has $\tau_p \propto L^z$ with dynamical critical exponent $z \leq 1$~\cite{block2022measurementinducedtransition}, hence $ \tau_p  \ll \tau_\star$ in the vicinity of MIPT, when $p^{\mathrm{MIPT}}_c <p^{\mathrm{APT}}_c $.

In Fig.~\ref{supfig:1}, we present the results for the quantum mutual information $I_3$ for  a quadripartition of equal (connected) intervals $l_{X_1}=l_{X_2}=l_{X_3}=l_{X_4}=L/4$, at various $\alpha$ in the (a, b) the short-range regime ($\alpha=3.5,2.5$) and the (c) the volume-to-power-law transition ($\alpha=1.75$). The entanglement entropy above the MIPT in our setup with feedback-control operations with exponent $\alpha$ behaves in the same manner as in generic circuits \cite{block2022measurementinducedtransition, sharma2022measurementinducedcriticality}: for $\alpha  \geq 2$ it follows an area-law, whereas for $\alpha < 2$, it scales algebraically with subsystem size.
In our setup, we find that the scaling function $-I_3=f((p-p_c)L^{1/\nu}$ captures the scaling of the data close to the critical point. Remarkably, in this model,  we find no scaling dimension for $I_3$  in contrast to the results in Ref.~\cite{sharma2022measurementinducedcriticality}. The optimal parameters $p_c$ and $\nu$ are obtained by minimizing the cost function proposed in~\cite{Kawashima1993}. Similarly, we have obtained the critical exponents for other values of $\alpha \in[1.5,3.5]$, see Fig.~\ref{supfig:2}(c). 

Extrapolating the critical measurement rate $p^{\mathrm{MIPT}}_c$ as a function of $\alpha$ towards $\alpha < 1.5$, we find that $p^{\mathrm{MIPT}}_c = p^{\mathrm{APT}}_c$ for $\alpha \equiv \alpha_c \approx 1.3$. This shows us that a MIPT between volume-law and sub-volume-law (with power-law scaling of the entanglement entropy) phases occurs in our system  for $\alpha \in [\alpha_c, 2]$. The transition is separated from the APT. In that case, APT is associated with an entanglement transition between the sub-volume law phase, which occurs for $p^{\mathrm{MIPT}}_c < p < p^{\mathrm{APT}}_c$, see Fig.~\ref{supfig:2}(d) and the zero-law phase at $ p > p^{\mathrm{APT}}_c$. Finally, for $\alpha <\alpha_c$, the system is in the volume-law phase provided that $p<p^{\mathrm{APT}}_c$. In that case, the only entanglement transition in the system is the dynamical transition between the volume-law phase and the zero-law phase, described in the Main Text where we took an exemplary value $\alpha=0.5$ (in the subsequent section, we present results for other choices as well).

To further confirm our results about the MIPT, we present two instances of the characteristic single-qubit purification time $\tau_P$ in Fig.~\ref{supfig:2}(a,b) for $\alpha=2.5$ (volume-to-area transition) and $\alpha=1.75$ (volume-to-power-law transition). The dynamical critical exponent $z$, the critical point $p_c$, and the correlation length exponent $\nu$ are again obtained through the minimization of a cost function~\cite{Kawashima1993}. Importantly, the analysis of $I_3$ and $\tau_P$ are independent and compatible, providing a robust crosscheck of the identified universality classes. 
A summary for the critical exponents  $\nu$ and $z$ is given in Fig.~\ref{supfig:2}(c), where we plot the critical exponents changing $\alpha$. The dashed line presents the conformal field theory expectation $z=1$ and the value $\nu=1.27$ observed in the short-range case \cite{sierant2022measurementinducedphase}, which are restored for $\alpha\ge 3$~\cite{block2022measurementinducedtransition,sharma2022measurementinducedcriticality}.

\subsection{Choice of control operation and robustness of the dynamical behavior at absorbing state phase transition}
This section discusses the criticality as encoded in the dynamics of the entanglement entropy for the absorbing state transition when it coincides with the measurement-induced transition ($\alpha\le \alpha_c \approx 1.3$) and when it does not ($\alpha> \alpha_c$).

\begin{figure}
    \centering
    \includegraphics[width=0.95\linewidth]{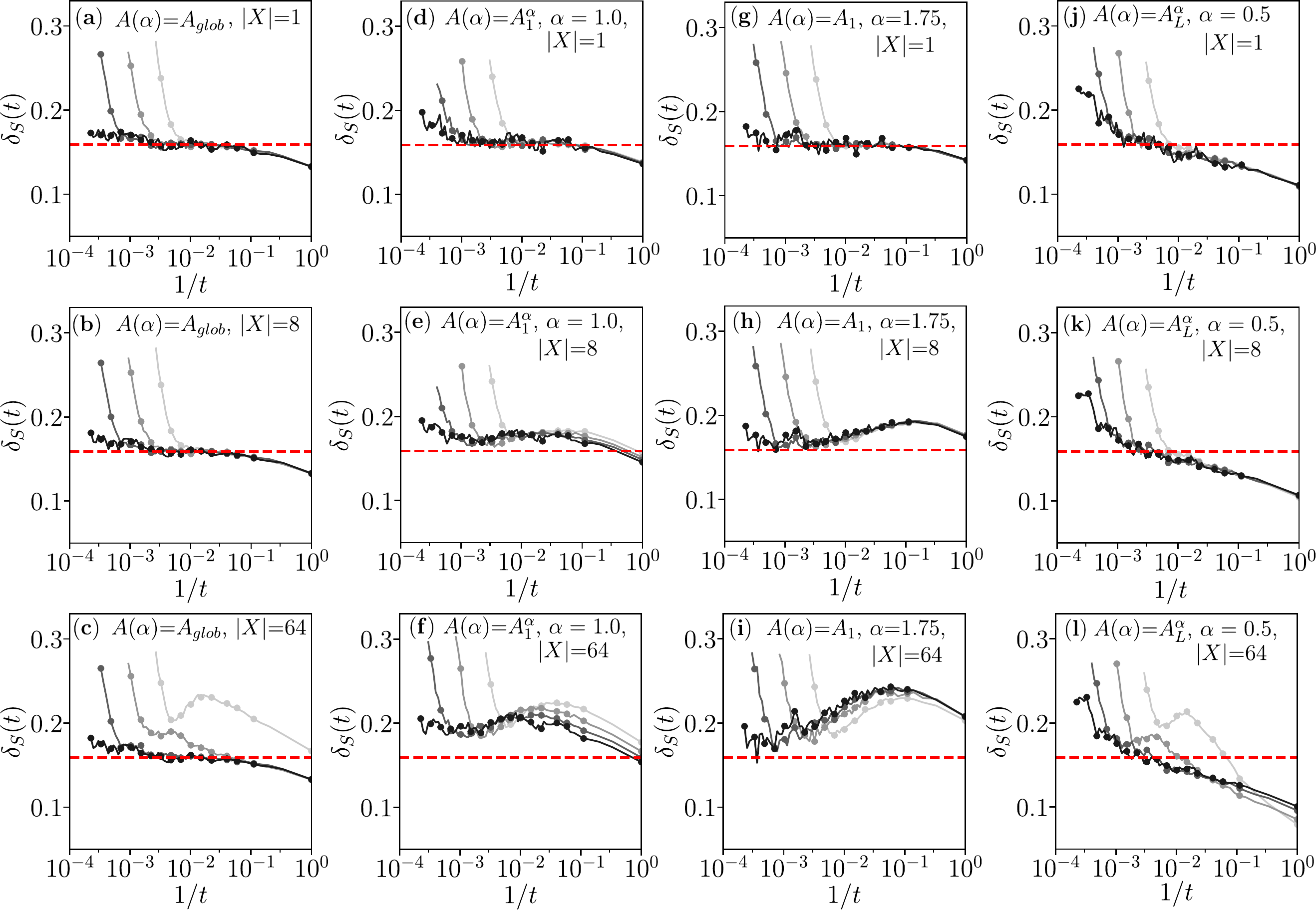}
    \caption{ Robustness of the observed behavior of the entanglement entropy, $S(t)\propto t^{-\delta_S(t)}$, at the critical point $p_c^\mathrm{APT}$ for different choices of the control operation $A(\alpha)$ and various $\alpha$. The results show the trend with increasing system sizes $L=128\div 1024$ (darker color mean increasing $L$), times $t$ and subsystem sizes $l_X$ flowing toward the expected direct percolation universality class $\delta_S(t) \to 0.1595(1)$, denoted by the red dashed line. 
    (\textbf{a,b,c}) Global gate $A_\mathrm{glob}$ acting as a random Clifford gate on the $N_\mathrm{def}$ defects. 
    The results for $A_1^\alpha$, the operation considered in the Main Text are shown for $\alpha=1.0$ in (\textbf{d,e,f}), and $\alpha=1.75$  in (\textbf{g,h,i}). Lastly, the case of $L$ two-body power-law distributed gates ($A_L^\alpha$) is considered in  (\textbf{j,k,l}) for $\alpha=0.5$. 
    It is important to note that, for all the subplots, there is a volume-law, $S\propto L t^{-\delta_S(t)}$, except for $\alpha=1.75$, where the transition is between sub-volume law ($S\propto L^\mu t^{-\delta_S(t)}$) to area-law. \label{supfig:3}
    }
\end{figure}

Specifically, we are considering $A(\alpha)$ to be either a global operation ($A(\alpha)=A_\mathrm{glob}$), or to be built out of power-law distributed $N_\mathrm{def}$ two-body gates ($A(\alpha)=A_1^\alpha$), or to be an action of $L$ two-body, power-law distributed gates ($A(\alpha)=A_L^\alpha$). 
As in the Main Text, we consider $S(t)\propto t^{-\delta_S(t)}$ at large times and compute the exponent $\delta_S(t)$. A saturating exponent represents a well-defined universality class. We recall that if $\delta_S(t)$ grows, increasing subsystem size, the entanglement entropy decays speeds up.

First, let us consider the $A(\alpha)=A_\mathrm{glob}$, which is a global $N_\mathrm{flag}$ qubit gate acting on sites at which $f_i=0$. In this case, as argued in the Main Text, MIPT and APT coincide, as do the critical exponents. 
In Fig.~\ref{supfig:3}(a,b,c), we show how increasing the system size ($L$) and subsystem size ($l_X \equiv |X|$), the exponent $\delta_S$ develops a plateau that is compatible with the  direct percolation (DP) universality class of the absorbing state transition. 
For this model, the entanglement entropy is directly proportional to the $n_\mathrm{def} \cdot l_X$, manifesting a clear volume-to-area transition for any timescale smaller than exponential in the system size. Notably, even in the simple case of the global operation $A_\mathrm{glob}$, we observe non-trivial finite size effects manifested by the approach of $\delta_S(t)$ towards the DP prediction for the largest subsystem size considered, $|X|=64$. Nevertheless, we observe that $\delta_S(t)$ converges towards the DP value for any $l_X \ll L$. This indicates that the universal DP behavior of $S(t)$ occurs at any $l_X$ in the thermodynamic limit $L\to \infty$. 

In Fig.~\ref{supfig:3}(d-i), we consider various data within the volume-to-area (d-f) and area-to-area (g-i) transitions for $A(\alpha)=A_1^\alpha$. These figures complement the analysis provided in the Main Text, where the results for $\alpha=0.5$ and $\alpha=3.5$ are shown in Fig. 3. 
Here, finite-size effects are considerably larger. Nevertheless, the exponent $\delta_S(t)$ flows towards the DP universality class prediction when the system size $L$ and time $t$ increase.

Lastly, in Fig.~\ref{supfig:3}(j-l), we consider a system with $L$ two-body gates selected through the power-law distribution $P(r)\sim r^\alpha$, acting on the unflagged sites, ($A(\alpha)=A_L^\alpha$). 
Our numerics cannot draw a conclusive answer in this case, as the trend of the critical exponent is not fully saturated with the system size. Nevertheless, the system seems to develop an intermediate plateau region at $\delta_S(t)\approx 0.1595$ for smaller subsystem sizes before decaying to the absorbing state. This agrees with our expectation that at sufficiently large times, the control operation $A_L^\alpha$ acts as a global gate on all unflagged sites. This occurs when $L$ (the number of gates in $A_L^\alpha$) is bigger than $N_{\mathrm{flag}}^2$, which is bound to happen at sufficiently late times since  $N_{\mathrm{flag}}/L \stackrel{t \to \infty}{\to} 0$ at $p=p^{\mathrm{APT}}_c$.

\end{document}